\documentclass[aps,rmp,showpacs,twocolumn,final,superscriptaddress,10pt]{revtex4-1}

\usepackage{amssymb,amsmath,latexsym,framed}
\usepackage{amsthm}
\usepackage[usenames,dvipsnames]{xcolor}
\usepackage{nicefrac}
\usepackage{enumitem}
\usepackage{graphicx}
\usepackage{mathrsfs}
\usepackage[sans]{dsfont} 
\usepackage{overpic}
\usepackage{subfigure}
\usepackage{rotating}
\usepackage{anyfontsize}
\usepackage{tcolorbox}
\usepackage[utf8]{inputenc}
\usepackage[T1]{fontenc}
\usepackage{physics}
\usepackage{epstopdf}

\usepackage[colorlinks = true,
             linkcolor = blue,
             urlcolor  = blue,
             citecolor = blue,
             anchorcolor = green,
]{hyperref}

\graphicspath{{./figure/}}

\begin{document}

\title{Secure quantum key distribution with realistic devices}

\author{Feihu Xu}
\affiliation{Hefei National Laboratory for Physical Sciences at Microscale and Department of Modern Physics, University of Science and Technology of China, Hefei 230026 China}
\affiliation{Shanghai Branch, CAS Center for Excellence and Synergetic Innovation Center in Quantum Information and Quantum Physics, University of Science and Technology of China, Shanghai 201315 China}

\author{Xiongfeng Ma}
\affiliation{Center for Quantum Information, Institute for Interdisciplinary Information Sciences, Tsinghua University, Beijing 100084 China}

\author{Qiang Zhang}
\affiliation{Hefei National Laboratory for Physical Sciences at Microscale and Department of Modern Physics, University of Science and Technology of China, Hefei 230026 China}
\affiliation{Shanghai Branch, CAS Center for Excellence and Synergetic Innovation Center in Quantum Information and Quantum Physics, University of Science and Technology of China, Shanghai 201315 China}

\author{Hoi-Kwong Lo}
\email{hklo@ece.utoronto.ca}
\affiliation{Center for Quantum Information and Quantum Control, Department of Physics and Department of Electrical \& Computer Engineering, University of Toronto, M5S 3G4 Toronto Canada}

\author{Jian-Wei Pan}
\email{pan@ustc.edu.cn}
\affiliation{Hefei National Laboratory for Physical Sciences at Microscale and Department of Modern Physics, University of Science and Technology of China, Hefei 230026 China}
\affiliation{Shanghai Branch, CAS Center for Excellence and Synergetic Innovation Center in Quantum Information and Quantum Physics, University of Science and Technology of China, Shanghai 201315 China}

\date{\today}

\begin{abstract}
In principle, quantum key distribution (QKD) offers information-theoretic security based on the laws of physics. In practice, however, the imperfections of realistic devices might introduce deviations from the idealized models used in security analyses. Can quantum code-breakers successfully hack real systems by exploiting the side channels? Can quantum code-makers design innovative counter-measures to foil quantum code-breakers? This article reviews theoretical and experimental progress in the practical security aspects of quantum code-making and quantum code-breaking. After numerous attempts, researchers now thoroughly understand and are able to manage the practical imperfections. Recent advances, such as the measurement-device-independent protocol, have closed the critical side channels in the physical implementations, paving the way for secure QKD with realistic devices.
\end{abstract}

\maketitle

\tableofcontents

\bigskip


\section{Introduction}\label{sec:introduction}

\subsection{Secure communication}\label{sec1:securecomm}
For thousands of years, code-makers and code-breakers have been fighting for supremacy. With the recent rise of Internet of Things, cyber security has become a hot topic. Cyber warfare that can undermine the security of critical infrastructures, such as smart power grids and financial systems, threatens the well-being of individual countries and the global economy.

In conventional cryptography, two distant parties, traditionally called Alice and Bob, share a communication channel and they would like to communicate privately in the presence of an eavesdropper, Eve. The Holy Grail of secure communication is information-theoretical security. It is known that one could achieve information-theoretically secure communication via the one-time-pad (OTP) method~\cite{vernam1926cipher}, if the two users, Alice and Bob, share a long random string that is kept secret from Eve. Note that, for the OTP scheme to be information-theoretically secure, it is
important not to re-use the key~\cite{shannon1949communication}. That is to say that the key has to be
as long as the message itself and can only be used once. How to distribute such a long key in the presence of Eve is called the key distribution problem. In fact, the key distribution problem is a central challenge in all kinds of encryption methods.

In principle, \emph{all} conventional key distribution schemes that rely on classical physics and mathmatics can only provide computational security, because in classical physics, there is nothing to prevent an eavesdropper from copying the key during the key distribution process. Now, if Eve and Bob have the same key, whatever Bob can decrypt, Eve can decrypt too.

Currently, the key distribution problem is often solved by public key cryptography. In public key cryptography, there are a pair of keys: a public key and a private key. An intended recipient Bob will publish the public key so that anyone, such as an intended sender Alice, can encrypt a message, called a plain text, with the public key and send the encrypted message, a cipher text, to Bob. On the other hand, only Bob with the private key can decrypt the cipher text to recover the plain text efficiently. The security of public key cryptography is based on computational assumptions. Given the public key, there is no efficient known algorithm for Eve to work out the private key or to recover the plain text, from the cipher text. For instance, the security of the best-known public key crypto-system, RSA~\cite{rivest1978method}, is based on the presumed hardness of factoring large integers. Unfortunately, public key cryptography is vulnerable to unanticipated advances
in hardware and software. Moreover, in 1994, Peter Shor then at AT\&T invented an efficient quantum algorithm for factorization~\cite{shor1999polynomial}. For this reason, if a large scale quantum computer is ever constructed, much of conventional cryptography will fall apart!

After more than two decades of intense theoretical and experimental efforts, primitive small scale quantum computers have already been built. Several big companies and a number of labs and start-ups are racing to build the world's first practical quantum computer. For instance, Google AI Quantum Laboratory\footnote{Google Q: research.google/teams/applied-science/quantum} has realized the quantum advantage (or supremacy) over state-of-the-art classical supercomputer for a specific computational task~\cite{arute2019quantum}, and plans to commercialize quantum computers within a few years~\cite{mohseni2017commercialize}; IBM Q has already put its sixteen-qubit quantum processor online for client use\footnote{IBM Q: www.research.ibm.com/ibm-q}; Rigetti has also provided the quantum cloud service\footnote{Rigetti: www.rigetti.com}; Chinese Academy of Sciences (CAS) and Alibaba have established the Quantum Computing Laboratory to advance the research of quantum computing\footnote{CAS-Alibaba: quantumcomputer.ac.cn/index.html}; Other companies, such as Intel, Microsoft, Baidu, Tencent, IonQ, Xanadu, Zapata and so forth, have also joined the international race to build a quantum computer. Moreover, China is building the National Laboratory for Quantum Information Science to support the revolutionary research in quantum information; The European Commission is planning to launch the flagship initiative on quantum technologies\footnote{ec.europa.eu/digital-single-market/en/news/quantum-europe-2017-towards-quantum-technology-flagship}; USA has already launched the National Quantum Initiative Act in 2018\footnote{www.congress.gov/bill/115th-congress/house-bill/6227}. All in all, the risk of successful construction of a quantum computer in the next decade can no longer be ignored.

Note that some data such as our DNA data and health data need to kept secret for decades. This is called \emph{long-term security}. However, cryptographic standards could take many years to change. An eavesdropper intercepting encrypted data sent in 2019 may save them for decades as they wait for the future successful construction of a quantum computer. The eavesdropper could then retro-actively successfully crack an encryption scheme, therefore cryptographic standards need to consider potential future technological advances of the next few decades. For instance, Canadian Census Data is required to be kept confidential for 92 years\footnote{www12.statcan.ca/English/census01/Info/chief.cfm} or until 2111. To ensure such security, we need to predict the future technology in the next century. As a comparison, the first general-purpose electronic computer, ENIAC, was formally dedicated in 1946, which was less than 92 years ago. This meant that general-purpose electronic computers did not even exist 92 years ago. Therefore, if history is any guide, we think that it is not realistic for one to predict with any confidence what types of technology would exist 92 years from now.

In 2015, the US National Security Agency (NSA) announced a plan for transition to quantum-safe crypto-systems. For instance, the US National Institute of Standards and Technology (NIST) has made a call for quantum-safe candidate algorithm nominations, which was due November 30, 2017\footnote{csrc.nist.gov/Projects/Post-Quantum-Cryptography}. Over the next few years, those candidate algorithms will be evaluated.

Broadly speaking, there are two approaches to a quantum-safe encryption scheme. The first approach is to use conventional cryptography and to develop alternative public-key encryption schemes, such as hash-based or code-based encryption schemes, in which known quantum attacks such as Shor's algorithm~\cite{shor1999polynomial} do not apply. This approach is called \emph{post-quantum cryptography} and it has the advantages of being compatible with existing crypto infrastructure while having high key rates that are available over long distances. Recently, Google has performed a test deployment of a post-quantum crypto algorithm in Transport Layer Security (TLS)\footnote{security.googleblog.com/2016/07/experimenting-with-post-quantum}. One drawback of post-quantum algorithms is that those conventional algorithms are only shown to be secure against \emph{known} quantum attacks. There is always a possibility that some smart conventional or quantum physicist or computer scientist might one day come up with clever algorithms for breaking them efficiently. As said, this would lead to a retroactive security breach in future for data transmitted today with potentially disastrous consequences.

The second approach is to use quantum cryptography~\cite{bennett1984quantum,ekert1991quantum},
particularly quantum key distribution (QKD). It has the advantage of promising information-theoretical security based on the fundamental laws of quantum physics, i.e., the security is independent of all future advances of algorithm or computational power.

Note however that quantum cryptography cannot replicate all the functionalities of public key cryptography. In future, quantum cryptography is likely to be combined with the post-quantum cryptography to form the infrastructure of quantum-safe encryption scheme. For instance, the post-quantum cryptography can be used to perform the initial authentication. This authentication is only required in a short time, and once it is done, the generated QKD key will be secure. Therefore, we believe that the two approaches---post-quantum cryptography and quantum cryptography---are \emph{complementary} to each other (rather than mutually exclusive).

\subsection{Quantum key distribution (QKD)}\label{sec1:QKD}
The main goal of QKD is to achieve information-theoretical security by harnessing the laws of physics~\cite{bennett1984quantum,ekert1991quantum}. The quantum no-cloning theorem dictates that an unknown quantum state cannot be cloned reliably~\cite{Wootters1982,dieks1982communication}. If Alice distributes a key via quantum (e.g., single-photon) signals, because there is only a single copy of the key to begin with, there is \emph{no} way for Eve to clone the quantum state reliably to produce two copies of the same quantum state. Therefore, if Eve tries to eavesdrop in QKD, she will unavoidably introduce disturbance to the quantum signals, which will then be detected by the users, Alice and Bob. Alice and Bob can then simply discard such a key\footnote{Note that a key is simply a random string of numbers and if a key is aborted, it will not be used. So, there is no loss in security in aborting.} and try the key distribution process again.

Note that an important advantage of QKD is that, since the communication is quantum, once a QKD session is over, there is \emph{no} classical transcript for Eve to keep. Therefore, an eavesdropper has to break a QKD session real-time or it will be secure forever. This is very different from conventional key distribution schemes.

\subsubsection{BB84 protocol}\label{sec1:BB84}
\begin{figure} [htbp]
\begin{center}
\includegraphics[scale=0.6]{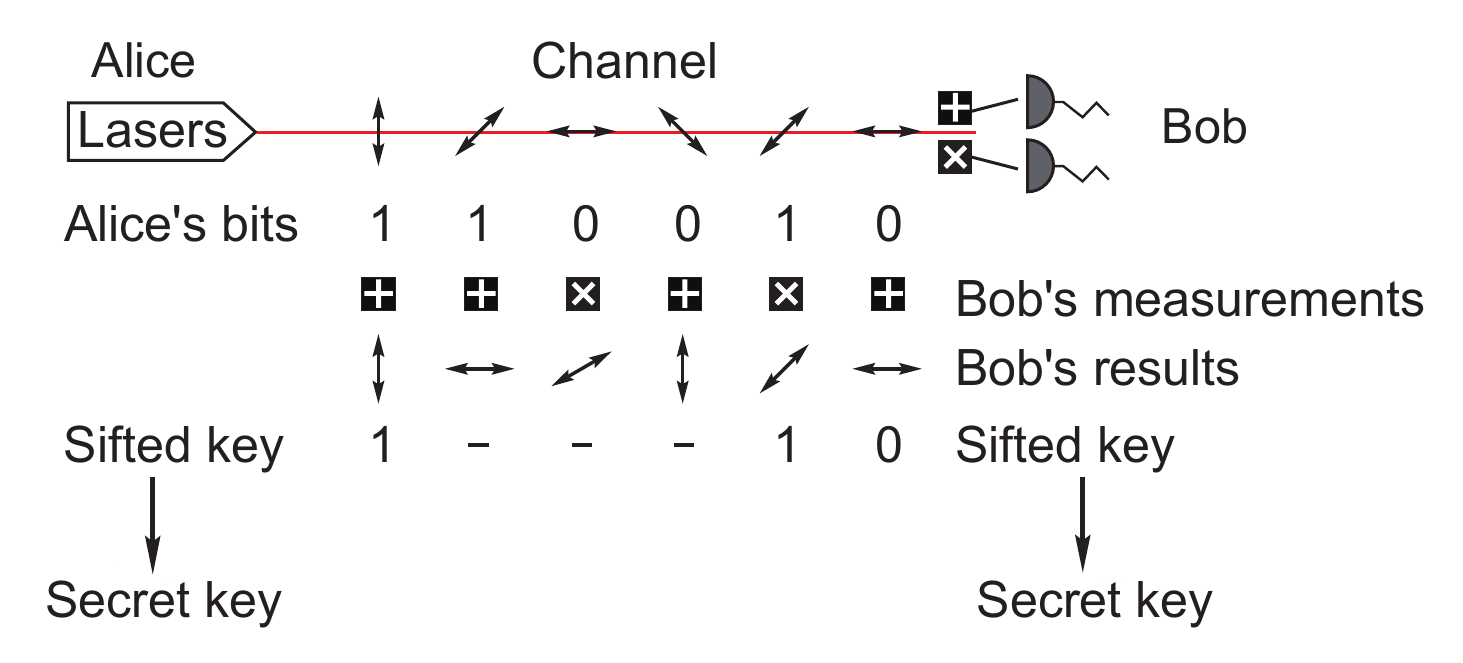}
\end{center}
\caption{Schematic diagram of the BB84 protocol. Alice encodes random bits on the polarization states of single photons. Bob randomly selects measurement bases, rectilinear ($+$) or diagonal ($\times$), to perform measurements using two detectors. They keep only those polarization data that have been encoded and detected in the same basis as the Sifted key, and perform additional classical postprocessing on the Sifted key to produce the final Secret key.
\label{Fig:bb84}}
\end{figure}

The best-known QKD scheme is the Bennett-Brassard-1984 (BB84) protocol~\cite{bennett1984quantum}. The BB84 protocol allows two users, Alice and Bob, who share a quantum channel (e.g., an optical fiber or free-space) and an authenticated conventional classical channel, to generate a secure key in the presence of an eavesdropper with unlimited quantum computing powers. In the BB84 protocol, a sequence of single photons carrying qubit states are sent by Alice to Bob through a quantum channel. A schematic diagram of the BB84 protocol is illustrated in Fig.~\ref{Fig:bb84}, and the steps of the protocol are listed in Box~\ref{tab:bb84}.

\begin{tcolorbox}[title = {Box I.B.1: BB84 protocol.}]
(1) For each signal, Alice randomly encodes a single photon with one of the four polarization states, namely, vertical, horizontal, 45-degree and 135-degree, and sends the photon through a quantum channel to Bob. \\
(2) For each signal, Bob chooses one of the two bases, rectilinear and diagonal, to perform a measurement on the polarization of a received photon. After detection, Alice and Bob publicly announces their basis choices through an authenticated conventional channel. \\
(3) Alice and Bob discard the polarization data that have been encoded and detected in different bases. They keep only those polarization data in the same basis. This remaining data forms the sifted key. Alice and Bob can choose a random sample of the sifted key bits and compare them to compute the quantum bit error rate (QBER).\\
(4) If the computed QBER is too high, they abort. Otherwise, they proceed with classical postprocessing such as error correction and privacy amplification to generate a secret key.
\end{tcolorbox}\label{tab:bb84}

\subsubsection{Intuition of security}\label{sec1:intuition}

The quantum no-cloning theorem guarantees that Eve cannot copy the unknown quantum state sent by Alice reliably~\cite{Wootters1982,dieks1982communication}. Furthermore, a key feature in quantum mechanics is the complementarity between the two conjugate bases, rectilinear and diagonal. Since the two measurements corresponding to the two bases do {\it not} commute with each other, there is no way to measure the two observables simultaneously without disturbing the state. Therefore, Eve who tries to eavesdrop and extract information on the polarization data will inevitably introduce disturbance to the state. Bob on the other hand, with the authenticated classical channel, has a fundamental advantage over Eve because he can compare his basis choice with Alice and determine the QBER for data that is encoded and detected in the same basis.

What happens if Eve attacks the quantum channel? A simple example of an eavesdropping strategy is the \emph{intercept-resend attack}~\cite{bennett1984quantum}. In this attack, for each photon sent from Alice, Eve performs a measurement in a randomly chosen basis and re-sends a new photon to Bob according to her measurement result. Let us focus on those cases when Alice and Bob happen to use the same basis since they will throw away the rest. If Eve happens to use the correct basis (50\%), then both she and Bob will decode Alice's bit value correctly. No error is introduced by Eve. On the other hand, if Eve uses the wrong basis (50\%), then both she and Bob will have random measurement results. This suggests that if Alice and Bob compare a subset of the sifted key, they will see a significant amount of errors. Here, for these bits, the photons will be passed on to Bob in the wrong basis, so regardless of Eve's measurement result, Bob will have a 50\% probability of measuring the opposite of Alice's bit value. In other words, Eve's attack will introduce 50\% QBER for half of the total bits, and thus a total of 25\% QBER. This example illustrates the basic principle behind QKD: \emph{Eve can only gain information at the cost of introducing disturbance}, which will expose her interference.

\subsubsection{Overview of recent developments} \label{sec1:overview}

\emph{Theoretical developments.} On the theoretical side, the first security proof of QKD was based on the uncertainty principle by Mayers~\cite{mayers2001unconditional}. Mayers's proof was put into a conceptually simple framework based on entanglement distillation by Lo and Chau~\cite{lo1999unconditional}, building on the earlier work of quantum privacy amplification \cite{deutsch1996quantum} and entanglement distillation \cite{Bennett1996Mixed}. Later on, Shor and Preskill employed the idea of the Calderbank-Shor-Steane (CSS) quantum error correcting code \cite{PhysRevA.54.1098,PhysRevA.54.4741} to simplify the entanglement-based proof to a prepare-and-measure protocol~\cite{shor2000simple}. See also~\cite{biham2000proof,Devetak2005Distillation,koashi2009simple} for security proofs of QKD.

Rather interestingly, the rigorous definition of secure keys was presented afterwards in 2000s \cite{ben2005universal,Renner2005Security}, where the composable security definition in conventional cryptography \cite{Canetti2001} was introduced to quantum cryptography \cite{ben2005universal}. A further development was the security proof for the consideration of finite-key effects in a more rigorous manner~\cite{renner2008security,scarani2008quantum,tomamichel2012tight}.

Device imperfections in practical systems were investigated in security analyses \cite{lutkenhaus2000security,inamori2007unconditional}, and the remarkable framework of the security analysis for realistic devices was established by Gottesman-Lo-L\"{u}tkenhaus-Preskill (GLLP) \cite{gottesman2004security}. Moreover, new protocols, such as the decoy-state~\cite{hwang2003quantum,lo2005decoy,wang2005beating}, differential-phase shift (DPS)~\cite{DPS:2002}, SARG-04~\cite{Scarani2004}, coherent-one-way (COW)~\cite{stucki2005fast}, measurement-device-independent (MDI)~\cite{lo2012measurement} [see also~\cite{braunstein2012side}] and round-robin DPS~\cite{sasaki2014practical}, were proposed to address the issues of device imperfections. In particular, the decoy-state protocol enables secure QKD with weak coherent pulses and the MDI protocol removes all side channels in the detection. Furthermore, device-independent QKD was proposed~\cite{mayers1998quantum,barrett2005no,acin2007device} to allow QKD with uncharaterized devices. Its security was proven effective against collective attacks~\cite{pironio2009device,masanes2011secure} [see also~\cite{hanggi2010efficient}] and later against general attacks~\cite{vazirani2014fully,Arnon2018Practical}.

\emph{Experimental developments.} After more than two decades of efforts~\cite{gisin2002quantum,Lo2014NP}, QKD developments include the first laboratory demonstration performed in 1992 over 32.5-cm free space~\cite{bennett1992experimental}, to the recent landmark accomplishment of quantum satellite QKD experiment in 2017 over 1200 km by China~\cite{Liaosate}, and 7600 km in 2018 between China and Austria~\cite{liao2018satellite}. Note that this is a seven order of magnitude of improvement in terms of the distance of QKD. There are also on-going efforts on satellite-based quantum communications by Europe, USA, Canada, Japan, and Singapore~\cite{joshi2018space}. In fiber, the distance has been pushed to 500-km ultra-low loss fiber~\cite{Fang2019surpassing,chen2019sending}.

In addition to long distances, high secret key rate is important for practical applications. Researchers have recently pushed the secret key rate of QKD from 1 Mbits/s over 50-km fiber~\cite{Luca2013} to more than 10 Mbits/s~\cite{Yuan:18,islam2017provably}. Commercial QKD systems are currently available on the market by several companies such as ID Quantique, Quantum CTek, Qasky and Toshiba Europe. Several institutes, e.g., European Telecommunications Standards Institute (ETSI), International Organization for Standardization (ISO), and International Telecommunication Union (ITU), have made great efforts to address the standardization issues in QKD.

Besides point-to-point link, a number of field-test QKD networks have been conducted in USA~\cite{elliott2005current}, Europe~\cite{peev2009secoqc,stucki2011long}, Japan~\cite{sasaki2011field}, China~\cite{chen2009field,chen2010metropolitan,wang2010field}, UK~\cite{dynes2019cambridge} and so forth. Based on trustful relays\footnote{In trusted-relay scenario, Alice and Bob respectively share a secret key with a relay in the middle, and then the relay announces the XOR results of both keys publicly. With the announced result, Alice and Bob can get each other's key via the XOR with her/his own key. The negative side for this method is that the relay must be trustful. However, the positive side is reducing the cost and complexity as compared to the all-connected point-to-point links, and extending the transmission distance.}, remote users can be connected beyond point-to-point links. Recently, China has successfully completed the 2000-km-long fiber-optic backbone link between Beijing to Shanghai~\cite{Yuao2019}. UK has launched the Quantum Communications Hub project that aims to build quantum networks in England\footnote{www.quantumcommshub.net/about-us/}. US is deploying their first dark fiber quantum network connecting Washington DC to Boston over 800-km\footnote{techcrunch.com/2018/10/25/new-plans-aim-to-deploy-the-first-u-s-quantum-network-from-boston-to-washington-dc/}.

\begin{figure*}
\begin{center}
\includegraphics[scale=0.5,angle=0]{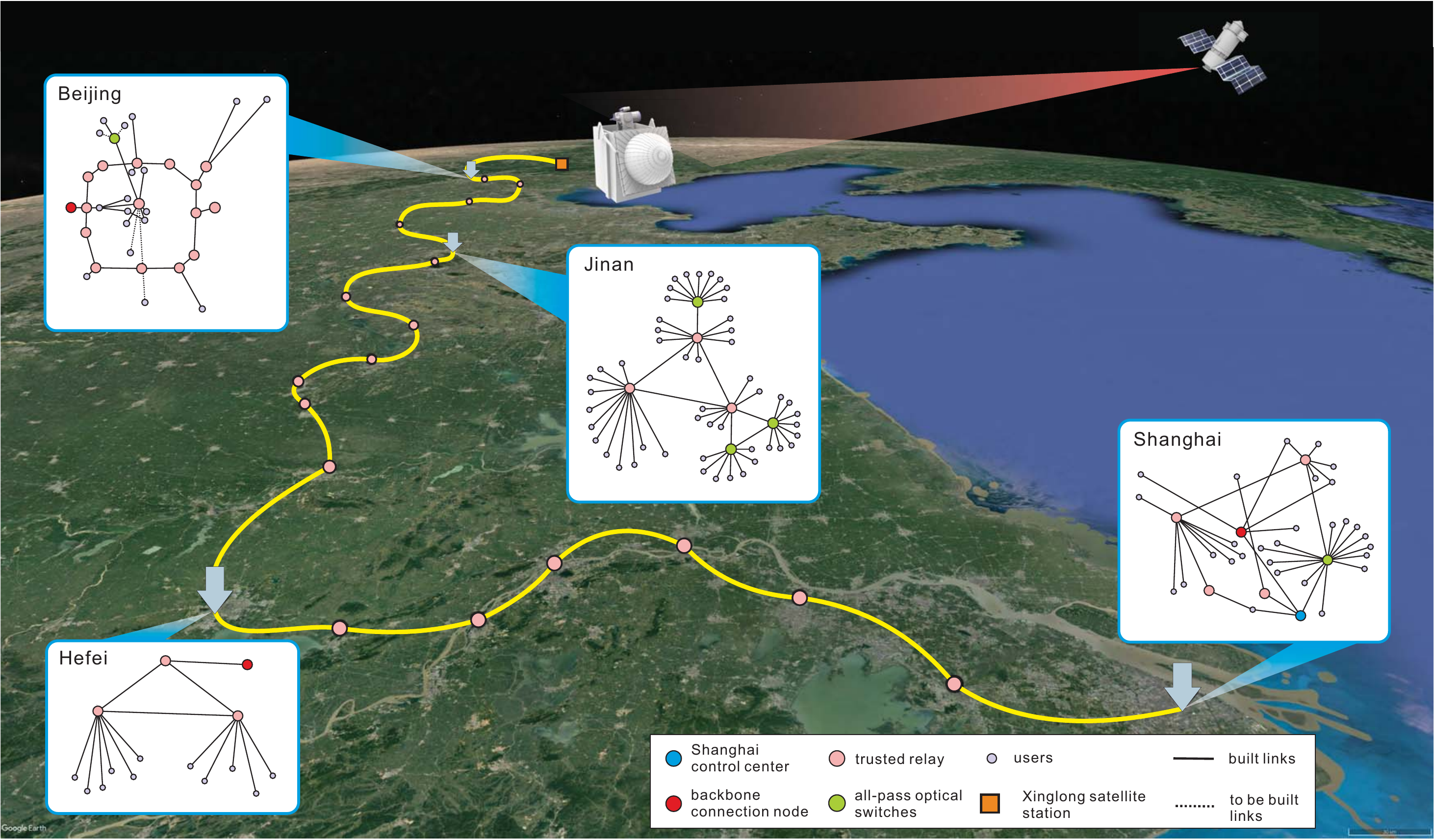}
\end{center}
\caption{(Color online) Schematic diagram of the space-ground integrated quantum network in China~\cite{Yuao2019}, consisting of four quantum metropolitan area networks in the cities of Beijing, Jinan, Shanghai, and Heifei, a backbone network over 2000 km, and ground-satellite links. There are three types of nodes in the network: user nodes, all-pass optical switches, and trusted relays. The backbone network is connected by trusted intermediate relays. The satellite is connected to a ground satellite station near Beijing, which can provide ultralong distance communications~\cite{liao2018satellite}.}\label{Fig:backbone}
\end{figure*}

Overall, QKD is already mature for several real-life applications~\cite{qiu2014quantum}. For instance, QKD was used to encrypt security communications in the 2007 Swiss election and the 2010 World Cup. In China, QKD is being widely used to ensure long-term security for numerous users in government, financial and energy industry~\cite{Yuao2019}, including the People's Bank of China, the China Banking Regulatory Commission, and the Industrial and Commercial Bank of China. Figure~\ref{Fig:backbone} shows a schematic diagram of the space-ground integrated quantum network~\cite{Yuao2019}, constructed already in China, which spans more than 2000~km coverage area and has more than 600 QKD links.


\subsection{Focus of this review}\label{sec1:motivation}
In the Code Book by Simon Singh~\cite{singh2000code}, the author boldly proclaimed that quantum cryptography achieves the Holy Grail of cryptography by offering unconditional security. Therefore, quantum cryptography presents the final stage of evolution of cryptography. After quantum cryptography, cryptography will no longer continue to evolve. Is this really true?

In principle, QKD promises unconditional security based on the laws of physics. In practice, however, the realistic devices display imperfections, which might seldom conform to idealized theoretical models used in the security analysis by theorists. The deviations might also be vulnerable by some special attacks, i.e., quantum hacking. For this reason, an arms race has been going on in quantum cryptography among quantum code-makers and quantum code-breakers. The main goal is to assess the deviations between the system and the ideal, thus establishing the \emph{practical security} for real QKD systems.

Table~\ref{Tab1} summarizes the quantum hacking strategies developed in the last two decades. See also~\cite{jain2016attacks} for an earlier review on the subject. Right after the QKD security proofs, in which ideal devices were presented, a well-known hacking strategy was proposed --- photon number splitting (PNS) attack~\cite{lutkenhaus2000security,brassard2000limitations} that targets practical QKD source. The source device imperfection severely undermines the performance of a QKD system, typically below 30-km fiber~\cite{lutkenhaus2000security,gottesman2004security,Ma2006low}. In order to close this side channel for a QKD source, the decoy state method has been proposed by quantum code-makers to make QKD practical with standard weak coherent pulses (WCPs) that are generated by attenuated lasers~\cite{hwang2003quantum,lo2005decoy,wang2005beating}. Decoy-state QKD presents dramatic performance improvement over the conventional security proofs~\cite{gottesman2004security}, and it has become a standard technique in current QKD experiments. Table~\ref{Tab2} provides a list of decoy-state QKD experiments.

\begin{table*}[ht!]
\scriptsize
\caption{List of quantum hacking strategies.} \label{Tab1}
\begin{tabular}{ l @{\hspace{0.3cm}} l  @{\hspace{0.3cm}} l  @{\hspace{0.3cm}} l @{\hspace{0.3cm}} l }
  \hline \hline
   \textbf{Attack} & \textbf{Source/Detection} & \textbf{Target component} & \textbf{Manner} & \textbf{Year} \\
   \hline
Photon-number-splitting~\cite{lutkenhaus2000security,brassard2000limitations} & Source & WCP (multi-photons) & Theory & 2000 \\
Detector fluorescence~\cite{kurtsiefer2001breakdown} & Detection & Detector & Theory & 2001 \\
Faked-state~\cite{makarov2005faked,Makarov2006} & Detection & Detector & Theory & 2005 \\
Trojan horse~\cite{vakhitov2001large,gisin2006trojan} & Source\&Detection & Backflection light & Theory & 2006 \\
Time shift~\cite{Qi2007,Zhao2008} & Detection & Detector & Experiment$^{*}$ & 2007 \\
Time side-channel~\cite{lamas2007breaking} & Detection & Timing information & Experiment & 2007 \\
Phase remapping~\cite{Fred2007,xu2010} & Source & Phase modulator & Experiment$^{*}$ & 2010 \\
Detector blinding~\cite{makarov2009controlling,Lars2010} & Detection & Detector & Experiment$^{*}$ & 2010 \\
Detector blinding~\cite{gerhardt2011full,gerhardt2011experimentally} & Detection & Detector & Experiment & 2011 \\
Detector control~\cite{lydersen2011controlling,wiechers2011after} & Detection & Detector & Experiments & 2011 \\
Faraday mirror~\cite{Sun2011}& Source & Faraday mirror & Theory & 2011\\
Wavelength~\cite{Li2011,huang2013quantum}& Detection & Beam-splitter & Experiment & 2011\\
Dead-time~\cite{Wei2011} & Detection & Detector & Experiment & 2011\\
Channel calibration~\cite{Jain2011}& Detection & Detector & Experiment$^{*}$ & 2011\\
Intensity~\cite{Jiang2012,Shihan2015} & Source & Intensity modulator & Experiment & 2012 \\
Phase information~\cite{Sun2012,Tang2013,Sun2015} & Source & Phase randomization & Experiment & 2012 \\
Memory attacks~\cite{barrett2013memory} & Detection & Classical memory & Theory & 2013 \\
Local oscillator~\cite{jouguet2013preventing,ma2013local}$^{**}$ & Detection & Local oscillator & Experiment & 2013 \\
Trojan horse~\cite{jain2014trojan,jain2015risk} & Source\&Detection & Backflection light & Experiment & 2014 \\
Laser damage~\cite{Audun2014,Vadim2016} & Detection & Detector & Experiment & 2014\\
Laser seeding~\cite{Sun2015} & Source & Laser phase/intensity & Experiment & 2015 \\
Spatial mismatch~~\cite{sajeed2015security,chaiwongkhot2019eavesdropper} & Detection & Detector & Experiment & 2015 \\
Detector saturation~\cite{qin2016quantum}$^{**}$ & Detection & Homodyne detector & Experiment & 2016 \\
Covert channels~\cite{curty2017quantum} & Detection & Classical memory & Theory & 2017 \\
Pattern effect~\cite{yoshino2018quantum} & Source & Intensity modulator & Experiment & 2018 \\
Detector control~\cite{qian2018hacking} & Detection & Detector & Experiment & 2018 \\
Laser seeding~\cite{Sun2015,huang2019laser,pang2019hacking} & Source & Laser & Experiment & 2019 \\
Polarization shift~\cite{wei2019implementation} & Detection & SNSPD & Experiment & 2019 \\
   \hline \hline
$*$Demonstration on commercial QKD system \\
$**$Continuous-variable QKD \\
\end{tabular}
\end{table*}

\begin{table*}[htb!]
\scriptsize
\caption{List of decoy-state QKD experiments and their performance.} \label{Tab2}
\begin{tabular}{l @{\hspace{0.3cm}} l  @{\hspace{0.3cm}} l @{\hspace{0.3cm}} l @{\hspace{0.3cm}} l @{\hspace{0.3cm}} l @{\hspace{0.3cm}} l }
  \hline \hline
  \textbf{Reference} & \textbf{Clock rate} & \textbf{Encoding} & \textbf{Channel} & \textbf{Maximal distance} & \textbf{Key rate (bps)} & \textbf{Year} \\
  \hline
  \cite{Yi2006,zhao2006simulation} & 5MHz & Phase & Fiber & 60km & 422.5 & 2006 \\
  \cite{Peng2007} & 2.5MHz & Polarisation & Fiber & 102km & 8.1 & 2007\\
  \cite{rosenberg2007long} & 2.5MHz & Phase & Fiber & 107km & 14.5 & 2007\\
  \cite{Tobias2007} & 10MHz & Polarisation & Free-space & 144km & 12.8$^{\ast}$ & 2007\\
   \cite{yuan2007unconditionally} & 7.1MHz & Phase & Fiber & 25.3km & 5.5K & 2007\\
  \cite{Yin2008} & 1MHz & Phase & Fiber & 123.6km & 1.0 & 2008\\
  \cite{Qin2008}$^{\ast\ast}$ & 0.65MHz & Phase & Fiber & 25km & 0.9 & 2008\\
  \cite{Dixon2008} & 1GHz & Phase & Fiber & 100.8km & 10.1K & 2008\\
  \cite{peev2009secoqc} & 7MHz & Phase & Fiber network & 33km & 3.1K & 2009\\
  \cite{Rose2009} & 10MHz & Phase & Fiber & 135km & 0.2 &2009\\
  \cite{Yuan2009} &1.036GHz & Phase & Fiber & 100Km & 10.1K & 2009\\
  \cite{chen2009field} & 4MHz & Phase& Fiber network & 20km & 1.5K & 2009\\
  \cite{liu2010decoy} & 320MHz & Polarisation &Fiber & 200km & 15.0 & 2010 \\
  \cite{chen2010metropolitan} & 320MHz & Polarisation & Fiber network & 130km & 0.2K &2010 \\
  \cite{sasaki2011field} & 1GHz & Phase & Fiber network & 45km & 304.0K & 2011\\
  \cite{Wang2013} & 100MHz & Polarisation & Free space & 96km & 48.0 &2013\\
  \cite{Bernd2013} & 125MHz & Phase & Fiber network & 19.9km &43.1K & 2013\\
  \cite{Luca2013} & 1GHz & Phase & Fiber & 80km &120.0K &2013\\
  \cite{frohlich2017long} & 1GHz& Phase & Fiber & 240km$^{\ddagger}$ & 8.4 & 2017\\
  \cite{Liaosate} & 100MHz & Polarisation & Free space& 1200km & 1.1K & 2017\\
  \cite{Yuan:18} & 1GHz & Phase & Fiber & 2dB & 13.7M & 2018\\
  \cite{boaron2018secure} & 2.5GHz & Time-bin & Fiber & 421km$^{\ddagger}$ & 6.5 & 2018\\
   \hline \hline
$^{\ast}$Asymptotic key rate\\
$^{\ast\ast}$Heralded single-photon source\\
$^{\ddagger}$Ultralow loss fiber \\
\end{tabular}
\end{table*}

After the decoy-state method, however, various quantum hacking attacks have been performed by quantum code-breakers against other components in practical QKD systems (see Table~\ref{Tab1}). To counter those attacks, a few important concepts have been proposed by quantum code-makers. One practical counter-measure against quantum hacking is the measurement-device-independent quantum key distribution (MDI-QKD)~\cite{lo2012measurement} [see also~\cite{braunstein2012side}]. MDI-QKD completely removes all security loopholes in the detection system and ensures a QKD network security with \emph{untrusted} relays. It is practical with current technology. Table~\ref{Tab3} summarizes the MDI-QKD experiments after its invention.

In addition, an efficient version of MDI-QKD --- twin-field (TF) QKD --- has the potential to greatly extend the secure distance. Table~\ref{Tab:tfqkd} summarizes the recent TF-QKD experiments. Also, we note some of the recent developments of continuous-variable (CV) QKD (see Table~\ref{Tab:cvqkd}), chip-based QKD (see Table~\ref{Tab:chipqkd}), and other QKD protocols and implementations (see Table~\ref{Tab:otherqkd}). We also summarize a list of some developments of recent quantum cryptographic protocols, besides QKD (see Table~\ref{Tab5}).

\begin{table*}[ht!]
\scriptsize
\caption{List of MDI-QKD experiments and their performance.} \label{Tab3}
\begin{tabular}{l @{\hspace{0.3cm}} l  @{\hspace{0.3cm}} l @{\hspace{0.3cm}} l @{\hspace{0.3cm}} l @{\hspace{0.3cm}} l @{\hspace{0.3cm}} l }
  \hline \hline
   \textbf{Reference} & \textbf{Clock rate} & \textbf{Encoding} & \textbf{Distance/loss} & \textbf{Key rate (bps)} & \textbf{Year} & \textbf{Notes} \\
   \hline
   \cite{Rub2013}$^{\ddagger}$ & 2MHz & Time-bin & 81.6km & $0.24^{\ast}$ & 2013 & Field-installed fiber \\
   \cite{Liu2013} & 1MHz & Time-bin & 50km & $0.12$ & 2013 & First complete demonstration \\
   \cite{Silva2013}$^{\ddagger}$ & 1MHz & Polarisation & 17km & $1.04^{\ast}$ & 2013 & Multiplexed synchronization\\
    \cite{Tang2014} & 0.5MHz & Polarisation & 10km & $4.7\times10^{-3}$ & 2014 & Active phase randomization\\
    \cite{yanlin2014} & 75MHz & Time-bin & 200km & $0.02$ & 2014 & Fully automatic system\\
    \cite{Yanlin2015} & 75MHz & Time-bin& 30km & $16.9$ & 2015 & Field-installed fiber\\
    \cite{Wang2015} & 1MHz & Time-bin & 20km & $8.3^{\ast}$ & 2015 & Phase reference free\\
    \cite{valivarthi2015measurement} & 250MHz & Time-bin & 60dB & $5\times10^{-2}$ & 2015 & Test in various configurations\\
    \cite{pirandola2015high}$^{\ddagger}$ & 10.5MHz & Phase & 4dB & 0.1 & 2015 & Continuous variable\\
     \cite{Yanlin2016} & 75MHz & Time-bin & 55km & $16.5$ & 2016 & First fiber network \\
      \cite{Yin2016} & 75MHz & Time-bin & 404km & $3.2\times10^{-4}$ & 2016 & Longest distance \\
      \cite{tang2016experimental} & 10MHz & Polarisation & 40km & 10 & 2016 & Include modulation errors \\
      \cite{Comandar}$^{\ddagger}$ & 1GHz & Polarisation & 102km & 4.6K & 2016 & High repetition rate\\
      \cite{kaneda2017quantum}$^{\ddagger}$ & 1MHz & Time-bin & 14dB & 0.85 & 2017 & Heralded single-photon source\\
      \cite{wang2017measurement} & 1MHz & Time-bin & 20km & $6.3\times10^{-3}$ & 2017 & Stable against polarization change \\
      \cite{valivarthi2017cost} & 20MHz & Time-bin & 80km & $100$ & 2017 & Cost-effective implementation \\
      \cite{liu2018polarization} & 50MHz & Time-bin & 160km & $2.6^{\ast}$ & 2018 & Phase reference free  \\
      \cite{liu2018experimental} & 75MHz & Time-bin & 100km & $14.5$ & 2019 & Asymmetric channels  \\
      \cite{wei2019high} & 1.25GHz & Polarisation & 20.4 dB & 6.2K & 2019 & Highest repetition/key rate\\
   \hline \hline
$^{\ast}$Asymptotic key rate \\
$^{\ddagger}$No random modulations \\
\end{tabular}
\end{table*}

\begin{table}[ht!]
	\scriptsize
	\caption{List of TF-QKD experiments.} \label{Tab:tfqkd}
	\begin{tabular}{l @{\hspace{0.2cm}}  l @{\hspace{0.3cm}} l @{\hspace{0.2cm}} l @{\hspace{0.2cm}} l }
		\hline \hline
		\textbf{Reference}  & \textbf{Distance/loss} & \textbf{Key rate (bps)} & \textbf{Year} \\
		\hline
		\cite{minder2019experimental} & 90.8dB & 0.045$^\ast$ & 2019  \\
		\hline
		\cite{wang2019beating} & 300km & 2.01k$^\ast$ & 2019 \\
		\hline
		\cite{LiuTF2019} & 300km & 39.2 & 2019  \\
		\hline
		\cite{ZhongTF2019} & 55.1dB &  25.6$^\ast$ & 2019  \\
		\hline
		\cite{Fang2019surpassing} & 502km$^{\ddagger}$ & 0.118 & 2019  \\
		\hline
		\cite{chen2019sending} & 509km$^{\ddagger}$ & 0.269 & 2019 \\
		\hline \hline
		$^{\ast}$Asymptotic key rate \\
        $^{\ddagger}$Ultra-low loss fiber \\
	\end{tabular}
\end{table}

\begin{table*}[ht!]
\scriptsize
\caption{List of some recent CV-QKD experiments and their performance.} \label{Tab:cvqkd}
\begin{tabular}{l @{\hspace{0.3cm}} l @{\hspace{0.3cm}} l @{\hspace{0.3cm}} l @{\hspace{0.3cm}} l @{\hspace{0.3cm}} l }
  \hline \hline
   \textbf{Reference} & \textbf{Clock rate} & \textbf{Distance/loss} & \textbf{Key rate (bps)} & \textbf{Year} & \textbf{Notes} \\
   \hline
   \cite{jouguet2013experimental} & 1MHz & 80.5km & $\sim$250 & 2013 & Full implementation   \\
   \hline
   \cite{qi2015generating} &25MHz  & --- & --- & 2015 & Local LO   \\
   \hline
   \cite{soh2015self} & 250KHz & --- & --- & 2015 & Local LO   \\
   \hline
   \cite{huang2015high} & 100MHz & 25km & 100K & 2015 &  Local LO  \\
   \hline
   \cite{pirandola2015high} & 10.5MHz & 4dB & 0.1  & 2015 & CV MDI-QKD  \\
   \hline
   \cite{Huang:15} & 50MHz & 25km & $\sim$1M & 2015 & High key rate   \\
   \hline
   \cite{Kumar_2015} & 1MHz & 75km &490 & 2015 &  Coexistence with classical  \\
   \hline
   \cite{Zhang_2019} & 5MHz & 50km& 5.8K & 2019 & Field test   \\
   \hline
   \cite{zhang2020long} & 5MHz & 202.8km$^{\ddagger}$ & 6.2 & 2020 & Long distance   \\
   \hline \hline
   $^{\ddagger}$Ultra-low loss fiber \\
\end{tabular}
\end{table*}

\begin{table*}[ht!]
\scriptsize
\caption{List of chip-based QKD experiments.} \label{Tab:chipqkd}
\begin{tabular}{l @{\hspace{0.3cm}} l @{\hspace{0.3cm}} l @{\hspace{0.3cm}} l @{\hspace{0.3cm}} l @{\hspace{0.3cm}} l }
  \hline \hline
   \textbf{Reference} & \textbf{Clock rate} & \textbf{Distance/loss} & \textbf{Key rate (bps)} & \textbf{Year} & \textbf{Notes} \\
   \hline
   \cite{ma2016silicon} &10MHz  & 5km & 0.95K & 2016 & Silicon, decoy-BB84   \\
   \hline
   \cite{sibson2017chip} & 1.72GHz & 4dB & 565K & 2017 &  InP, DPS  \\
   \hline
   \cite{sibson2017integrated} & 1.72GHz & 20km & 916K & 2017 &  Silicon, COW  \\
   \hline
   \cite{bunandar2018metropolitan} & 625MHz & 43km & 157K & 2018 &  Silicon, decoy-BB84  \\
   \hline
   \cite{ding2017high} & 5KHz & 4dB & $\sim$7.5 & 2018 & Silicon, high-dimension   \\
   \hline
   \cite{zhang2019integrated} & 1MHz & 16dB & 0.14K & 2019 & Silicon, CV-QKD   \\
   \hline
   \cite{paraiso2019modulator} & 1GHz & 20dB & 270K & 2019 & InP, modulator-free \\
   \hline
   \cite{wei2019high} & 1.25GHz & 140km & 497 & 2019 & Silicon, MDI-QKD   \\
   \hline \hline
\end{tabular}
\end{table*}

\begin{table*}[ht!]
\scriptsize
\caption{List of recent experiments of other QKD protocols.} \label{Tab:otherqkd}
\begin{tabular}{l @{\hspace{0.3cm}} l @{\hspace{0.3cm}} l @{\hspace{0.3cm}} l @{\hspace{0.3cm}} l @{\hspace{0.3cm}} l }
  \hline \hline
   \textbf{Reference} & \textbf{Clock rate} & \textbf{Distance/loss} & \textbf{Key rate (bps)} & \textbf{Year} \\
   \hline
   Quantum access network~\cite{Bernd2013} & 125MHz & 19.9km & 259 & 2013  \\
   \hline
   Centric network~\cite{hughes2013network} & 10MHz & 50km & --- & 2013    \\
   \hline
   RR-DPS~\cite{guan2015experimental} & 500MHz & 53km & $\sim$118.0 & 2015   \\
   \hline
   RR-DPS~\cite{takesue2015experimental} & 2GHz & 20km & 2.0K & 2015     \\
   \hline
   RR-DPS~\cite{wang2015experimental} & 1GHz & 90km & $\sim$800 & 2015    \\
   \hline
   RR-DPS~\cite{li2016experimental} & 10KHz & 18dB & 15.5 & 2016   \\
   \hline
   High-dimension~\cite{lee2014entanglement} & 8.3MHz & --- & 456  & 2014   \\
   \hline
   High-dimension~\cite{zhong2015photon} & CW & 20km & 2.7M & 2015  \\
   \hline
   High-dimension~\cite{mirhosseini2015high} & 4KHz & --- & 6.5 & 2015    \\
   \hline
   High-dimension~\cite{sit2017high} & --- & 0.3km  & $\sim$30K & 2017   \\
   \hline
   High-dimension~\cite{islam2017provably} & 2.5GHz & 16.6dB  & 1.07M & 2017   \\
   \hline
   Coherent-one-way~\cite{korzh2015provably} & 625MHz & 307km & 3.2 & 2015    \\
   \hline
   Modulator-free~\cite{yuan2016directly} & 1GHz & 40dB & $\sim$10 & 2018   \\
   \hline \hline
\end{tabular}
\end{table*}

\begin{table*}[ht!]
\scriptsize
\caption{List of recent developments of other quantum cryptographic protocols beyond QKD.} \label{Tab5}
\begin{tabular}{l @{\hspace{0.3cm}} l  @{\hspace{0.3cm}} l}
  \hline \hline
   \textbf{Protocol} & \textbf{Theory/Experiment} & \textbf{Notes} \\
   \hline
   Noisy quantum storage~\cite{konig2012unconditional,wehner2008cryptography,damgaard2008cryptography} & Theory & Unconditional security \\
   Oblivious transfer~\cite{erven2014experimental} & Experiment & Noisy-storage model \\
   Bit commitment~\cite{ng2012experimental} & Experiment & Noisy-storage model \\
   Bit commitment~\cite{kent2012unconditionally} & Theory & Relativistic assumption \\
   Bit commitment~\cite{lunghi2013experimental,liu2014experimental} & Experiment & Relativistic assumption \\
   Bit commitment~\cite{lunghi2015practical,chakraborty2015arbitrarily,verbanis2016hour} & Experiment & Long commitment time \\
   Digital signature~\cite{clarke2012experimental} & Experiment & First demonstration  \\
   Digital signature~\cite{collins2014realization,dunjko2014quantum} & Experiment & No quantum memory \\
   Digital signature~\cite{donaldson2016experimental,yin2017experimental} & Experiment & Insecure channel \\
   Coin flipping~\cite{berlin2011experimental,pappa2014experimental} & Experiment & Loss tolerance \\
   Data locking~\cite{Fawzi2013low,seth2013quantum,lupo2014robust} & Theory & Loss tolerance \\
   Data locking~\cite{liu2016experimental,lum2016quantum} & Experiment & Loss tolerance \\
   Blind quantum computing~\cite{broadbent2009universal,barz2012demonstration} & Theory,Experiment & No quantum memory \\
   Blind quantum computing~\cite{reichardt2013classical,huang2017experimental} & Theory,Experiment & Classical clients \\
\hline  \hline
\end{tabular}
\end{table*}

Note that the side channels are common problems to \emph{any} cryptosystems, i.e., not only to quantum cryptography but also to conventional cryptographic systems. For instance, the power consumption of the CPU performing encryption and decryption and the timing of the signals are common side channels, which can threaten implementations of both quantum and conventional cryptographic systems~\cite{kocher1999differential,brumley2005remote}. Therefore, closing the side channels are essentially required in all cryptographic technologies. It is only through painstaking battle-testing that the security of a practical crypto-system could be established with confidence. The arms race between code-makers and code-breakers will continue in cryptographic systems.

Nonetheless, QKD is a physics-based crypto system and its security is working on the physical layer. Compared to the conventional mathematical-based cryptography, QKD can provide an accurate description of the physical realization of a cryptographic system, and the security can be proved based on this description. More importantly, QKD has the fundamental advantage of promising information-theoretical security, which is independent of all future advances of computational power. Furthermore, the recent advances, such as MDI-QKD, have closed the critical side channels in the detection of physical implementations, paving the way for secure QKD with realistic devices. Therefore, we believe that QKD does represent an important chapter in the history of code-making. We hope that QKD will play an important role in the quantum-safe encryption infrastructure for real applications, and it will bring us one step closer to the dream of information-theoretical security.

\subsection{Outline of this review}\label{sec1:scope}
This review will focus mainly on the practical security of realistic QKD systems. We begin with a discussion of security analysis in Section~\ref{Sc:Security} and the basic implementation of QKD in Section~\ref{Sc:Implement}. In Section~\ref{sec:3}, we review various quantum hacking attacks against QKD implementations. In Section~\ref{sec:4}, we review the security of a practical QKD source. In particular, we focus on the decoy-state protocol which is a standard method for secure QKD with attenuated lasers. In Section~\ref{sec:5}, we turn to detector security. We primarily review the MDI-QKD protocol and how it automatically foils all attacks on the detection system. In Section \ref{sec:CVQKD}, we review the developments of CV-QKD schemes and their practical security aspects. Section~\ref{sec:6} contains a review of other quantum cryptographic protocols. In Section~\ref{sec:9}, we present some concluding remarks.

For those readers who want to learn further basics of QKD, we refer to the two earlier reviews published in Review of Modern Physics, one by Gisin \emph{et al.} that introduces the basic experimental elements and systems~\cite{gisin2002quantum} and the other one by Scarani \emph{et al.} that discusses the basic security analysis tools of various QKD protocols~\cite{scarani2009security}. An early review on the first stage of development of QKD can be found in the book by Sergienko~\cite{sergienko2018quantum}. An earlier review on quantum attacks can be seen in~\cite{jain2016attacks}. A brief overview of the implementation security of QKD can be found in a survey article in~\cite{Lo2014NP} and an ETSI white paper\footnote{https://www.etsi.org/images/files/ETSIWhitePapers/} by Lucamarini \emph{et al.} A short overview of the practical challenges associated with QKD can be found in~\cite{Diamanti2016}. Moreover, the entropy uncertainty relation, an important tool to analyze the security of QKD, can be seen in~\cite{coles2017entropic}, and the quantum random number generator, a basic element in a practical QKD system, can be found in~\cite{Ma2016,herrero2017quantum}. A review on various techniques of single-photon detectors can be seen in~\cite{Hadfield:2009,zhang2015advances}. Furthermore, we may not cover too much on some important topics, but we refer the readers to other review articles on the topics of CV-QKD~\cite{weedbrook2012gaussian,diamanti2015distributing,laudenbach2018continuous}, high-dimensional QKD~\cite{xavier2020quantum}, quantum repeaters~\cite{sangouard2011quantum,pan2012multiphoton,munro2015inside}, quantum Internet~\cite{kimble2008quantum,wehner2018quantum}, Bell nonlocality and device-independent protocols~\cite{brunner2014bell}, and blind quantum computing~\cite{fitzsimons2017private}. These related review articles are summarized in Table~\ref{Tab6}.

\begin{table*}[ht!]
\scriptsize
\caption{List of related reviews to QKD.} \label{Tab6}
\begin{tabular}{l @{\hspace{1.2cm}} l }
  \hline \hline
   \textbf{Reference} & \textbf{Subject} \\
   \hline
   \cite{gisin2002quantum} & Experimental basics of QKD \\
   \cite{scarani2009security} & Theoretical basics of QKD \\
   \cite{Lo2014NP,Diamanti2016,Zhang:18} & Practical challenges of QKD \\
   \cite{jain2016attacks} & Quantum hacking attacks \\
   \cite{Xu:IEEE} & Measurement-device-independent QKD \\
   \cite{Hadfield:2009,zhang2015advances} & Single-photon detector \\
   \cite{Ma2016,herrero2017quantum} & Quantum random number generator \\
   \cite{coles2017entropic} & Entropy uncertainty relation \\
   \cite{weedbrook2012gaussian,diamanti2015distributing,laudenbach2018continuous} & Continuous-variable QKD \\
   \cite{sangouard2011quantum,pan2012multiphoton,munro2015inside} & Quantum repeaters \\
   \cite{kimble2008quantum,wehner2018quantum} & Quantum internet \\
   \cite{brunner2014bell} & Bell nonlocality/device-independent QKD \\
   \cite{fitzsimons2017private} & Blind quantum computing \\
   \cite{xavier2020quantum} & High-dimensional QKD \\
\hline  \hline
\end{tabular}
\end{table*}


\section{Security analysis} \label{Sc:Security}
We review the security aspects of QKD, including the security definition, various security proofs and implementation assumptions. We present a general framework to address device imperfections in security analysis. While we mainly focus on the widely implemented BB84 protocol, most of the results can be extended to other QKD protocols. We shall leave the MDI-QKD case in Section~\ref{sec:5:mdi}, and the DI-QKD case in Section~\ref{sec:diQKD}.

\subsection{Security definition} \label{Sc:SecurityDef}
To prove the security of QKD, one needs to define the security criteria first. Ideally, a secure key satisfies two requirements. First, the key bit strings possessed by Alice and Bob need to be identical, i.e., be \emph{correct}. Second, the key bit string should be uniformly distributed to anyone (say Eve) other than Alice and Bob, i.e., be \emph{secret}. Due to practical issues, such as the finite data size and non-ideal error correction, Alice and Bob cannot generate an ideal key. In reality, it is reasonable to allow the key to have a small failure probability $\epsilon$. For some $\epsilon_{\text{cor}}$ and $\epsilon_{\text{sec}}$, we say that the QKD protocol is $\epsilon$-secure with $\epsilon=\epsilon_{\text{cor}}+\epsilon_{\text{sec}}$, if it is $\epsilon_{\text{cor}}$-correct and $\epsilon_{\text{sec}}$-secret~\cite{ben2005universal,Renner2005Security}.

We define $K_A$ and $K_B$ (with the same length $m$) to be the key bit strings obtained by Alice and Bob, respectively. The secret key can be correlated to a quantum state $\rho_E$ held by Eve. The joint state $\rho_{ABE}$ is a \emph{classical-classical-quantum (c-c-q)} state,
\begin{equation} \label{eq:rho_ABE}
\rho_{ABE} = \sum_{k_A, k_B} Pr(k_A,k_B) \ket{k_A}\bra{k_A} \otimes \ket{k_B}\bra{k_B} \otimes \rho_E^{(k_A, k_B)},
\end{equation}
where $k_A, k_B \in \{0,1\}^{m}$ are the bit values. In particular, an ideal key state held by Alice and Bob is described by the private state,
\begin{equation} \label{eq:rho_ideal}
\rho_{ABE}^{ideal} = 2^{-m} \sum_{k} \ket{k}_A\bra{k} \otimes \ket{k}_B\bra{k} \otimes \rho_E,
\end{equation}
where $k_A=k_B=k$ implies that Alice and Bob hold the same string, and $\rho_E$ is independent of $k$, i.e., Eve has no information on the key string variable $K$.

A QKD protocol is defined to be $\epsilon_{\text{cor}}$-correct, if the probability distribution $Pr(k_A, k_B)$ of the final state $\rho_{ABE}$ in Eq.~\eqref{eq:rho_ABE} satisfies,
\begin{equation}
Pr(k_A \neq k_B) \leq \epsilon_{cor}.
\end{equation}
A QKD protocol is defined to be $\epsilon_{sec}$-secret~\cite{Renner2005Security}, if the state $\rho_{AE}$ is close in trace distance to the single-party private state $\rho_{AE}^{ideal}$
\begin{equation}
\min_{\rho_E} \dfrac{1}{2}(1-p_{\text{abort}})|| \rho_{AE} - \rho_{AE}^{ideal} ||_1 \leq \epsilon_{sec},
\end{equation}
where $p_{\text{abort}}$ is the probability that the protocol aborts, $\rho_{AE}^{ideal}\equiv 2^{-m} \sum_{s} \ket{s}_A\bra{s} \otimes \rho_E$, and $||A||_1 \equiv Tr[\sqrt{A^\dag A}]$ is the trace norm. It turns out that the security definition from the trace-distance metric owns a composable security property \cite{ben2005universal,Renner2005Security}.

In general, following the definition from Ben-Or \emph{et al.}'s work \cite{ben2005universal}, a QKD protocol can be defined to be $\epsilon$-secure, if the final distilled \emph{c-c-q} state $\rho_{ABE}$ is \emph{$\epsilon$-close} to the ideal key state $\rho_{ABE}^{ideal}$ given in Eq.~\eqref{eq:rho_ideal} with a proper chosen $\rho_E$
\begin{equation} \label{eq:tracedis}
\min_{\rho_E} \dfrac{1}{2}(1-p_{\text{abort}})|| \rho_{ABE} - \rho_{ABE}^{ideal} ||_1 \leq \epsilon.
\end{equation}

Note that if a distilled state is $\epsilon$-close to the ideal key state, then the guessing probability for Eve on the final key is also bounded by $\epsilon$. Here, we want to emphasize that one should not interpret the security parameter $\epsilon$ used in the above definition as the guessing probability. In fact, the statement, \emph{a key is $\epsilon$-close to the ideal key}, is much stronger than the statement, \emph{Eve's guessing probability on a key is bounded by $\epsilon$}. Let us show a simple example. Denote $l=-\log\epsilon$ and $l<m$. We consider an $m$-bit key $K_{bad}$, which concatenates a uniformly distributed $l$-bit string with $m-l$ bit of 0's. Obviously, this key $K_{bad}$ does not satisfy the trance-distance (statistical distance in this case since everything is classical here) $\epsilon$-security definition used in Eq.~\eqref{eq:tracedis}, because the statistical distance between $K_{bad}$ and $K_{ideal}$ is close to 1 when $m\gg l$. However, the guessing probability of Eve on the key $K_{bad}$ is bounded by $\epsilon$. Clearly, the guessing probability alone is \emph{not} a proper security parameter definition. This is a common mistake for those who are confused about the security foundation of (quantum) cryptography, see for example \cite{Yuen7403842}. This common mistake has also been pointed out and clearly explained by Renner in \cite{Renner2012Reply}.

\subsection{Security proofs} \label{sub:SecurityProof}

\subsubsection{Lo-Chau security proof}
In the Lo-Chau security proof \cite{lo1999unconditional}, the joint quantum state shared by Alice and Bob before the final key measurement is one of the Bell states,
\begin{equation} \label{eq:4Bell}
\begin{aligned}
\ket{\Phi^{\pm}} &= \frac1{\sqrt2}(\ket{00}\pm\ket{11}) \\
\ket{\Psi^{\pm}} &= \frac1{\sqrt2}(\ket{01}\pm\ket{10}).
\end{aligned}
\end{equation}
To see how security of QKD is related to entanglement, consider the case where Alice and Bob share $m$-pairs of perfect EPR pairs $\ket{\Phi^+}^{\otimes m}$. It is not hard to verify that if both of them perform the local $Z$ measurement, $M_{zz}$, on their halves of $m$ pairs, they will share the ideal key state $\rho_{ABE}^{ideal}$ in Eq.~\eqref{eq:rho_ideal}. In other words, the amount of distillable entanglement from quantum transmission would give a lower bound on the key generation rate.

The main job for a security analysis is to make sure that Alice and Bob eventually share (almost perfect) EPR pairs before they make the final $ZZ$ measurement to obtain secure key bits. The procedure to extract perfect EPR pairs from imperfect ones is called entanglement distillation \cite{Bennett1996Mixed}. The main idea of the Lo-Chau security proof lies on \emph{quantum error correction} \cite{lo1999unconditional}, which proved the security of an entanglement-based QKD protocol. Let us recap the Bennett-Brassard-Mermin-1992 (BBM92) \cite{bennett1992quantum} protocol, an entanglement version of BB84 in Box~\ref{tab:procedure}. For the simplicity of description, we assume
Alice and Bob own quantum memories, which will be removed shortly in the Shor-Preskill security proof~\cite{shor2000simple}.

\begin{tcolorbox}[title = {Box II.B.1: BBM92 protocol with quantum memories, an entanglement version of BB84.}]
(1) Alice prepares an EPR pair, $\ket{\Phi^+}$, stores one half of it locally, and sends the other half to Bob. \\
(2) Upon receiving a qubit, Bob stores the half of the EPR pair in a quantum memory. If the qubit is lost in the channel or the quantum storage fails, they discard the pair. \\
(3) Repeat the above two steps many times until Alice and Bob store $N$ pairs of qubits. \\
(4) With the help of pre-shared perfect EPR pairs, Alice and Bob apply a quantum error correcting code to correct all the errors in the $N$ pairs. \\
(5) After a random hashing test, Alice and Bob share almost perfect EPR pairs. They return the cost of pairs in the previous step and measure the rest in the local $Z$ basis to obtain the final key.
\end{tcolorbox}\label{tab:procedure}

The (quantum) random hashing test happens in the two conjugate bases separately. In each basis, Alice and Bob can compare the parities of the qubits. Comparison of each parity will cost Alice and Bob an EPR pair. Once they agree on an enough number of parities, the states are stabilized by the operations, $X\otimes X$ and $Z\otimes Z$, with a small failure probability. This step comes from the error verification in classical error correction~\cite{Fung2010Finite}. There are a few notes on this scheme.
\begin{enumerate}
\item
This scheme is source-device-independent, which means that the source can be fully untrusted~\cite{Koashi2003BasisInd}. In the first step, the state preparation can be done by Eve. Then, Eve prepares qubits pairs (designed to be EPR pairs) and sends to Alice and Bob who store the quantum states in memories. The rest steps, 4 and 5, are the same.

\item
After quantum transmission, Alice and Bob share $N$ EPR pairs. Due to channel disturbance or Eve's interference, these $N$ EPR pairs are generally imperfect and might be entangled with each other and Eve's system. Here, we consider the most general coherent attacks.

\item
In a security proof, it is crucial to evaluate the number of EPR pairs cost in Step 4.
\end{enumerate}

When Alice and Bob both measure in the local $Z$ basis, an error occurs when the outcomes are different. We call it a \emph{bit error}. Similarly, when they both measure in the $X$ basis, a \emph{phase error} occurs when the outcomes are different. Denote the bit and phase error rates to be $e_b$ and $e_p$, respectively,
\begin{equation} \label{eq:biterror}
\begin{aligned}
e_b &= \frac{\text{\# of bit errors}}{N}, \\
e_p &= \frac{\text{\# of phase errors}}{N}. \\
\end{aligned}
\end{equation}
Since we are considering the most general coherent attacks, the errors are in general not independent but correlated. Note that bit and phase errors can be defined in any two complementary bases in the qubit case. For quantum signals measured in a particular basis, where the bit error is defined, the phase error denotes the hypothetical error if these signals were measured in its complementary basis. For higher dimension cases, such definitions would be slightly trickier with more than one types of phase errors.

\begin{figure}[hbt!]
\centering
\includegraphics[width=8.5cm]{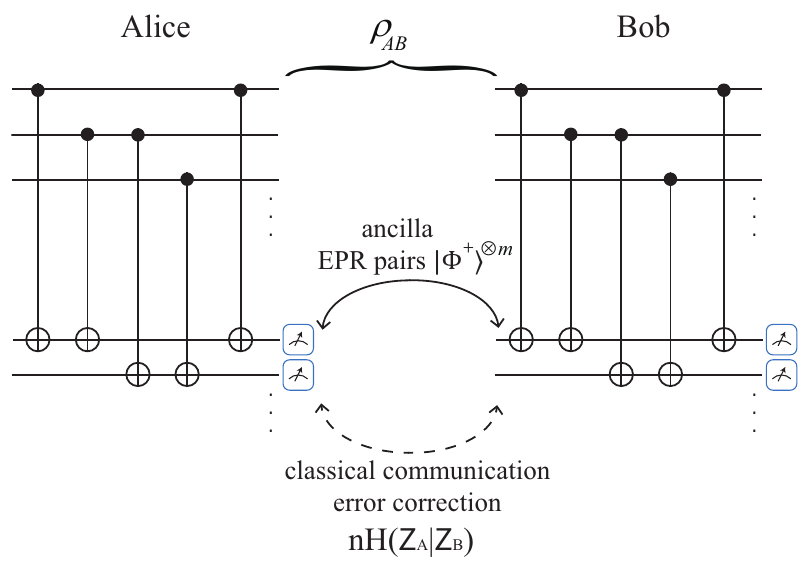}
\caption{Illustration of bit error correction. By adding Hadamard gates, the circuit can also be used for phase error correction.}
\label{fig:BitPhaseEC}
\end{figure}

In order to distill perfect EPR pairs from imperfect ones with errors defined in Eq.~\eqref{eq:biterror}, Alice and Bob can employ quantum error correction. Entanglement distillation can be done in two steps via bit and phase error correction. In bit error correction, Alice hashes her qubits in the $Z$ basis by applying \emph{Control-NOT} (C-NOT) to ancillary perfect EPR pairs, as shown in Fig.~\ref{fig:BitPhaseEC}. Alice sends the measurement results of ancillary qubits to Bob, which serves as error syndrome in error correction. In the infinite data size limit, the number of perfect EPR pairs cost in this procedure is given by the Shannon entropy, $N H(e_b)$. By applying Hadamard gates, one can switch between bit and phase spaces. Then, similarly, the phase error correction will cost additional $N H(e_p)$ EPR pairs.

Finally, the net rate of EPR pairs generated is given by,
\begin{equation} \label{eq:ShorPreskillR}
\begin{aligned}
r \ge 1-H(e_b)-H(e_p), \\
\end{aligned}
\end{equation}
where $H(e)=-e\log e-(1-e)\log(1-e)$ is the binary Shannon entropy function. Note that this formula is not tight in general. If two-way classical communication is allowed in quantum error correction, more key can be distilled \cite{Gottesman2003twoway,Chau2002twoway}.

In reality, when the data size is finite, the entanglement distillation might fail with a small failure probability, which can be understood as the failure probability of quantum error correction. In original security proofs \cite{lo1999unconditional,shor2000simple,koashi2009simple}, the fidelity between the key state $\rho_{ABE}$ to the ideal state $\rho_{ABE}^{ideal}$ is often used as an intermediate measure to finally bound the mutual information between the final key and Eve's system. In fact, this definition is not composable. In order to make the security parameter composable~\cite{ben2005universal,Renner2005Security}, one can apply the connections between fidelity and trace distance using a general inequality relating them [see Section III in~\cite{Fung2010Finite}].

\subsubsection{Shor-Preskill: reduction to prepare-and-measure schemes}
In general, this quantum error correction based entanglement distillation procedure, which is the essence of the Lo-Chau security proof, requires quantum memories and quantum computers. However, these quantum memories and quantum computers are not available with the current technology~\cite{lo1999unconditional}. In order to remove this quantum memory or quantum computer requirement, one can move the final measurement ahead of the two error correction steps. The bit error correction becomes classical error correction, and the phase error correction becomes privacy amplification \cite{shor2000simple}. There are a few steps for this permutation of operations to work.
\begin{enumerate}
\item
Quantum bit and phase error correction operations commute, as shown in Fig.~\ref{fig:BitPhaseEC}. This is due the fact that Alice and Bob use EPR pairs as ancillary qubits.

\item
The $Z$-basis measurement on the ancillary EPR qubits commutes with all the operations for error correction. This is straightforward to see since there are only two possible operations on ancillary qubits, $I$ and $X$ (from C-NOT), both of which commute with the $Z$ measurement.

\item
The $Z$-basis measurement on the Alice and Bob qubits commutes with the bit error correction. This is true since the $Z\otimes Z$ measurement commutes with C-NOT. After moving the $Z$ measurement ahead, the C-NOT operation becomes regular \emph{exclusive-OR} ($XOR$) on the two outcome bits.

\item
The $Z$-basis measurement on the Alice and Bob qubits commutes with the phase error correction. This relies on the usage of EPR pair ancillary states.

\item
In phase error correction, after locating the errors, phase error correction does not affect the values of final key measurement in the $Z$ basis. Thus, no ``correction'' operation is needed. Of course, the EPR pairs are still cost in this step.

\item
Then, all the quantum operations become classical bit operations, essentially, hashing.

\item
In order to perform privacy amplification, one still needs to estimate the phase error rate $e_p$. Now, let us focus on the case that the key bits are measured in the $Z$ basis. The phase error rate can be estimated by measuring the key bits in the $X$ basis. Of course, in order for this estimation to work, one needs to make sure the sampling is fair, which raises the critical assumptions in security proof discussed in Section \ref{sec:assumptions}.
\end{enumerate}

After considering the permutation of quantum error correction and measurement, Alice and Bob can directly measure the EPR pairs once they receive them. Suppose Alice prepares the original EPR pairs, measures halves of the pairs, and sends the rest halves to Bob. Conditioned on Alice's measurement outcomes, the states sent from Alice to Bob are pure. It is equivalent for Alice to prepare these states directly and send to Bob. Now, the entanglement-based protocol is reduced to a prepare-and-measure one.

Reduction from quantum bit error correction to classical error correction is easy to understand. Let us take Fig.~\ref{fig:BitPhaseEC} for example. Alice and Bob need to compare the ancillary qubit measurement results. Since the final $Z\otimes Z$ measurement commute with C-NOT operation, one can measure all the qubits in the $Z$ basis first and $XOR$ the bit values of all the measurement outcomes of the control qubits to the target qubits. The CNOT links shown in Fig.~\ref{fig:BitPhaseEC} can be understood as a hashing matrix, meaning it is equivalent to construct a matrix and multiply with the raw bit string. Of course, such error correction is linear. In general, any error correcting code can be applied, once bit and phase error correction can be decoupled.

\begin{figure}[hbt]
\begin{center}
\includegraphics[scale=0.55,angle=0]{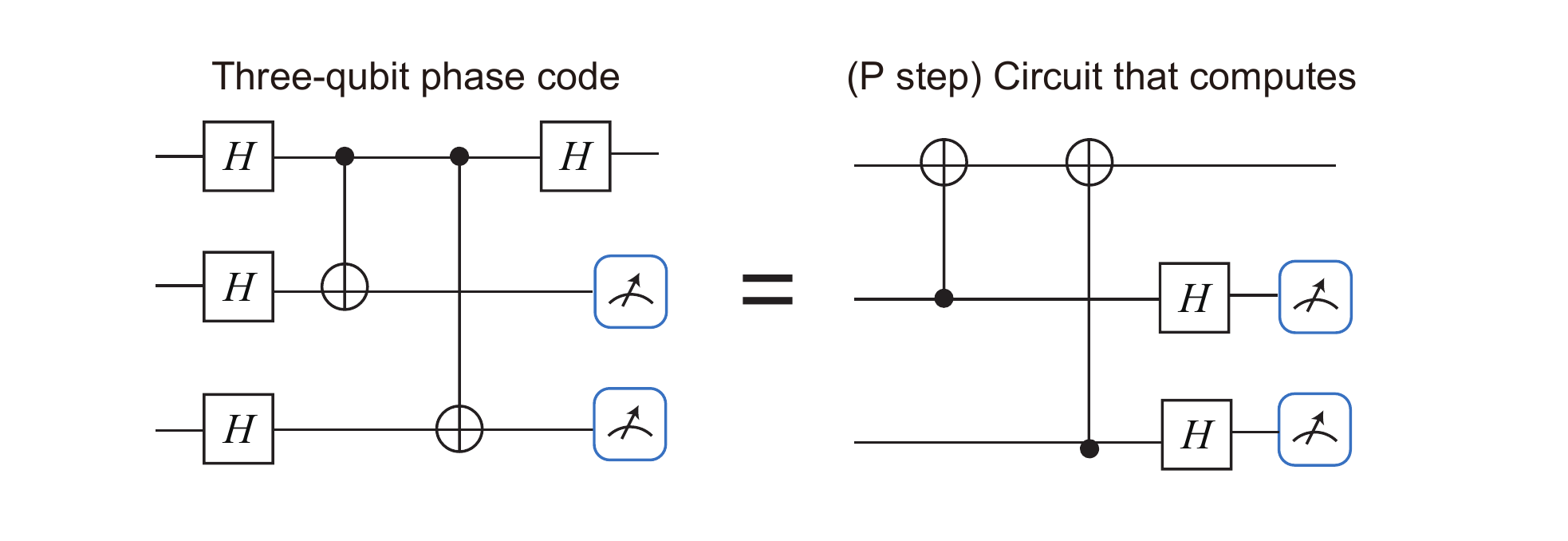}
\end{center}
\caption{Equivalence between the phase error correction and privacy amplification. The outcome of a simple 3-qubit repetition phase error correcting code is same as the hashing of three bit values.}\label{Fig:PEPA}
\end{figure}

Reduction from quantum phase error correction to privacy amplification is trickier. In general, after Hadamard gates, C-NOT operation does not commute $Z$ measurement any more. In fact, those two operations become anti-commute. In this case, Alice and Bob can design phase error correcting code such that it commutes with the $Z$ measurement. Again, let us take the linear code as an example. Certain number of parity bits need to be exchanged for error correction. Assuming universal hashing, Alice sends $N H(e_p)$ bits to Bob and Bob corrects the phase errors. Note that final key measurement must commute with this hashing. Then, they can use the null space of the hashing matrix as for the final key space. The equivalence between the phase error correction and privacy amplification is illustrated in Fig.~\ref{Fig:PEPA}. This can also be understood as a random number extraction. Alice and Bob use phase error rate to estimate the randomness in the key and apply universal hashing to extract out true randomness.

In quantum error correction, we assume Alice and Bob to use ancillary EPR pairs.  As shown in Fig.~\ref{fig:BitPhaseEC}, with EPR pairs, bit and phase error correction operations commute with each other. That is, one can decouple these two error correction steps \cite{Lo2003Method}. In the Shor-Preskill security proof \cite{shor2000simple}, no ancillary EPR pairs are employed. Instead, the CSS quantum error correcting code \cite{PhysRevA.54.1098,PhysRevA.54.4741} is used to decouple these two steps.

After the reduction to prepare-and-measure schemes, the data postprocessing can be divided into two steps: error correction and privacy amplification. Error correction is a step to reconciliate Alice's and Bob's sifted key. If we allow one-way key reconciliation, the cost in this step is $H(A|B)$, where $A$ and $B$ represent the random variables of Alice's and Bob's sifted key, respectively. In a symmetric channel, where the detected numbers of bits 0's and 1's are the same, the cost per bit is given by $H(e_b)$ as shown in Eq.~\eqref{eq:ShorPreskillR}. It turns out that the cost can be reduced if we allow two-way key reconciliation. The optimal key rate is an open question even in the classical key agreement case \cite{748999}.

Privacy amplification is a procedure for Alice and Bob to distill a common private key from a raw key about which Eve might have partial information \cite{bennett1995generalized}. The concept of privacy amplification is closely related to the randomness extraction problem \cite{doi:10.1137/100813683,Ma2013Extractor}. The difference is that in privacy amplification local randomness is a free resource, which in randomness extraction any randomness is valuable. Note that initially, the classical treatment on privacy amplification~\cite{bennett1995generalized} is only applicable to QKD under restrictive assumptions, i.e., the adversary has no quantum memory. Later on, however, this treatment has been generalized to the case where the adversary has quantum memory~\cite{renner2008security}. Details of data postprocessing, which distills a secure key from the raw data measured in quantum transmission (see Fig.~\ref{fig:flowchart}), are presented in Section \ref{sc:postprocessing}.

In the end, Eq.~\eqref{eq:ShorPreskillR} gives the key rate of the BB84 protocol. Considering the symmetric case where the bit and phase error rates are the same, it is not hard to see that the tolerable error rate of Eq.~\eqref{eq:ShorPreskillR} is 11\%, comparing to 7\% in the Mayers's proof \cite{mayers2001unconditional}. Similar to the entanglement distillation case, this formula is not tight. With two-way classical communication, one can achieve advantage distillation \cite{Gottesman2003twoway} using bit-flip error {\it detection}. Nonetheless, phase error detection remains forbidden in the absence of quantum computers. From the QKD postprocessing point of view, the bit and phase errors might be correlated. Alice and Bob can perform some preprocessing to reduce the total amount of key cost in error correction and privacy amplification. For example, they can group bits into pairs and compare parities and discard the ones with different parities. In this way, one can reduce the errors in remaining bits. This is called B step \cite{Gottesman2003twoway}. It terms out that, such prepossessing is useful in practical QKD processing \cite{Ma2006Twoway}. With two-way classical communication, one can also increase the tolerable error rates \cite{Gottesman2003twoway,Chau2002twoway}. Also, with the six-state protocol \cite{PhysRevLett.81.3018,lo2001proof}, it has been shown that the tolerable error rate is higher. We list all the tolerable error rates in Table \ref{Tab:tolerable}. Apparently, there are gaps between the lower (tolerable) and upper error rate bounds. This is an open question in QKD as well as in entanglement distillation for many years, which is also related to key agreement problem in classical communication case \cite{748999}.

\begin{table}[ht!]
\caption{List of tolerable error rate bounds for different schemes and proofs. The upper bounds are evaluated by simple individual (intercept-and-resend) attacks \cite{Gottesman2003twoway}.} \label{Tab:tolerable}
\begin{tabular}{l @{\hspace{0.6cm}} l @{\hspace{0.6cm}} l @{\hspace{0.6cm}} l}
  \hline \hline
  \textbf{Scheme} & \textbf{One-way} & \textbf{Two-way} & \textbf{Upper bound} \\
  \hline
  BB84 & 11.0\% &  20.0\% & 1/4 \\
  Six-state & 12.7\% &  27.6\% & 1/3 \\
  \hline \hline
\end{tabular}
\end{table}

\subsubsection{Koashi's complementarity approach}
The aforementioned security analyses by Lo-Chau and Shor-Presill based on quantum error correction complication certainly enjoy the strong intuition from entanglement to privacy. In fact, it turns out that entanglement (or a quantum channel that is capable of transmitting entangled state) is a precondition for secure QKD \cite{curty2004}. The main drawback of this approach is its complication of introducing a virtual entanglement-based protocol. Although the bit and phase error correction can be decoupled in postprocessing by employing the CSS quantum error correcting code~\cite{shor2000simple} or ancillary EPR pairs \cite{Lo2003Method}, these two steps always mix together in security proofs. Sometimes, constructing a virtual entanglement protocol could be highly nontrivial \cite{PhysRevLett.90.167904,PhysRevA.69.032316,fung2009security}. Error correction and privacy amplification are very different procedures in conventional cryptography. The former is to guarantee that Alice and Bob share an identical key, while the latter is to make sure they share a private key. One key observation is that error correction step is not directly related to quantum laws in the security analysis. That is, if Alice and Bob only want to share an identical key, they can just transmit classical states to do the job. From this observation, Koashi has developed a simplified security proof framework based on \emph{complementarity} \cite{koashi2009simple}.

In Koashi's approach, error correction and privacy amplification are decoupled from the beginning. Alice and Bob perform error correction first to make sure that the two bit strings are the same. Now, the problem becomes how many private key bits can be distilled from Alice's (same as Bob's) error corrected key. In this case we only need to deal with two parties, Alice and Eve. Denote the length of Alice's key string to be $N$. Alice's $N$-bit string can be regarded as the $Z$-basis measurement outcome of $N$ qubits $\rho_A\in \mathcal{H}_{2^N}$. Note that, under the most general coherent attacks, these $N$ qubits are correlated (or even entangled) with each other. The key idea is that in a virtual protocol, if each qubit of $\rho_A$ is measured in the complementary $X$-basis measurement and only $+1$ results are obtained, then $\rho_A=\ket{+}^N$, where $\ket{+}$ is the eigenstate of $X$ with the eigenvalue $+1$. In this ideal scenario, no one (including Eve) can predict Alice's key bits without accessing the measurement results directly. Like EPR pairs discussed in the Lo-Chau security proof, this ideal case will render perfect privacy. Interestingly, this unpredictability on the (computational) $Z$-basis is quantified by the coherence measure in the resource theory \cite{Yuan2015Coherence}, which is recently connected to the security of QKD \cite{Ma2019Operational}.

In general, $\rho_A$ is not a product of $\ket{+}$ states. In this framework, the phase error rate $e_p$ is defined as the ratio of getting $-1$ eigenstates of the complementary $X$-basis measurement on $\rho_A$. The parameter $e_p$ can be estimated differently in different QKD protocols. For instance, in BB84, essentially Alice randomly chooses some qubits to be measured in the $X$-basis and use random sampling to estimate $e_p$. Details of random sampling for parameter estimation will be discussed in Section \ref{sc:postprocessing}. Now, Alice can perform a virtual phase error correction on her $N$ qubits by the similar means discussed in the Lo-Chau phase error correction. Alice can hash the $X$-basis measurement outcomes and find the error syndrome. After phase error correction, Alice's state becomes close to $\ket{+}^N$, again, measured by fidelity or trace distance.

The key difference between Lo-Chau and Koashi security proofs lies on the definition of the phase error rate $e_p$. In Lo-Chau security proof, $e_p$ is defined on Bob's system relative to Alice's, while in Koashi's, it is defined on Alice's (or Bob's) side locally depending on protocols and Bob (Alice) can have an arbitrary system (irrelevant for security). In Koashi's approach, the complementary basis can be chosen arbitrary as long as $e_p$ can be estimated accurately. Meanwhile, along the line of complementarity approach, security proofs based on entropic uncertainty relations~\cite{coles2017entropic} have been developed \cite{Koashi_2006,berta2010uncertainty}.

In summary, the Lo-Chau, Shor-Preskill and Koashi security proofs are all based on the phase error correction. Note that, in this line of approach, the estimation of $e_p$ is the core of the security analysis. Sometimes, more sophisticated tools like semi-definite programming is employed to upper bound the phase error rate \cite{wang2019characterising}. Recently, there is an effort to make a connection between Shor-Preskill's type security proof \cite{shor2000simple,koashi2009simple} and Entropic approach~\cite{renner2008security} by Tsurumaru~\cite{tsurumaru2018leftover}.

So far, the security proof we reviewed here focuses on the BB84 protocol. Obviously, the security proof based on phase error correction can be extended to other protocols, like Bennett-1992 (B92) \cite{PhysRevLett.68.3121,PhysRevLett.90.167904} and six-state protocols \cite{PhysRevLett.81.3018,lo2001proof}. Meanwhile, this technique can also be employed to general qudit systems \cite{1412037}. Note that there is also security proof based on the idea of twisted states~\cite{PhysRevLett.100.110502,4529275}. Intuitively twisted states include shields. It allows the phase error correction syndrome to be hidden in the shield and thus become unaccessible to Eve. In principle, a virtual conceptual measurement on the joint state of Alice and Bob's shield will allow them to extract the missing phase error correction syndrome to complete the quantum error correction process. In practice, Alice and Bob do not need to perform such a virtual measurement.

\subsubsection{Entropic approach}
There is another line of security analysis~\cite{renner2008security,renner2005information,scarani2008quantum,tomamichel2012tight,coles2017entropic} that originates from the communication complexity and quantum memory approach \cite{BenOr2002Security,renner2008security}. Based on the entanglement distillation idea, a framework has been established for a general $\rho_{AB}$ by Devetak and Winter~\cite{Devetak2005Distillation}. In this quantum-entropy based framework, Alice and Bob share many i.i.d.~copies of $\rho_{AB}$, on which they perform measurements to obtain key bits. The Devetak-Winter key rate formula is given by
\begin{equation} \label{eq:DevetakWinter}
\begin{aligned}
r &= S(A|E)-H(A|B), \\
S(A|E)&=S(\rho_{AE})-S(\rho_{E}) \\
&=S(\rho_{B})-S(\rho_{AB}), \\
\end{aligned}
\end{equation}
where $S(A|E)$ is conditional quantum entropy, $S(A)=-\Tr(\rho_A\log\rho_A)$ is the von Neumann entropy. In the derivation, we assume the worst case scenario where $\rho_{ABE}$ is pure. In fact, the privacy amplification term can also be written in a relative entropy form \cite{coles2016numerical},
\begin{equation} \label{eq:DevetakWinterRel}
\begin{aligned}
S(A|E)=D(\rho_{AB}||\Delta_z(\rho_{AB})),
\end{aligned}
\end{equation}
where $\Delta_z(\rho_{AB})=\sum_{i}\dyad{i}_A\rho_{AB}\dyad{i}_A$ is the partial dephasing operation on system A in the $Z$ basis and the relative entropy function $D(\rho||\sigma)=\Tr\rho\log\rho-\Tr\rho\log\sigma$. This allows us to give an operational interpretation of coherence in QKD \cite{Ma2019Operational}.

The density matrix information $\rho_{AB}$ is unknown to Alice and Bob due to Eve's interference. They have to monitor $\rho_{AB}$ in real time, say via tomography. Thus, the Devetak-Winter analysis is normally applied to the i.i.d.~case, where Eve interferes all rounds of QKD identically and independently, i.e., collective attack~\cite{renner2008security,scarani2008quantum}. Nonetheless, the security analysis can be extended to the case of coherent attack by further analysis~\cite{tomamichel2012tight,coles2017entropic}, such as the de Finetti theorem \cite{renner2007symmetry}, the post-selection technique \cite{PhysRevLett.102.020504} and uncertainty relation for smooth entropies \cite{tomamichel2011uncertainty}. Another advantage of this approach is that the security of a complicated QKD scheme can be analyzed numerically \cite{coles2016numerical,Winick2018reliablenumerical}.

\subsection{Security assumptions} \label{sec:assumptions}
We will discuss the security assumptions made in general security proofs. We focus on the BB84 protocol, but most of the discussions can be applied to other protocols, such as B92, BBM92, and the six-state protocols. In security proofs \cite{lo1999unconditional,shor2000simple,koashi2009simple}, as shown in Section \ref{sub:SecurityProof}, we assume Alice sends ideal qubit states in $\{\ket{0},\ket{1},\ket{+},\ket{-}\}$ and Bob performs ideal qubit $Z$-basis and $X$-basis measurement. The channel, on the other hand, is assumed to be under a full control of Eve.

Nevertheless, in actual experiments these assumptions can be problematic. Table~\ref{Tab:secassumptions} summarizes the main differences between the security assumptions of security proofs and typical experimental setups. These differences, if unnoticed, might essentially open the security issue of \emph{basis dependency} between $Z$-basis and $X$-basis, thus causing the problem of quantum hacking attacks (see Table~\ref{Tab1}).

\begin{table}[ht!]
\scriptsize
\caption{Security assumptions and actual setup for BB84.} \label{Tab:secassumptions}
\begin{tabular}{l @{\hspace{0.3cm}} l @{\hspace{0.3cm}} l @{\hspace{0.3cm}} l}
  \hline \hline
   \textbf{Component} & \textbf{Security assumption} & \textbf{Practical setup} \\
   \hline
   Photon source & Ideal single photon & Coherent laser \\
   Encoding state & Two-dimension & Arbitrary-dimension  \\
   Encoding state & Basis-independent & Source flaws \\
   Measurement & Two-dimension & Arbitrary-dimension  \\
   Measurement & Basis-independent & Measurement flaws   \\
   Photon detection & Ideal SPD & Threshold detector \\
\hline  \hline
\end{tabular}
\end{table}

\subsubsection{Source}
First, let us relax the requirement on source by considering a more general source. In a prepare-and-measure QKD protocol, Alice randomly prepares system $B$ on one of the four states, $\{\rho_{x0},\rho_{x1},\rho_{z0}, \rho_{z1}\}$, and sends it to Bob. These four states can be denoted as $\rho_{\beta \kappa}$, where $\beta\in\{X, Z\}$ represents the encoding basis, and $\kappa\in\{0,1\}$ represents the encoding key bit. Here, we consider four states with two bases, but such scenario can be easily extended to more general cases with an arbitrary number of states and bases.

The prepare-and-measure protocol can be linked to the entanglement-based one as follows. Define the purification of state $\rho_{\beta \kappa}$ as $\ket{\psi_{\beta\kappa}}_{A_0 B}$, where system $A_0$ is an ancillary system. From an entanglement-based view of protocol, Alice sends out state $\rho_{\beta \kappa}$ is equivalent to her preparing
\begin{equation} \label{eq:StatePrepare}
\begin{aligned}
\ket{\Psi_\beta}_{A A_0 B} = \dfrac{1}{\sqrt{2}}\sum_{\kappa} \ket{\beta_\kappa}_A \ket{\psi_{\beta\kappa}}_{A_0 B},
\end{aligned}
\end{equation}
then measuring system $A$ in the $\beta$-basis, and sending out system $B$ according to measurement result $\kappa$. Here, system $A$ is a qubit system, $\ket{\beta_\kappa}_A$ is the $\beta$-basis eigenstate whose eigenvalue is $\kappa$. For the ideal BB84 protocol, there is no ancillary system $A_0$ (or $A_0$ is just a detached trivial system), since all encoding states $\rho_{\beta\kappa}$ are pure. Then, the states sent by Alice are,
\begin{equation} \label{eq:StateB}
\begin{aligned}
\rho_{\beta\kappa} &= Tr_{A_0}(\ket{\psi_{\beta\kappa}}\bra{\psi_{\beta\kappa}}_{A_0 B}),
\end{aligned}
\end{equation}
the four BB84 states.

To send out $\rho_{x\kappa}$, in the entanglement-based equivalent protocol, Alice  prepares $\ket{\Phi^+}_{AB} = \dfrac{1}{\sqrt{2}} (\ket{++} + \ket{--})_{AB}$, measures system $A$ on the $X$ basis, and obtains the measurement result $\kappa$. Similarly, to send out $\rho_{z\kappa}$, Alice first prepares $\ket{\Phi^+}_{AB} = \dfrac{1}{\sqrt{2}} (\ket{00} + \ket{11})_{AB}$, measures $A$ on the $Z$ basis, and obtains the measurement result $\kappa$. No matter which basis Alice wants to send, the initial entangled states prepared are the same. Let us denote the $X$-basis state and $Z$-basis state are
\begin{equation}
\begin{aligned}
\rho_x
&= \dfrac{1}{2}(\rho_{x0} + \rho_{x1}), \\
\rho_z
&= \dfrac{1}{2}(\rho_{z0} + \rho_{z1}), \\
\end{aligned}
\end{equation}
which are the quantum state transmitted given Alice and Bob choose the $X$ and $Z$ bases, respectively. Thus, in the ideal BB84 source case, the state sent out by Alice is independent of the basis choice,
\begin{equation} \label{eq:basisindep}
\begin{aligned}
\rho_x=\rho_z. \\
\end{aligned}
\end{equation}
We call this kind of source \emph{basis-independent} \cite{Koashi2003BasisInd,Ma2007Entangled}.

In the original proposal of the BB84 protocol, the basis choice is assumed to be unknown to Eve. This is also a crucial assumption in security proofs \cite{lo1999unconditional,shor2000simple}, as shown in Section \ref{sub:SecurityProof}. This is important for phase error estimation. If the source is basis-dependent, one cannot simply use one basis information to estimate the other. This is guaranteed by Eq.~\eqref{eq:basisindep}. In fact, as long as the source is basis-independent, it can be in arbitrary dimension or state. It can even be assumed to be under the control of Eve.

In practice, it is hard to construct single-qubit sources. Instead, entangled photon sources are widely used as a basis-independent source. Notice that, for entangled photon sources to work as a basis-independent source, the measurement for heralding has to be basis-independent. Later in Section~\ref{sub:GLLP}, we will show that the security can be guaranteed once the source contains a certain amount of basis-independent components.

In some QKD schemes, such as BBM92~\cite{bennett1992quantum}, Alice and Bob choose basis after the quantum signals transmitted through the channel. In BBM92 protocol, Alice prepares an entangled source, holds one part by herself and sends another to Bob. Alice measures her own part in some basis to realize the basis choice and encoding. In these schemes, the quantum source can even be assumed to be in the possession of Eve. Then for these schemes, Eq.~\eqref{eq:basisindep} can be guaranteed by the experimental setting.

\subsubsection{Measurement}
The requirement on measurement is very similar. Again, take the BB84 protocol as an example. There are four measurement outcomes labeled by two bits, $\beta, \kappa$. The corresponding four POVM elements are $M_{\beta\kappa}$,
\begin{equation} \label{eq:MxMz}
\begin{aligned}
M_x &= M_{x0} + M_{x1}, \\
M_z &= M_{z0} + M_{z1}. \\
\end{aligned}
\end{equation}
Here, $\{M_{x0}$, $M_{x1}\}$ form the $X$-basis measurement, while $\{M_{z0}, M_{z1}\}$ form the $Z$-basis measurement. We also require the measurement to be basis-independent,
\begin{equation} \label{eq:Mbasisindep}
\begin{aligned}
M_x=M_z. \\
\end{aligned}
\end{equation}

On the measurement side, the requirement is more strict. For the security proof presented in Section \ref{sub:SecurityProof}, it must be qubit measurements in the $X$ and $Z$ bases. Such requirement can be extended to more general projection measurements.

In practice, a \emph{squashing model} is widely employed \cite{gottesman2004security,Beaudry2008squashing,Fung2011universal}. In a squashing model, an arbitrary quantum state (from the channel) is projected to a qubit or vacuum. Then the $X$ or $Z$ measurement is performed. It has been proved that a typical threshold detector model adapts to the squashing model \cite{Beaudry2008squashing,Tsurumaru2008Security}.

Now, one can see that the assumptions on the source and measurement are quite different. For source, one only needs to guarantee its basis-independent property in Eq.~\eqref{eq:basisindep}. On measurement, it must be specific projection measurements. In practice, the source requirement is easier to meet comparing to the measurement requirement. Hence, there are more practical security issues on measurement than on source. A full security analysis needs to take account of these measurement deviations. We present it in Section~\ref{sub:GLLP}. This problem is finally resolved by MDI-QKD (see Section~\ref{sec:5:mdi}).

\subsubsection{Channel}
In security proofs, the channel is assumed to be under the full control of Eve. Thus, in principle, we do not put any requirement on channel. In fact, if any implementation deviation from the ideal QKD protocol can be put into the channel, it will not cause any security problems. For example, detectors normally have a finite efficiency. The loss caused by detectors can be moved to channel. Then, a detector can be replaced by a 100\% efficiency one in security analysis.

Now, the question is what kind of implementation deviations can be moved to channel. The implementation deviation can be regarded as some deviation operation acting on an ideal implementation. The key requirement is that the deviation operation must commute with basis switch operation. Alice and Bob each uses a basis switching device (say, a phase modulator in phase-encoding schemes). The channel is defined as the operation on the quantum signals in between the two basis switching devices.

\subsection{Practical security analysis}
In practice, there are two issues that need to be addressed: device imperfection and statistical fluctuation. In Section \ref{sec:assumptions}, we review the assumptions in the security proofs. In reality, these assumptions might be fully satisfied. Implementation devices might be (slightly) deviated from the ideal case used in the security proofs. When the deviation is small enough, we expect that a secure key can still be generated. In Section \ref{sub:GLLP}, we review the quantification of device imperfections and its effects on the security analysis.

In principle, the error rates defined in Eq.~\eqref{eq:biterror} cannot be obtained accurately since they are measured in complementary bases. In the security proofs reviewed in Section \ref{sub:SecurityProof}, we employ random sampling to estimate the error rates. When the data size goes to infinite, the error rates approach error probabilities, which can be estimated accurately. In a finite data size, such parameter estimation would render a finite confidence interval. In Section \ref{sec:randomsampling}, we review the parameter estimation with random sampling by calculating the failure probabilities and parameter bounds.

In Section \ref{sc:postprocessing}, we review the classical postprocessing of QKD and explain how can Alice and Bob distill secure keys in the raw bit strings from quantum measurement to final secure keys, with the help of public discussions. Note that some of the discussions need to be encrypted and/or authenticated.

\subsubsection{GLLP} \label{sub:GLLP}
There exist deviations between realistic QKD systems and the ideal QKD protocol. In order to achieve practical security of a QKD system, Alice and Bob need to characterize these device deviations or imperfections carefully and take them into account in the security analysis. Based on previous works on the topics \cite{lutkenhaus2000security,inamori2007unconditional}, Gottesman, Lo, L\"utkenhaus, and Preskill (GLLP) established a general framework for security analysis with realistic devices \cite{gottesman2004security}.

First, Alice and Bob need to characterize their devices to see how much deviation from the ideal ones used in the security proofs. One can employ typical distance measures, like fidelity and trace distance, to quantify the deviation. In principle, Alice and Bob can perform some virtual measurement on the devices for each run in realtime to see whether it works the same as the ``ideal device" or its ``orthogonal case". Then, they can tag the sifted key bit as ``good" if the virtual measurement projects to the ideal case, and tag ``bad" if it is ``orthogonal case". Of course, in reality, Alice and Bob do not know the virtual measurement result. Instead, they know the ratio of these two cases. Both source and measurement imperfections can fit into this scenario. The GLLP security analysis essentially tells us how to extract secure bits when the good bits are mixed with bad ones. So far, the discussion is rather abstract, in the following discussions, we take the source imperfection as an example.

The framework is generic. Here, let us take the BB84 protocol for example. In reality, a weak coherent state photon source is widely used as an approximate single photon source. With phase randomization, one can treat it as a mixture of Fock states \cite{lo2005decoy}. The vacuum and single-photon components are basis-independent, whereas the multiphoton components are not. In principle, Alice can measure the photon number to tag each encoded state being basis-independent or not (this is the aforementioned virtual measurement part). Denote the ratio of Bob's detected bits from the basis-independent source (good part, e.g. vacuum and single-photon component in the BB84 protocol) to be $1-\Delta$, and the rest (bad parte.g., multiphoton components in the BB84 protocol) is $\Delta$. Details of the source model and its security analysis will be presented in Section \ref{Sub:Source}.

With Alice's tagging information (photon number in this example), she can sort the sifted key bit string $k_A$ into two sub-strings, $k_{good}$ and $k_{bad}$, where
\begin{equation} \label{eq:kgoodbad}
\begin{aligned}
|k_{good}| &= (1-\Delta)N, \\
|k_{bad}| &= \Delta N. \\
\end{aligned}
\end{equation}
Following the phase error correction security proof, the underlying phase error rate of $k_{good}$ is $e_p$, which can be estimated accurately via complement measurements. The phase error rate of $k_{bad}$ is unknown. In the worst case scenario, the phase error rate of the string $k_{bad}$ could be as high as 1/2. The main idea of the GLLP security analysis is that if Alice and Bob employ linear privacy amplification, such as the matrix hashing introduced in Section \ref{sc:postprocessing}, they can still distill secure keys from $k_{good}$ by accessing $k_A$ only.

Denote $k_{good}'$ to be the bit string if Alice modifies the sifted key bit string by setting the bad bit positions to be 0. Similarly, denote $k_{bad}'$ to be the bit string if Alice sets the good bit positions to be 0. Then,
\begin{equation} \label{eq:kprime}
\begin{aligned}
|k_{good}'|=|k_{bad}'|&=|k_A|, \\
k_{good}'\oplus k_{bad}'&=k_A. \\
\end{aligned}
\end{equation}
Suppose a hashing matrix $T$ can distill secure bits from $k_{good}$. That is, $T k_{good}$ is a secure key. Then, it is not hard to show that $T' k_{good}'$ results in the same secure key if one extends the matrix $T$ to $T'$ by inserting new columns corresponding to the bad positions of $k_{good}'$. That is, $T$ is a sub-matrix of $T'$ by taking certain column vectors. Here comes the clever trick of GLLP: since $T' k_{good}'$ is private, the $XOR$ result
\begin{equation} \label{eq:Tkprime}
\begin{aligned}
T' k_{good}'\oplus T' k_{bad}' &=T' (k_{good}'\oplus k_{bad}') \\
&=T' k_A \\
\end{aligned}
\end{equation}
is also private even Eve knows everything about $T' k_{bad}'$. Note that the new added columns from $T$ to $T'$ can be arbitrary. In practice, Alice can just pick up a universal hashing matrix $T'$ to do privacy amplification and its sub-matrix $T$ will automatically a smaller-size universal hashing matrix.

Therefore, the secure key rate formula of GLLP is given by
\begin{equation} \label{eq:GLLP}
\begin{aligned}
r \ge -H(E)+(1-\Delta)[1-H({e_p})].
\end{aligned}
\end{equation}
where $E$ is the total QBER. This key rate formula can be viewed as an extension to Eq.~\eqref{eq:ShorPreskillR}. Furthermore, we do not need to restrict ourselves to two tags case, good and bad. In principle, Alice and Bob can label sifted key bits with an arbitrary dimensional tag $g$ and for each $g$ they can derive its corresponding phase error rate $e_p^g$. With the same argument as above, we can extended the GLLP formula \cite{Ma2008PhD},
\begin{equation} \label{eq:GLLPext}
\begin{aligned}
r \ge -H(E)+\sum_g q_g[1-H(e_p^g)],
\end{aligned}
\end{equation}
where $q_g$ is the ratio of sifted key bits with the tag $g$ and $\sum_g q_g=1$. Here, we assume Alice and Bob cannot access the tag $g$ in reality and hence they have to do the error correction part for all bits together. If they can really read out tags for each run, they can divide this error correction part as well.

\subsubsection{Random sampling and finite-data size} \label{sec:randomsampling}
The infinite data size limit ($N\rightarrow\infty$) is used for the key-rate formula Eq.~\eqref{eq:ShorPreskillR} and \eqref{eq:GLLPext}. When the data size is finite, the phase error rate, $e_p$, used for evaluate the amount of privacy amplification cannot be measured accurately. Instead, Alice and Bob can bound $e_p$ via certain complementary measurements.

In the BB84 protocol, the phase error probability in the $Z$-basis is the same as the bit error probability in the $X$-basis. In the following discussions, we assume Alice and Bob obtained the sifted key in the $Z$-basis measurement and want to estimate the underlying phase error rate $e_p$. Thus, by sampling the qubits in the $X$-basis, Alice and Bob can bound $e_p$. This is a typical \emph{random sampling} problem. Given a certain number of phase error rates in $n_x+n_z$ positions, Alice and Bob randomly sample $n_x$ positions for phase error testing and find $n_xe_{bx}$ errors. The sampling problem lies on bounding the phase error rate $e_{pz}$ in the remaining $n_z$ positions. The (upper) bound is related to the failure probability by a hypergeometric function \cite{Fung2010Finite}.

Specifically, the main objective is to evaluate the deviation $\theta$ of the phase error rate from the tested value, the bit error rate in the complementary basis, due to the finite-size effect. Here we recap the results from Section IX of \cite{Fung2010Finite} and list the variables in Table \ref{Tab:phaseest}. The phase error rate $e_{pz}$ is bounded by
\begin{equation} \label{Finite:Phbound}
\begin{aligned}
e_{pz} \le e_{bx}+\theta,
\end{aligned}
\end{equation}
with a failure probability of
\begin{equation} \label{Finite:PhFailX}
\begin{aligned}
\varepsilon_{ph} &\le \frac{\sqrt{n_x+n_z}}{\sqrt{e_{bx}(1-e_{bx})n_xn_z}} 2^{ -(n_x+n_z)\xi(\theta) }, \\
\end{aligned}
\end{equation}
where $\xi(\theta)=H(e_{bx}+\theta-q_x\theta)-q_xH(e_{bx})-(1-q_x)H(e_{bx}+\theta)$. If we take the Taylor expansion of Eq.~\eqref{Finite:PhFailX}, one can obtain the first order approximation essentially the same as the Gaussian limit used in the Shor-Preskill security proof \cite{shor2000simple}.

Another approach to deal with the problem of finite-size effect is by employing the \emph{smooth min-entropy}~\cite{renner2008security}, which is a valid measure of randomness in the non-asymptotic cases, and degenerates to Shannon entropy in the i.i.d.~limit. This approach has been applied to QKD to prove the finite-key security with almost tight bounds~\cite{tomamichel2011uncertainty,tomamichel2012tight}. Moreover, the smooth min-entropy approach is rather general to deal with non-i.i.d.~case and can be applied to other quantum information processing protocols, such as one-shot coherence resource theory~\cite{Zhao2018One} and device-independent QKD~\cite{Arnon2018Practical}. Note that for the security analysis of QKD systems with realistic devices, the finite data size effects are much more complicated. We shall review it in Section \ref{sec4:decoy}.

\begin{table}[htb!]
\caption{List of notations in phase error estimation.} \label{Tab:phaseest}
\begin{tabular}{c|c}
  \hline \hline
  \textbf{Notation} & \textbf{Definition} \\
  \hline
  $n_z$ & number of bits measured in the $Z$-basis \\
  $n_x$ & number of bits measured in the $X$-basis \\
  $e_{bx}$ & bit error rate in the $X$-basis \\
  $e_{pz}$ & phase error rate in the $Z$-basis \\
  $q_x$ & sampling ratio $n_x/(n_x+n_z)$ \\
  $\theta$ & deviation of the phase error rate \\
  $\varepsilon_{ph}$ & failure probability of phase error estimation \\
  \hline \hline
\end{tabular}
\end{table}

\section{QKD Implementation} \label{Sc:Implement}
In practice, security of a QKD system is often related to its implementation. A QKD implementation is composed of three parts: source, channel, and detection. In a rigorous security proof, the channel is assumed to be under the full control of Eve, who can replace the channel with any quantum operation she desires. In the security proof model, no implementation assumption is required on the channel. As a result, the security of the system does not depend on the physical realization of the quantum channel. Therefore, the practical security for the channel is not an issue. For the quantum source and detection, on the other hand, a security proof normally requires some assumptions on practical realization.

Photons are most widely used for communication, due to their robustness against decoherence due to noisy environment and fast traveling speed. Hence, we mainly focus on the quantum optical realization of QKD systems. We first discuss the encoding and decoding methods, then briefly introduce the practical source, channel and detection devices, and finally the classical postprocessing. Here, we primarily review the practical components of a discrete-variable QKD (DV-QKD) system, while the discussions for continuous-variable QKD (CV-QKD) can be found in Section~\ref{sec:CVQKD}.

\subsection{Encoding and decoding}
Different encoding and decoding methods are reflected on source, channel, and detection. For discrete-variable QKD schemes, Alice needs to figure out an efficient method to encoding her qubit (or qudit) into the quantum states. Accordingly, Bob needs to develop an efficient method to read out the quantum information encoded by Alice.

In general, for qubit-based QKD, the quantum information can encoded into two quantum modes, $s$ and $r$, and their relative phases. Normally, the two modes are assumed to be orthogonal, say, using orthogonal polarizations or distinct time bins. Then, for a photon, the states $\{\ket{10}_{sr}, \ket{01}_{sr}\}$ form a Hilbert space, named $Z$ basis. Here, ``0" and ``1" refer to the photon number in a mode. Two complementary bases, $X$ and $Y$, are defined with the relative phases. The $X$ and $Y$ basis states can be written as, $\{ \ket{10}_{sr} \pm \ket{01}_{sr} \}$ and $\{ \ket{10}_{sr} \pm i \ket{01}_{sr} \}$.

\begin{figure*}[hbt]
\centering
\includegraphics[width=14cm]{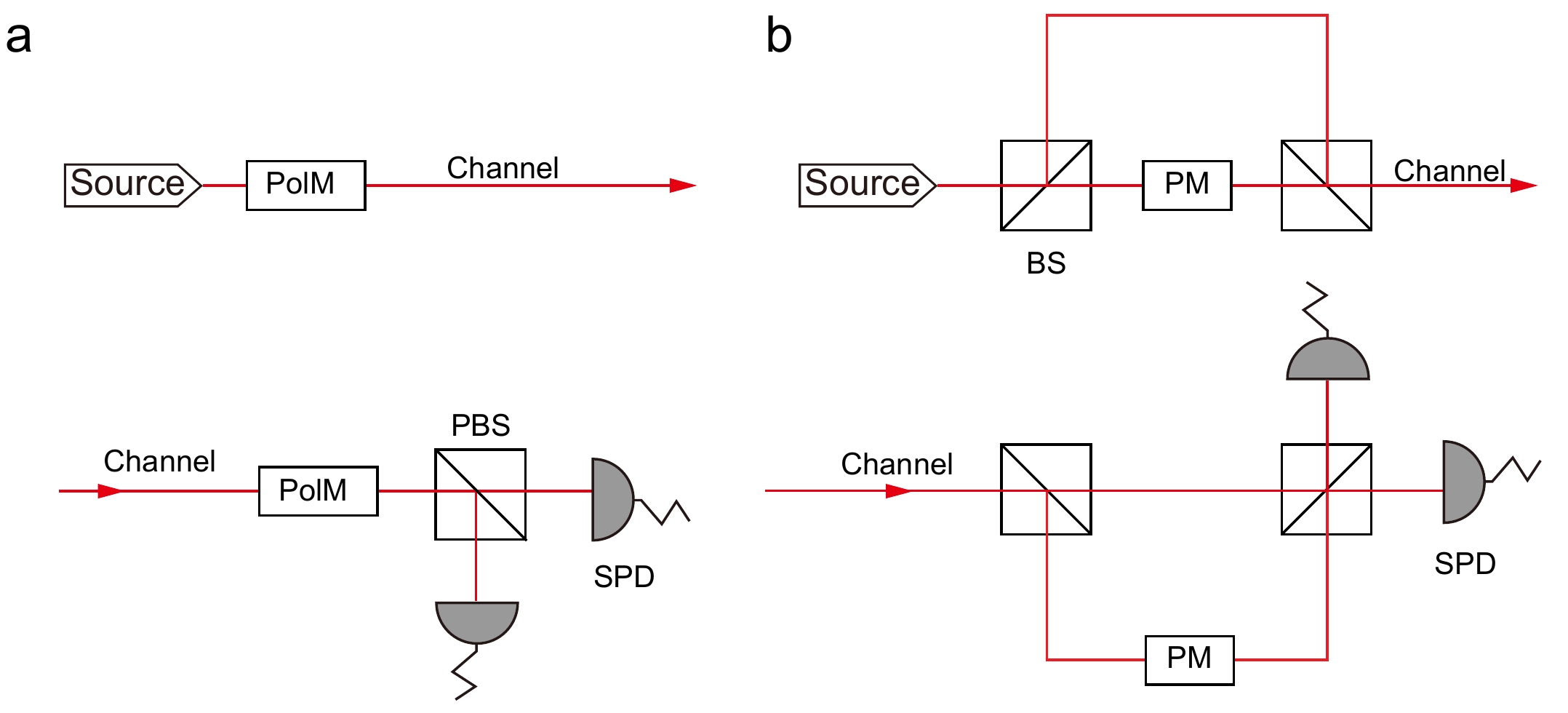}
\caption{Illustration of optical device of \textbf{a}, polarization encoding, and \textbf{b}, relative phase encoding. The top shows the encoding, while the bottom shows the decoding. PolM: polarization modulator; PM: phase modulator; PBS: polarization beam splitter; SPD: single photon detector; BS: beam splitter.}
\label{fig:encoding}
\end{figure*}

In reality, a widely applied method is polarization encoding, which utilizes the polarization modes. The horizontal and vertical polarizations of a photon, denoted by $\ket{10}_{HV}$ and $\ket{01}_{HV}$, are used for the $Z$ basis encoding. Then, the $X$ basis states, $\{\ket{10}_{HV} \pm \ket{01}_{HV} \}$, denote the linear polarization modes along the directions of $\pm 45^\circ$, respectively. The $Y$ basis states, $\{ \ket{10}_{HV} \pm i \ket{01}_{HV} \}$, denote the left- or right-handed circular polarizations. In the decoding process, the basis choice is realized by a polarization controller (Fig.~\ref{fig:encoding}), and the polarization measurement is realized by polarization beam splitter (PBS) connected with single photon detectors.

Another common method is time-bin phase encoding, where Alice chooses two pulses, a signal pulse and a reference pulse, for two encoding modes, denoted by $s$ and $r$, respectively. Similar to the polarization encoding, for a single photon, the two time-bin modes form the $Z$ basis, $\{\ket{10}_{sr}, \ket{01}_{sr} \}$. Here, the qubit in the $Z$ basis determines whether the photon stays in the signal time bin or the reference time bin. The $X$ and $Y$ basis states, $\{ \ket{10}_{sr} \pm \ket{01}_{sr} \}$ and $\{ \ket{10}_{sr} \pm i \ket{01}_{sr} \}$, denote the photons with a relative phase $0,\pi$ and $\pi/2, 3\pi/2$ between the signal and reference pulses, respectively. In the decoding process, an interferometer (Fig.~\ref{fig:encoding}) is employed to extract the phase information.

For qudit-based QKD, Alice and Bob need to find $d$ orthogonal modes and the encoding and decoding are similar. For example, the orbital angular momentum is the freedom of photons on spatial distribution, which contains a large Hilbert space. By encoding the high-dimensional key information into the orbital angular momentum, one can enhance the performance of QKD \cite{Simon2006,Cerf2002}. Another example is to encode with multiple time bins. In differential phase shift (DPS) QKD, the relative phase or each time-bin pulse is only $0$ or $\pi$, and the key is encoded into relative phase of two neighbouring pulses. Round-Robin-DPS QKD \cite{sasaki2014practical} encoding and decoding the phase difference circularly.

\subsection{Photon sources} \label{Sub:Source}
Here, we mainly discuss various practical photon sources for QKD: weak coherent-state source, thermal source, heralded single photon source, and entangled photon source. For most of prepare-and-measure QKD protocols, a single photon source is preferred. However, it is experimentally challenging to realize a high-quality and high-performance single photon source. We will discuss the photon sources according to different QKD schemes.

\subsubsection{Prepare-and-measure}
In a standard prepare-and-measure scheme like BB84, the common way is to employ other practical weak light sources to approximate the single photon source. In general, they are modulated to be a Fock state mixture,
\begin{equation} \label{eq:FockSource}
\begin{aligned}
\rho &= \sum_{n=0}^{\infty} P(n) \ket{n}\bra{n}, \\
\end{aligned}
\end{equation}
where $P(n)$ is the photon number distribution and $\ket{n}$ is the $n$-photon number state. For different types of sources, the photon number distribution will also be different. Normally, the single photon component, $\ket{1}\bra{1}$, is required to be dominant comparing to higher order components.

The weak coherent-state source is the most widely employed in QKD, which can be easily realized by attenuating laser lights. The light generated by a laser can be regarded as a coherent pulse $\ket{\alpha}$ within the coherence time, where $\alpha$ is a complex number, and $\mu = |\alpha|^2$ is the average photon number. The coherent state can be expanded in the Fock basis as
\begin{equation} \label{eq:coherentstate}
\begin{aligned}
\ket{\alpha} = e^{-|\alpha|^2/2} \sum_{n=0}^{\infty} \dfrac{\alpha^n}{\sqrt{n!}} \ket{n}.
\end{aligned}
\end{equation}
The phase of $\alpha$ reflects the relative phase between different photon number components. To realize a photon source in the form of Eq.~\eqref{eq:FockSource}, Alice can randomize the phase of coherent pulses, and make it a mixture of photon number states \cite{lo2005decoy},
\begin{equation} \label{eq:coherentFock}
\begin{aligned}
\rho_\mu &= \dfrac{1}{2\pi} \int_{0}^{2\pi} d\phi \ket{\alpha e^{i\phi}} \bra{\alpha e^{i\phi}} \\
&= \sum_{n=0}^{\infty} P_{\mu}(n) \ket{n}\bra{n},
\end{aligned}
\end{equation}
where the photon number follows a Poisson distribution, $P_{\mu}(n) = e^{-\mu} \dfrac{\mu^n}{n!}$. In many QKD protocols, such as BB84, only single photon component is secure for key distribution. Thus, the light intensity $\mu$ is typically in the single photon level, $\mu=O(1)$.

The thermal source is a Fock state mixture, expanded by
\begin{equation} \label{eq:thermalFock}
\begin{aligned}
\rho_{th} &=\sum_{n=0}^\infty  P_{th}(n)\ket{n}\bra{n} \\
&=\sum_{n=0}^\infty \dfrac{ \mu^n }{ (\mu + 1)^{n+1} }\ket{n}\bra{n},
\end{aligned}
\end{equation}
where $\mu$ is the average photon number and the photon number follows a thermal distribution $P_{th}(n)$. Note that for a small average photon number $\mu\le2$, the single photon component ratio is bigger in a Poisson distribution than in a thermal distribution. This is the reason why the weak coherent state source normally can outperform the thermal one in QKD \cite{Curty2010PassiveDecoy}.

\subsubsection{Entanglement-based}
For the entanglement-based QKD protocol, such as BBM92 \cite{bennett1992quantum}, an entangled photon source via the parametric down-conversion (PDC) process is normally adopted. In a PDC process, a high frequency photon is converted to a pair of low frequency photons. A PDC source emits a superposition state of different number of photon pairs \cite{Walls2008Quantum,Ma2008PDC},
\begin{equation}\label{eq:SPDC}
\begin{aligned}
|\Psi\rangle=(\cosh\chi)^{-1}\sum_{n=0}^{\infty}(\tanh\chi)^n\ket{n,n},
\end{aligned}
\end{equation}
where $\chi$ is the nonlinear parameter for the down-conversion process, $\mu=\sinh^2\chi$ is the average photon pair number, and $\ket{n,n}$ represents $n$ photon pairs in two optical modes.

The PDC process is widely used to generate photon pairs. In this case, four optical modes are used. For example, a typical PDC photon source emits photon pairs in two directions. In each direction, the photon can be in $H$ or $V$ polarization. The two optical modes are entangled in polarization. Comparing to Eq.~\eqref{eq:SPDC}, due to different collection means, the amplitudes of photon pair numbers are slightly different from the one in Eq.~\eqref{eq:SPDC} \cite{Kok2000Postselected,Ma2007Entangled},
\begin{equation}\label{eq:entPDC}
\begin{aligned}
\ket{\Psi}=(\cosh\chi)^{-2}\sum_{n=0}^{\infty}\sqrt{n+1}\tanh^n\chi\ket{\Phi_n},
\end{aligned}
\end{equation}
where $\chi$ is the nonlinear parameter for the down-conversion process, $\mu=2\sinh^2\chi$ is the average number of entangled photon pairs, and $|\Phi_n\rangle$ is the state of an $n$-entangled-photon pair,
\begin{equation}\label{Model:PDCn}
\begin{aligned}
|\Phi_n\rangle=\frac{1}{\sqrt{n+1}}\sum_{m=0}^{n}(-1)^m|n-m,m\rangle_a|m,n-m\rangle_b.
\end{aligned}
\end{equation}
In the aforementioned example, $a$ and $b$ represent two direction of the light, and $\ket{n-m,m}_a$ represents $n-m$ photons in the $H$ polarization and $m$ photons in the $V$ polarization. The number of the entangled-photon pairs follows a Super-Poissonian distribution, slightly different from the thermal distribution,
\begin{equation}\label{eq:entPn}
\begin{aligned}
P(n)=\frac{(n+1)(\frac{\mu}{2})^n}{(1+\frac{\mu}{2})^{n+2}}.
\end{aligned}
\end{equation}

Notice that the PDC source can also be used as a heralded photon source in the prepare-and-measure scheme. If we only focus on one of the optical modes (normally called signal mode), tracing out the other (normally called idle mode), the photon number follows the thermal distribution, $P_{th}(n)$ given in Eq.~\eqref{eq:thermalFock}. A typical usage of a PDC source for heralded photon involves measuring the idle optical mode locally as a trigger and encoding the signal mode for QKD. In this case, once Alice obtains a trigger locally, she can largely rule out the vacuum component in the signal mode. In fact, conditioned on whether or not a detection clicks on the idle mode, the photon number distribution is different on the signal mode. Such source can be used as a passive decoy-state source \cite{Adachi2007Simple,Ma2008PDC}. Note that when $\mu$ is very small, such heralded photon source can well approximate a single photon source, which is widely used in multi-photon processing \cite{Pan2000GHZ}.

\subsection{Channel}
Theoretically, we put no assumption on the quantum channel used for QKD. However, in the real-world implementation, we will build the QKD channel with mature optical communication technology to enhance the performance of QKD protocol. There are two widely adopted channels for QKD: fiber and free space. The most common channel used in QKD is built with commercial optical fiber. For a standard commercial single mode fiber (SMF), losses depend exponentially on the channel distance $l$ as $10^{-\alpha l/10}$, where the loss rate $\alpha$ is roughly $0.2$ dB/km for telecom wavelength at around 1550 nm. The loss rate can be remarkable if we extend the transmission distance to over $400$ km fiber~\cite{Yin2016,Fang2019surpassing}. Besides loss, a fiber-based QKD implementation should also solve several problems, such as chromatic dispersion, polarization mode dispersion, birefringence and so forth~\cite{gisin2002quantum}.

Free space channel features some advantages compared to optical fiber. There are several atmospheric transmission
windows, such as 780–850 nm and 1520–1600 nm, which have a low loss with an attenuation less than 0.1 dB/km in clear
weather~\cite{bloom2003understanding}. More importantly, the attenuation is even negligible in the outer space above the Earth's atmosphere, which enables long-distance QKD over $1000$ km between ground and satellite~\cite{Liaosate}. Furthermore, the decoherence of polarization or of any other degree of freedom is practically negligible. However, there are also some drawbacks concerning the free space. For instance, the weather conditions influence the losses of free space heavily. The effective apertures of the sending and receiving telescopes, influenced by alignment, movements and atmospheric turbulence, contribute coupling losses and affect the performance of free space QKD.

\subsection{Detection}
For DV-QKD schemes, single photon detection is realized with threshold detectors which can only distinguish the vacuum (zero photon) from single photon or multi-photon cases. Besides, some imperfections may exist in the single-photon detector (SPD): the detector efficiency $\eta$ is not $100\%$, which means some non-vacuum signals will not cause a click on the SPD; there exists a dark count factor $p_d$, which means some vacuum signals will incorrectly cause a click. This will affect the performance of QKD systems.

The measurement model is based on the threshold SPDs mentioned above. For the single-photon subspace, the detection here can be regarded as $X/Y$ basis qubit measurement. However, there is multi-photon component in the final signal, and the behavior of the measurement device will be different from the required $Z$-basis and $X$-basis measurement in DV-QKD. For example, there will be double-click signals caused by multi-photon component, which will not happen in the ideal $X/Y$ basis detection. To address this issue, the squashing model of measurement is proposed, combined with the random-assignment of double-clicked signals \cite{Beaudry2008squashing,Fung2011universal}.

In 2012, the MDI-QKD scheme \cite{lo2012measurement} was proposed to fill the detection loophole. The design of measurement devices in MDI-QKD is similar to the one in point-to-point QKD protocol. In discrete variable MDI-QKD scheme, the measurement device, assumed to be manipulated by the adversary, can be divided into two categories, single detection and coincidence detection. The coincident detection MDI-QKD schemes \cite{Ma2012Alternative} is based on the schemes in which the two communication parties, Alice and Bob, encode their key information into a single photon, and build correlation between their key value by a Bell state projection. The single detection MDI-QKD scheme~\cite{Lucamarini2018TF,ma2018phase} can be regarded as the detection on the coherent states rather than the single photon. They both build correlations between Alice's and Bob's bit values by Bell state projections.


\subsection{Postprocessing} \label{sc:postprocessing}
\begin{figure} [htbp]
\includegraphics[scale=0.68,angle=0]{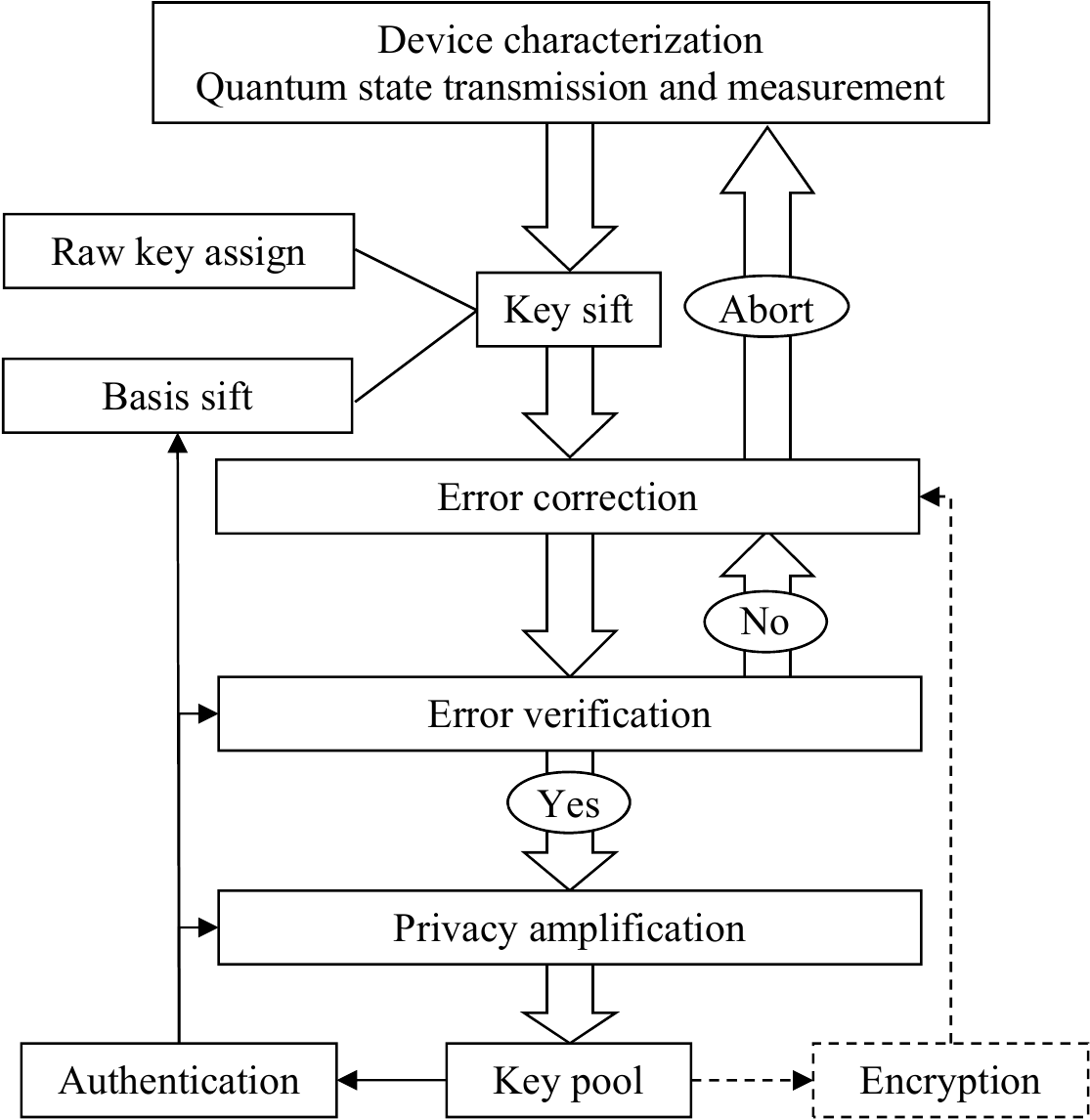}
\caption{Flow chart of data postprocessing procedures. The resource cost and the failure probabilities in encryption and authentication is listed in Table \ref{table:resources}. The encryption is optional for error correction depending on whether or not more privacy application is performed later. [Figure adapted from \cite{Fung2010Finite}].} \label{fig:flowchart}
\end{figure}

Postprocessing is a procedure for Alice and Bob to distill a secure key from the raw data measured in quantum transmission with the help of public discussions. The flow chart of QKD postprocessing is shown in Figure \ref{fig:flowchart}.

There are a few practical aspects need to be taken into consideration when the number of signals are finite, i.e., the finite-key effect. For example, the error correction efficiency may not reach to the Shannon limit; depending on the data size, normally a factor is applied. Also, on the privacy amplification side, there will be a small failure probability. Some public communication between Alice and Bob need to be authenticated and/or encrypted. Table~\ref{table:resources} summarizes the resource cost and the failure probabilities in the various steps.

The first step is raw key assignment, which depends on different schemes. For example, in the commercial BB84 implementation, Alice and Bob discard all the no-clicks and randomly assign double-clicks. In the MDI-QKD, this step is based on the announcement of the measurement site. In DI-QKD, Alice and Bob can perform arbitrary assignment. Of course, any improper assignment will reduce the key rate.

During the public discussions, some of the classical communication need to be authenticated, as listed in Table \ref{table:resources}. In the security proofs we reviewed, we assume the encryption of classical communication in error correction. Such encryption can be avoided in other security proofs. In this case, there may be some restriction on the error correction procedure and more follow-up privacy amplification is required. For example, in the original Shor-Preskill security proof, such encryption is not necessary when the CSS code is employed. In practice, there is an advantage to using error correction without encryption, since if Alice and Bob abort the QKD procedure after error correction, no pre-shared secret bits will be lost due to encryption. In the following discussions, we assume the number of bits communicated in error correction is counted for later privacy amplification. Thus, in privacy amplification, the extraction ratio will be $r$ given in Eq.~\eqref{eq:GLLPext} without considering the finite data size effects. If the one-time pad encryption is used for error correction, the privacy amplification ratio will be higher (by removing the error correction term in Eq.~\eqref{eq:GLLPext}). After that, certain amount of secure key bits need to return to the keep pool for encryption consumption. In the end, the final key rate will still be same in the encryption and non-encryption cases.

\begin{table} [htbp!]
\caption{List of resource cost and the failure probabilities in the various steps. The numbers of consumed secret key bits are denoted with $k$ while the failure probabilities with $\varepsilon$. Alice sends out $N$ signals and Bob detected $n$ of them in the $Z$-basis measurement. The final key output length is $l$. The tag length refers to authentication tag and - means no authentication is required. In error verification, no message but only an encrypted (authentication) tag is transmitted. The cost of error correction $k_{ec}$ is given by $n f H(E)$. No communication is required in phase error estimation. [Table adapted from \cite{Fung2010Finite}]} \label{table:resources}
\begin{tabular}{ l|ccc}
\hline \hline
\textbf{Procedures} & \textbf{Message} & \textbf{Tag} & \textbf{Failure Prob} \\
\hline
1. Raw key assign & $N$ & - & - \\
2. Basis sift & $n$ & $k_{bs}$ & $\varepsilon_{bs}$ [Eq.~\eqref{eq:failauthen}] \\
3. Bit error correction & $k_{ec}$  & - & -  \\
4. Error verification & - & $k_{ev}$ & $\varepsilon_{ev}$ [Eq.~\eqref{eq:failauthen}] \\
5. Phase error estimation & - & - & $\varepsilon_{ph}$ [Eq.~\eqref{Finite:PhFailX}] \\
6. Privacy amplification & $(n+l-1)$ & $k_{pa}$ & $\varepsilon_{pa}$ [Eq.~\eqref{eq:failurePA}] \\
\hline \hline
\end{tabular}
\end{table}

\subsubsection{Error correction}
For practical error correction, normally an efficiency factor $f>1$ is put before the $H(E)$ term Eq.~\eqref{eq:GLLPext}, which means the actual cost is larger than the theoretical Shannon limit. Previously, a widely used error correction protocol for QKD is Cascade \cite{Cascade1994}. The Cascade protocol is simple and highly efficient that is able to achieve an error correction factor of around $1.1\sim1.2$ for a large QBER range from 0\% to beyond 11\%. In the Cascade protocol, Alice and Bob divide their sift key bit strings into blocks and compare parities of each block to look for errors. They perform a binary search to locate the error when the parity of a block is different. The process repeats for a few times with different block sizes and permutation to ensure all the error bits are corrected. The Cascade protocol is highly interactive because the binary search requires $1+\log_2(n)$ communications and successful error correction often requires several passes. Later, several improved protocols have been proposed to reduce the interaction rounds \cite{buttler2003fast,nakassis2004expeditious,elliott2005current}.

Another family of error correcting codes is forward error correction, which only needs to send one syndrome from Alice to Bob. Due to its light classical communication load, the forward error correction is widely implemented in commercial QKD systems. One outstanding example is the low-density parity-check (LDPC) code \cite{mackay1996near}. The LDPC code works well for QKD due to its the high error correction efficiency and very limited communication requirement. The design and optimization of LDPC codes in QKD postprocessing is similar to the classical case, which can be divided into three steps.
\begin{enumerate}
  \item The first step is to find a good degree distribution \cite{mackay1999comparison,richardson2001capacity} for the target error rate.
  \item The second step is to generate a good parity-check matrix. Like in classical communication, the small cycles may contribute to localized information transmitted in decoding. Thus, a good parity-check matrix generation algorithm should yield a relatively large girth. Progressive edge-growth is one of the most successful algorithms to generate parity-check matrix eliminating small cycles \cite{hu2005regular}.
  \item The third step is to decode with Bob's key string and the received syndrome. The brief-propagation algorithm \cite{fossorier1999reduced}, which is also known as the sum-product algorithm, is highly efficient in decoding.
\end{enumerate}

The standard LDPC algorithm is only optimum at its designed rate for the designed QBER. But the actual QBER is fluctuating from round to round. The rate compatible LDPC can solve this problem with puncturing and shorting \cite{ha2006rate}. The main technology here is to select a mother code close to the target rate, and then to adjust the code rate with puncturing. The puncturing operation can be done multi-times, in order to find a best code rate suitable for the actual error rate. This method has been employed in QKD \cite{elkouss2009efficient,martinez2010interactive}. Besides efficiency, another important factor of error correction is the throughput. The limitation for the Cascade code is the highly interactive communications, and for LDPC is the computational cost in iterative decoding. It was reported the throughput with both the Cascade \cite{pedersen2013high} and the LDPC \cite{dixon2014high} codes can be higher than 10 Mbps. Note that the computing in decoding LDPC is always assisted with GPU acceleration.

\subsubsection{Error verification and authentication}
Before error correction, Alice and Bob sample the sifted key bits to roughly estimate the error rates. Then they perform error correction. After error correction, Alice and Bob can perform error verification to make sure they share the same key \cite{lutkenhaus1999estimates,Finite:Short:11}. Then, the failure probability for error correction is reflected in error verification step when finite data size is considered. It is not hard to see that the two jobs, error verification and message authentication, are very similar. In both cases, Alice and Bob want to make sure the bit strings on the two ends are the same. The only difference is that authentication tag might reveal information about the message, but error verification should not. This difference can be overcome by encrypting the tag, which has already been done in most of information-theoretically secure authentication schemes. If we employ the LFSR-based Toeplitz matrix construction, the relation between the tag length (the same as the key cost) and the failure probability is given by
\begin{equation} \label{eq:failauthen}
\begin{aligned}
\varepsilon &= n2^{-k+1},
\end{aligned}
\end{equation}
where $n$ is the message length and $k$ is the key cost.

After error correction and error verification, Alice and Bob are almost sure that they have located all the errors. Then, they can accurately count the number of bit errors and hence the rate $e_b$ defined in Eq.~\eqref{eq:biterror}. If Alice and Bob choose not to encrypt error correction, they can count the amount of classical communication used in the error correction, $k_{ec}$. Then they perform addition amount of privacy amplification. For example, in the ideal devices case of Eq.~\eqref{eq:ShorPreskillR} and infinite data limit, $k_{ec}=nH(e_b)$. The final key output length is given by $l=r n$.

\subsubsection{Privacy amplification}
Practical privacy amplification turns out to be very efficient in terms of finite data size effect, once the necessary parameters, such as the phase error rates, are estimated as reviewed in Section \ref{sec:randomsampling}. Denote the error corrected bit strings for Alice and Bob to be $k_A=k_B$ with a length of $n$ and the output length to be $l$. In the infinite key limit, $l/n=r$ given in Eq.~\eqref{eq:GLLPext} if error correction is not encrypted. In the privacy amplification procedure, Alice randomly chooses a universal hashing matrix $T\in\{0,1\}^{l\times n}$ and sends it to Bob via a public classical channel. The final key will be given by $T k_A=T k_B$, with a small failure probability.

Privacy amplification works for general classes of two-universal hash functions~\cite{tomamichel2011leftover}. In particular, the universal hashing function based on Toeplitz matrices is widely used for privacy amplification. An $l\times n$ Toeplitz matrix is a Boolean matrix with the structure of
\begin{equation} \label{eq:Toeplitz}
\begin{aligned}
T=
\begin{pmatrix}
a_0 & a_{-1} & a_{-2} & \cdots & a_{-n+1}\\
a_1 & a_0 & a_{-1} & \ddots \\
a_2 & a_1 & \ddots && \vdots\\
\vdots & \ddots \\
a_{l-1} & & \cdots & & a_{l-n}
\end{pmatrix},
\end{aligned}
\end{equation}
where $a_i\in\{0,1\}$ for $-n+1\le i\le l-1$. The Toeplitz matrix can also be concisely written as $T_{(i,j)}=a_{i-j}$ where $T_{(i,j)}$ is the $(i,j)$ element of $T$. Apparently, an $l\times n$ Toeplitz matrix can be specified by $N+K-1$ bits, as opposed to $N\times K$ bits for completely random matrices. The main advantage of Toeplitz matrix hashing is that the computational complexity for $T k_A$ is $O(n\log n)$ by the fast fourier transform (FFT).

Following the security proofs reviewed in Section \ref{sub:SecurityProof}, the matrix $T$ should be related to the phase error correction. In order to ensure the phase error correction to commute with the final key measurement, we require the null space of $H$ to be capable of correcting the underlying phase errors. For universal hashing functions, such error correcting capability can be evaluated with certain failure probabilities. Details of derivation can be found in Section X of \cite{Fung2010Finite}. The failure probability for privacy amplification with Toeplitz hashing is given by,
\begin{equation} \label{eq:failurePA}
\begin{aligned}
\varepsilon_{pa} &= 2^{-t_{pa}}. \\
t_{pa} &= n r-l, \\
\end{aligned}
\end{equation}
If Alice transmit the Toeplitz matrix to Bob, then she needs to authenticate that communication as well, which would add an extra term of Eq.~\eqref{eq:failauthen} to $\varepsilon_{pa}$. Here in privacy amplification, by sacrificing $t_{pa}$ extra bits in privacy amplification, one can obtain a failure probability of $2^{-t_{pa}}$. More general discussions for hash functions besides Toeplitz hashing can be seen in~\cite{tomamichel2011leftover}.

Note that message authentication can be done more efficiently by piling up classical communication data and authenticate them at once. That is, the authentication terms listed in Table \ref{table:resources} can be done once with one authenticated tag and one failure probability. The main drawback of this saving data and authenticating approach is that it might require lots of local data storage. In QKD system design, it is normally preferred that each procedure of postprocessing is isolated.

From the simulation results \cite{Fung2010Finite}, we learn that the failure probabilities for authentication, error verification, and privacy amplification are not the main contributions to the total system one. In fact, the one in phase error rate estimation, Eq.~\eqref{Finite:PhFailX}, is the dominate term. The summation of failure probabilities evaluated here can be converted to the trace-distance measure in Eq.~\eqref{eq:tracedis}.

\subsubsection{Finite-key length}
When the failure probability of the postprocessing procedure is $\epsilon$, the final key is $\sqrt{\epsilon(2-\epsilon)}$-secure in accordance with the composable security definition given in Eq.~\eqref{eq:tracedis}. Finally, by including the finite-data statistics for parameter estimation (see Sec.~\ref{sec:randomsampling}) and the postprocessing costs (see Table~\ref{table:resources}), we have the finite-key length $NR$ for the finite-size security of QKD, which can be written as~\cite{Fung2010Finite},
\begin{equation} \label{Finite:KeyFinal}
\begin{aligned}
NR &\ge l-k_{bs}-k_{ec}-k_{ev}-k_{pa} \\
\end{aligned}
\end{equation}
with a failure probability of
\begin{equation} \label{Finite:FailFinal}
\begin{aligned}
\varepsilon\le\varepsilon_{bs}+\varepsilon_{ev}+\varepsilon_{ph}+\varepsilon_{pa} \\
\end{aligned}
\end{equation}
where $l$ is given by,
\begin{equation} \label{Finite:PAkey}
\begin{aligned}
l &= n_x[1-H(e_{bz}+\theta_z)]+n_z[1-H(e_{bx}+\theta_x)], \\
\end{aligned}
\end{equation}
where the variables can be found in Table \ref{Tab:phaseest}.

Notice that one can also utilize the smooth min-entropy approach to obtain the finite-key length~\cite{renner2008security,scarani2008quantum} or the tight bounds~\cite{tomamichel2012tight}. Note that for QKD systems with realistic devices, the finite-key length is slightly complicated, we refer to \cite{lim2014concise} for decoy-state QKD, \cite{curty2014finite} for measurement-device-independent QKD, \cite{lorenzo2019tight} for twin-field QKD, \cite{Arnon2018Practical} for device-independent QKD and~\cite{furrer2012continuous,leverrier2013security} for continuous-variable QKD.

\section{Quantum Hacking}\label{sec:3}
In theory, it is traditional to divide Eve's hacking strategy to three main classes: \emph{individual}, \emph{collective} and \emph{coherent} (or general) attack. Individual attack means that Eve interacts with each secure qubit in the channel separately and independently; Collective attack means that Eve prepares independent ancilla, interacts each qubit independently, but can perform a joint measurement on all the ancilla; Coherent attack means that Eve can prepare an arbitrary joint (entangled) state of the ancilla, which then interact with the qubits in the channel before being measured jointly. The last one does not limit Eve's capabilities beyond what is physically possible. Any QKD system
aiming to implement an informational-theoretically secure protocol therefore has to be proven secure against coherent attacks. Another aspect which cannot be neglected is security in a finite size scenario. No key transmission session can be endless and the resulting statistical fluctuations have to be taken into account~\cite{scarani2009security}.

In this section, different from the theory attacks, we focus on the practical attacks which exploit the device imperfections in QKD systems. Specifically, Eve may try to exploit the imperfections in real QKD systems and launch the so-called \emph{quantum hacking} not covered by the original security proofs. Researchers have demonstrated several quantum hacking attacks in practical QKD systems. An earlier review on quantum hacking attacks can be seen in~\cite{jain2016attacks}. Here we will provide a review for the quantum attacks to both the source and the detection. The detection attacks are similar to those reviewed in~\cite{jain2016attacks}, but we provide more details for the attacks at source which exploit the multiple photons, timing or phase information of the laser source. Also, some new attacks after~\cite{jain2016attacks} will also be mentioned. Table~\ref{Tab1} summarizes a list of the attacks developed from early 2000 to the present.

\subsection{Attacks at source}

In the standard QKD scheme, it is assumed that Alice (state preparation) is placed in a protected laboratory and she prepares the required quantum state correctly. Unfortunately, imperfect state preparation may leak information about the secret key. Indeed, practical preparation may introduce some errors due to imperfect devices or Eve's disturbance~\cite{lutkenhaus2000security,brassard2000limitations,Fred2007,xu2010,Sun2012,Tang2013,Sun2015}. To steal the information about the states, Eve can also actively perform the Trojan-horse attack~\cite{gisin2006trojan,jain2014trojan,jain2015risk} on intensity modulators and phase modulators. This section will review some examples of attacks at source.

\subsubsection{Photon-number-splitting attack} \label{sec:3:PNS}
The first well known kind of hacking strategy that was considered is the photon-number-splitting (PNS) attack~\cite{lutkenhaus2000security,brassard2000limitations} aiming at the imperfect photon source.  As described in section~\ref{Sub:Source}, because of technological challenge, weak coherent pulses (WCPs) generated by a highly attenuated laser are widely used in QKD implementations. Since the photon number of a phase-randomized WCP follows the Poisson distribution (Eq.~\eqref{eq:coherentFock}), there is a non-zero probability for multiple-photon pulses, i.e., those pulses containing two or more photons. Consequently, Eve may exploit the multiple-photon pulses and launch the PNS attack. In this attack, for each WCP, Eve first utilizes a quantum non-demolition (QND) measurement to obtain the photon number information. Conditioned on the QND measurement result, Eve either blocks the one-photon pulse or splits the multiple-photon pulse into two. She stores one part of the multiple-photon pulse and sends the other part to Bob. Later on, during the basis-reconciliation process of the BB84 protocol, Eve can get the secret key information for the multiple-photon pulse without introducing any errors. By doing so, Alice and Bob could not notice Eve's attack.

The PNS attack restricts the secure transmission distance of QKD typically below 30 km~\cite{gottesman2004security}. Actually, in early 2000s, there were not many research groups working on QKD experiments~\cite{Hughes2000quantum,ribordy2000fast,gobby2004quantum}. Researchers in the field had a doubt on the future of QKD, and they generally thought that QKD may be impractical with WCP source. This concern severely limits the development of QKD at that time. Fortunately, the discovery of the decoy state method perfectly resolved the problem of PNS attack and made QKD practical with standard WCP~\cite{hwang2003quantum,lo2005decoy,wang2005beating}. More details on the decoy state method will be discussed in section~\ref{sec:4}.

\subsubsection{Phase-remapping attack}

\begin{figure}[!t]
\centering \resizebox{8cm}{!}{\includegraphics{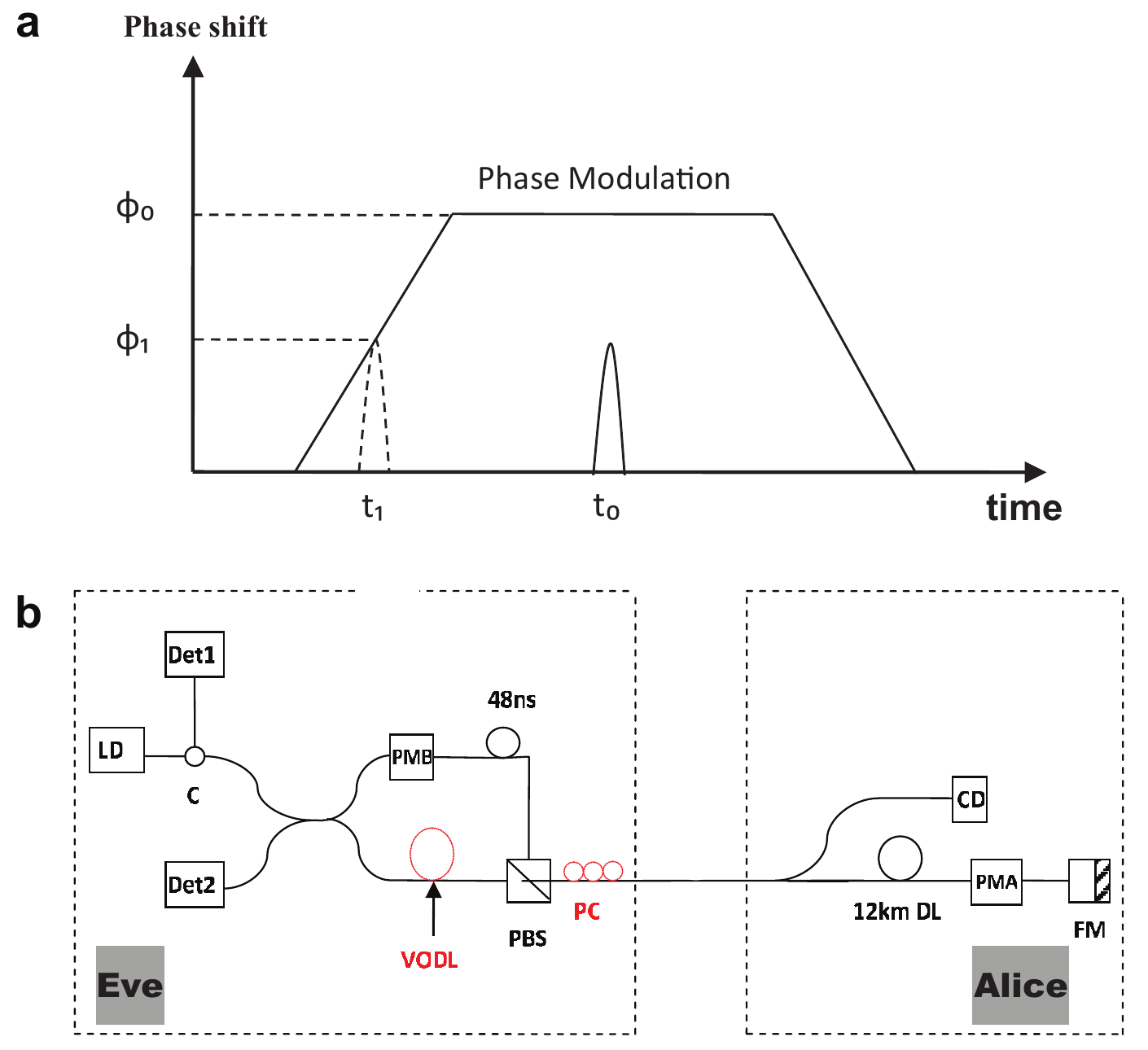}}
\caption{ Phase-remapping attack~\cite{xu2010}. \textbf{a}, Diagram of phase modulation signal. $t_{0}$ is the original time location where the signal pulse is properly
modulated to have phase $\phi_{0}$. Eve can time-shift the signal pulse from $t_{0}$ to $t_{1}$, where the pulse will undergo a new modulated phase $\phi_{1}$. \textbf{b}, Implementation of the phase-remapping attack in a commercial IDQ QKD system. Original QKD system: LD, laser diode. Det1/Det2, single photon detector; PMB/PMA, phase modulator; C, circulator. PBS, polarization beam splitter; FM, Faraday mirror; CD, classical detector; DL, delay line. Eve's modifications: VODL, variable optical delay line; PC, polarization controller. [Figure reproduced from~\cite{xu2010}].} \label{Fig:remapping}
\end{figure}

Phase modulators are commonly used to encode random bits in the source of phase-coding QKD systems~\cite{gisin2002quantum}. In practice, a phase modulator has finite response time, as shown in Fig.~\ref{Fig:remapping}a. Ideally, the pulse passes through the phase modulator in the middle of the modulation signal and undergoes a proper modulation (time $t_{0}$ in Fig.~\ref{Fig:remapping}a). However, if Eve can change the arrival time of the pulse, then the pulse will pass through the phase modulator at a different time (time $t_{1}$ in Fig.~\ref{Fig:remapping}a), and the encoded phase will be different. This phase-remapping process allows Eve to launch an intercept-and-resend attack, i.e., phase-remapping attack~\cite{Fred2007}. The phase-remapping attack is a particular threat for the bidirectional QKD schemes, such as the plug-and-play QKD structure~\cite{stucki2002quantum}.

In 2010, the phase-remapping attack was successfully demonstrated in a commercial ID-500 plug-and-play QKD system (manufactured by ID Quantique\footnote{https://www.idquantique.com/})~\cite{xu2010}, as shown in Fig.~\ref{Fig:remapping}b. In this experiment, Eve utilized the same setup as Bob to launch her attack. Eve modified the length of the short arm of her Mach-Zehnder interferometer by adding a variable optical delay line (VODL in Fig.~\ref{Fig:remapping}b) to shift the time delay between the reference pulse and the signal pulse. To remap the phase small enough into the low QBER range, Eve shifted the forward signal pulse out and only the backward signal pulse in the phase modulation range by using VODL, and properly aligned the polarization direction of the backward signal pulse orthogonal to the principal axis of the phase modulator by using a polarization controller (PC in Fig.~\ref{Fig:remapping}b). The experiment demonstrated that Eve could get full information and only introduce a QBER of 19.7\%, which is much lower than the well-known 25\% error rate for an intercept-and-resend attack in BB84.

\subsubsection{Nonrandom-phase attack}

Phase randomization is a basic assumption in most security proofs of QKD~\cite{gottesman2004security,hwang2003quantum,lo2005decoy,wang2005beating}. Although the security of QKD with non-random phase had been proven~\cite{Lo2007Nonrandom}, the performance is very limited in distance and key rate. By assuming that the overall phase is uniformly distributed in [0, $2\pi$], a coherent state with intensity can be reduced into a classical mixture of photon number states, i.e., Eq.~\eqref{eq:coherentFock}. This can greatly simplify the security proofs and allow one to apply classical statistics theory to analyze quantum mechanics. In practice, however, the phase randomization assumption may be violated in practice, thus resulting in various attacks~\cite{Sun2012,Tang2013,Sun2015}.

The first example is the USD attack demonstrated in~\cite{Tang2013}. When the phase of WCPs is not properly randomized, the quantum state will be a pure state. Then in decoy state QKD~\cite{hwang2003quantum,lo2005decoy,wang2005beating}, it is possible for Eve to distinguish the signal state and decoy state with an unambiguous state discrimination (USD) measurement. Hence, Eve first measures each of Alice's WCPs to distinguish between signal state and decoy state by performing a USD measurement, which is combined with positive operator-valued measurement (POVM) operators without disturbing the quantum state sent by Alice. After the USD, Eve performs the PNS attack. Since Eve knows which state the pulse belongs to (signal or decoy), she could do different strategies for signal state and decoy state. As a result, the key assumption in decoy state QKD~\cite{lo2005decoy} -- a decoy state and a signal state have the same characteristics -- is violated.

The second example is the laser seed-control attack which was proposed and demonstrated in~\cite{Sun2015}. Semiconductor laser diode (SLD) is normally used as a single-photon source in most commercial and research QKD systems. In the
interdriven mode, the semiconductor medium of the SLD is excited from loss to gain by each driving current pulse. A laser pulse is generated from seed photons originating from spontaneous emission. The phase of the laser pulse is determined by the seed photons. Since the phase of the seed photons is random, the phase of each laser pulse is random inherently. However, if a certain number of photons are injected from an external source into the semiconductor medium, these photons will also be amplified to generate laser pulses. Consequently, the seed photons consist of two parts: one from spontaneous emission and the other part from the external source. Both parts will affect the phase of the resulting laser pulse. If the injected photons greatly outnumber the photons from spontaneous emission, the phase of the output laser pulse is largely determined by the phase of the injected photons. Therefore, Eve can control the phase of Alice’s signal laser by illuminating the SLD from an external control source and successfully violate the phase randomization assumption~\cite{gottesman2004security}.

\subsection{Attacks at detection}
The detection component is much more vulnerable to quantum hacking attacks than the source. Since Eve controls the channel and can send any signals (e.g strong optical pulses combined with X-ray and neutrinos) to Bob and Bob has no choice but to receive Eve's signal and any filters used by Bob may be imperfect, it may be hard for Bob to isolate his lab and avoid side channels or detector control from/by Eve. For instance, a significant number of attacks have been proposed to hack single-photon detectors (SPDs)\footnote{The vulnerabilities of SPDs are mainly due to their complex working mechanism: the detection is affected by incoming light and the control electronic circuits. Therefore, Eve can manipulate the intensity, the time, or the wavelength of incoming light to control the responses of SPDs.} \cite{makarov2009controlling,Lars2010, gerhardt2011full,gerhardt2011experimentally,wiechers2011after,Seb2011}. SPDs are regarded as the ``Achilles heel" of QKD by C. H. Bennett\footnote{C. H. Bennett, “Let Eve do the heavy lifting, while John and Won-Young keep her honest,” http://dabacon.org/pontiff/?p=5340}. This section will review some examples of attacks at detection. The first two examples, double-click attack and fake-state attack, were proposed only in theory. The last two examples, time-shift attack and detector-blinding attack, were successfully demonstrated in experiment.

\subsubsection{Double-click attack}
Since QKD systems require the detection of two different bit values, bit 0 and bit 1, they require at least two SPDs. The double-click event refers to the case where both SPDs detect signals. The double-click event will introduce a QBER of 50\% when either one of the two bits is selected. A naive strategy is to determine double-click events as abnormal events and discards these events so as to minimize the QBER. However, this strategy results in the problem of double-click attack. In this attack, Eve simply floods Bob's polarization beam splitter with multiple photons or a strong pulse of the same polarization. Then, when Bob makes a measurement using a conjugate basis different from that of Eve, a double-click event occurs and it is discarded; when the receiver makes a measurement using the same basis as Eve's, a normal event is detected. Consequently, Alice and Bob finally share the same information with Eve. To solve this problem, L\"utkenhaus has proposed that double-click events are not discarded and bit 0 or bit 1 is randomly allocated by Bob whenever a double-click event occurs~\cite{lutkenhaus1999estimates,lutkenhaus2000security}.

\subsubsection{Fake-state attack}
\begin{figure}[!t]
\centering \resizebox{8.1cm}{!}{\includegraphics{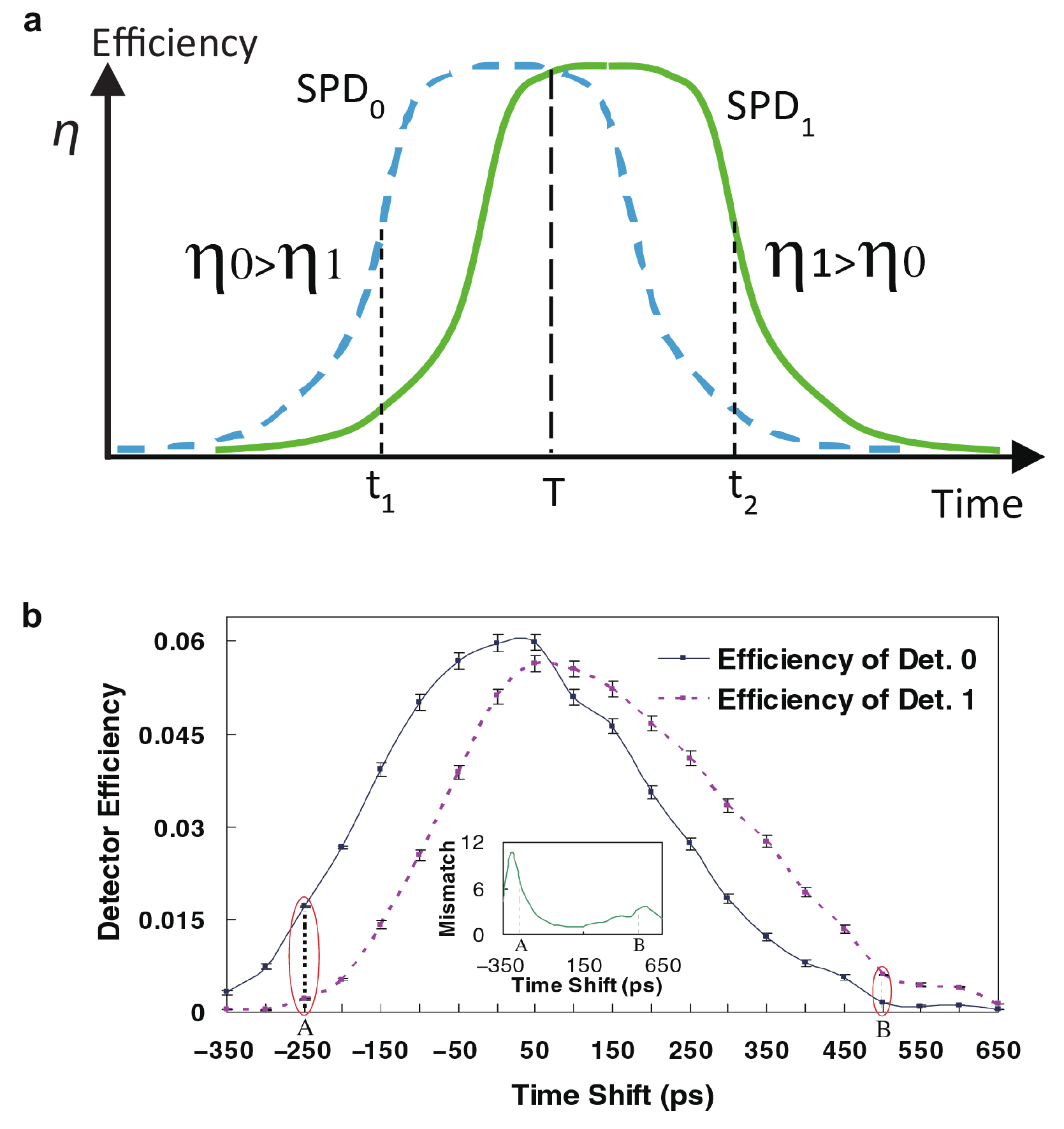}}
\caption{Schematic of detection efficiency mismatch~\cite{Makarov2006} and time-shift attack~\cite{Qi2007,Zhao2008}. \textbf{a}, SPD, single-photon detector. At the expected arrival time T, the detection efficiencies of SPD$_{0}$, $\eta_{0}$, for the event of bit 0, and SPD$_{1}$, $\eta_{1}$, for the event of bit 1 are the same. However, at time $t_{1}$, SPD$_{0}$ is more sensitive to the incoming photon than SPD$_{1}$, while at time $t_{2}$, SPD$_{1}$ is more sensitive to the incoming photon than SPD$_{0}$. \textbf{b}, Real detector efficiencies of the two SPDs characterized on a commercial QKD system (manufactured by IDQ) by Zhao et al.~\cite{Zhao2008}. [Fig.b reproduced from~\cite{Zhao2008}]. } \label{fig:efficiencymismatch}
\end{figure}

In 2005, Makarov et al. proposed a faked-state attack, which exploits the efficiency mismatch of two detectors in a practical QKD system~\cite{makarov2005faked,Makarov2006}. In practice, the standard SPDs such as Si/InGaAs APDs are often operated in a gated mode. Therefore, the detection efficiency of each detector
is time-dependent. Since QKD systems require the detection of two different bit values, 0 and 1, they often employ at least two SPDs. It
is inevitable that finite manufacturing precision in the detector and the electronics, and difference in optical path length will slightly misalign the two detector gates, and cause detector-efficiency mismatch. This is illustrated in Fig.~\ref{fig:efficiencymismatch}a. At the expected arrival time T, the detection efficiencies of the two detectors are identical. However, if the signal is chosen to arrive at some unexpected times (such as $t_{1}$ and $t_{2}$ in Fig.~\ref{fig:efficiencymismatch}a), it is possible that the detector efficiencies of the two detectors, $\eta_0$ and $\eta_1$, differ greatly. This problem often exists in practical QKD systems, and it will leave a back door for Eve to attack the system.

The faked-state attack is an intercept-and-resend attack. For each signal, Eve randomly chooses one of the two BB84 basis to perform a measurement and obtain a measurement result. Then, she re-sends the opposite bit value from her measurement result in the opposite basis, at a time when the detector for the opposite bit has a lower detection efficiency than the other detector. As analyzed in~\cite{Makarov2006}, Eve introduces less than 11\% QBER if the detection efficiency $\eta\leq$6.6\%. The faked-state attack, while conceptually interesting, is hard to implement in a real-life QKD system. This is because it is an intercept-resend attack and as such involves finite detection efficiency in Eve's detectors and precise synchronization between Eve and Alice-Bob's system. A typical countermeasure against detector efficiency mismatch is the four-state QKD protocol~\cite{Makarov2006}.

\subsubsection{Time-shift attack}

Motivated by the faked-state attack, in 2007, Qi et al. \cite{Qi2007} proposed the time-shift attack. This is also based on the detection-efficiency mismatch for gated SPDs in the time domain, but is much easier to implement. Let us suppose Fig.~\ref{fig:efficiencymismatch}a illustrates the detection efficiencies of the two gated SPDs in a real-life QKD system. Eve can simply shift the arrival time of each pulse sent from Alice by employing a variable optical delay line. For example, Eve randomly shifts the pulse from Alice to arrive at $t_{1}$ or $t_{2}$ through a shorter path or a longer path of optical line. This shifting process can partially
reveal the bit value of Bob: if the pulse arrives at $t_{1}$ (or $t_{2}$) and Bob announces receipt, the bit value is more likely to be 0 (1). Moreover, Eve can carefully set how many bits should be shifted forward and how many should be shifted backward to ensure that the distribution of bit 0 and bit 1 received by Bob is balanced. Hence, the time-shift attack does not make any measurement on the quantum state, and quantum information is not destroyed.

Since Eve does not need to make any measurement or state preparation, the time-shift attack is practically feasible with current technology. In 2008, it has been successfully implemented on a commercial QKD system by Zhao et al.~\cite{Zhao2008} as shown in Fig.~\ref{fig:efficiencymismatch}b. This is one of the first successful demonstrations of quantum hacking on a widely-used commercial QKD system. In their experiment \cite{Zhao2008}, Eve got an information-theoretical advantage in around 4\% of her attempts. The successful implementation of the quantum attack shows that a practical QKD system has non-negligible probability to be vulnerable to the time-shift attack.

\subsubsection{Detector-control attack}

\begin{figure}[hbt]
\centering \resizebox{8.2cm}{!}{\includegraphics{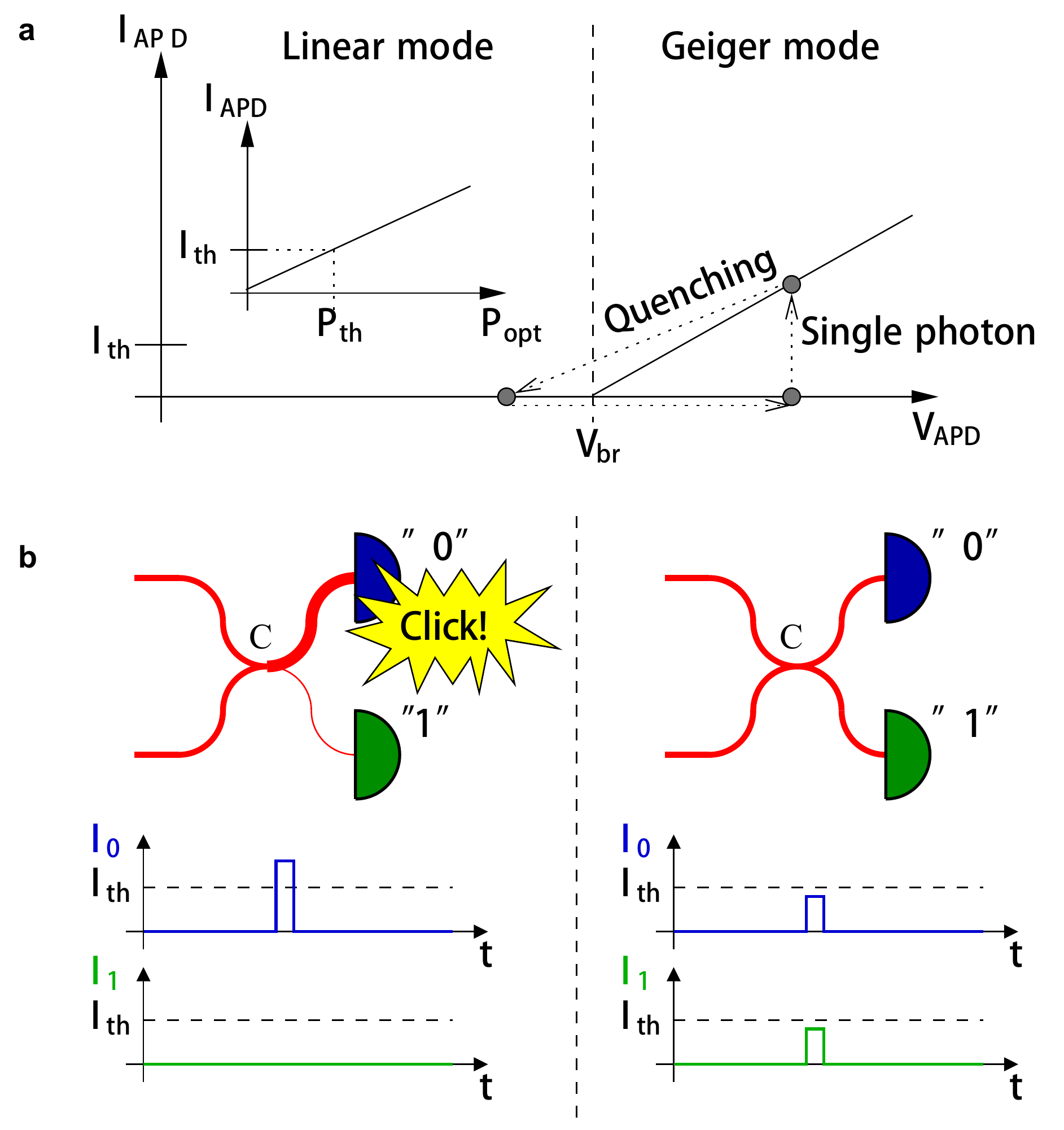}}
\caption{ Schematic illustration of the detector blinding attack~\cite{Lars2010}. \textbf{a}, Linear-mode and Geiger-mode APD operation. When the APD is reverse-biased above its breakdown voltage $V_{br}$, a single photon can cause a large current $I_{APD}$ to flow, and register this as photon detection (a `click'). After that, an external circuit quenches the avalanche by lowering the bias voltage below $V_{br}$, and then the APD goes into a linear mode. In the linear mode, $I_{APD}$ is proportional to the incident bright optical power $P_{opt}$. \textbf{b}, Eve sends Bob a tailored light pulse that produces a `click' in one of his detector only when Bob uses the same measurement basis as Eve. Otherwise, no detector `clicks'. [Figure reproduced from~\cite{Lars2010}].} \label{fig:blinding}
\end{figure}

The detector-control attack is the most powerful attack and it has been successfully demonstrated on several types of practical QKD systems~\cite{makarov2009controlling,Lars2010}. In general, the detector-control attacks can be divided into three categories: (i) detector-blinding attack \cite{makarov2009controlling,Lars2010,lydersen2011controlling,huang2016testing}, where Eve illuminates bright light to control detectors; (ii) detector-after-gate attack \cite{wiechers2011after}, where Eve just sends multi-photon pulses at the position after the detector gate; (iii) detector-superlinear attack~\cite{lydersen2011superlinear,qian2018hacking}, where Eve exploits the superlinear response of single-photon detectors during the rising edge of the gate.

Most available SPDs are InGaAs/InP APDs operating in a Geiger mode \cite{Hadfield:2009}, in which they are sensitive to a single photon. The working principle of this type of APDs is shown in Fig.~\ref{fig:blinding}a. In the detector blinding attack, by sending a strong light to Bob, Eve can force Bob's SPDs to work in a Linear mode instead of Geiger mode, as shown in Fig.~\ref{fig:blinding}a. In the Linear mode, the SPD, such as the one based on InGaAs APD, is only sensitive to bright illumination. This detector operation mode is called ``detector blinding". After blinding the detectors, Eve sends a bright pulse with tailored optical power such that Bob's detector always reports a detection event from the bright pulse, but never reports a detection event from a pulse with half power. This is illustrated in Fig.~\ref{fig:blinding}b. Consequently, Eve can successfully launch an intercept-and-resend attack without increasing QBERs. For example, when Eve uses the same basis as Bob to measure the quantum state from Alice, Bob gets a detection event as if there were no eavesdropper. But if Eve uses the opposite basis from Bob to measure the quantum state from Alice, her bright pulse will strike each of Bob's detectors with half power, and neither detector will report a detection event. In practice, a simple detector blinding attack will introduces a 50\% total loss. However, Eve can place her intercept-unit close to Alice's laboratory while compensating the loss in the remaining fiber by re-sending brighter states.

The detector-control attack is applicable to various types of SPDs, such as gated APDs~\cite{Lars2010,huang2016testing}, passively or actively quenched APDs~\cite{makarov2009controlling,Seb2011}, superconducting nanowire single-photon detectors (SNSPDs)~\cite{lydersen2011controlling}, and so forth. A full field implementation of the attacking strategy has been investigated in~\cite{gerhardt2011full}. The blinding attack was also demonstrated to fake the violation of Bell's inequality~\cite{gerhardt2011experimentally}. How to remove the detector-control attacks is a challenge in the field of QKD. One proposed countermeasure is carefully operating the single-photon detectors inside Bob's system and monitoring the photocurrent for anomalously high values~\cite{yuan2010avoiding2,yuan2011resilience}. This work also highlights that mis-operation of QKD devices allows the loophole to be exploited, which is related to the best-practice criteria for all QKD devices in QKD implementations~\cite{koehler2018best}. Recently, Qian et al. propose another countermeasure against the detector-control attacks by introducing a variable attenuator in front of the detector~\cite{qian2019robust}. However, these countermeasures may seem \emph{ad hoc} and lead away from provable security models of QKD and can often be defeated by advanced hacking technologies. A practical and promising solution is the MDI-QKD protocol, which will be reviewed in section~\ref{sec:5:mdi}.

\subsection{Other attacks} \label{sec:3:otherattacks}

\begin{figure}[hbt]
\centering \resizebox{7.8cm}{!}{\includegraphics{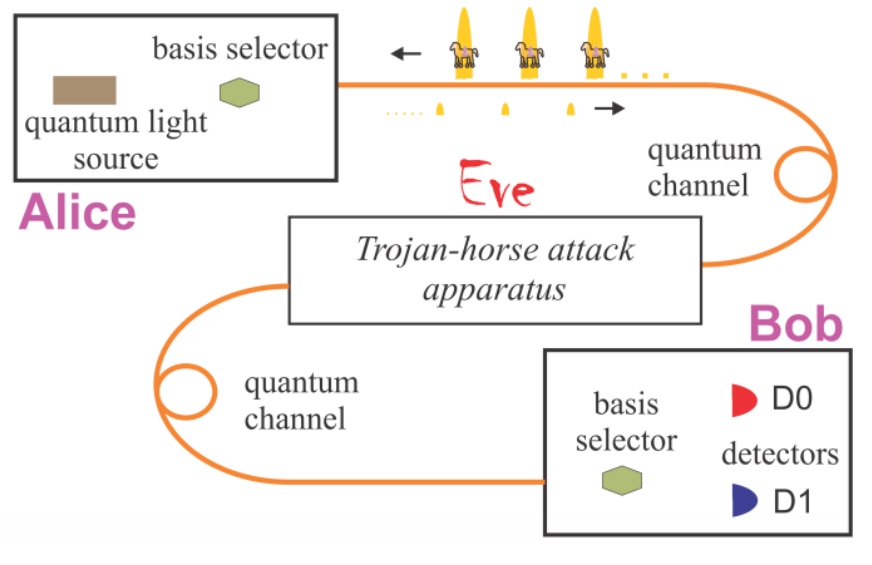}}
\caption{Schematic illustration of the Trojan-horse attack~\cite{jain2015risk}. Eve attacks Alice by sending bright Trojan-horse pulses to know Alice's selected basis information. This information is carried by the back-reflected pulses coming out of Alice. [Figure reproduced from~\cite{jain2015risk}].} \label{fig:trojan}
\end{figure}

Another well-know hacking strategy is the Trojan-horse attack (THA), as shown in Fig.~\ref{fig:trojan}, in which Eve sends a probe light to Alice or Bob and reads their information from the backscattered probe light. In 2001, Vakhitov et al. proposed the large pulse attack~\cite{vakhitov2001large} and Kurtsiefer et al. analyzed the possibility of THA by detecting the detector fluorescence of Si-based avalanche photodiodes~\cite{kurtsiefer2001breakdown}. Gisin \emph{et al.} studied the problem of THA in the QKD implementations where light goes two ways~\cite{gisin2006trojan}. Later, Jain \emph{et al.} performed a comprehensive analysis of the risk of THA against typical components in standard QKD systems~\cite{jain2014trojan,jain2015risk}. Recently, the backflash photons caused by detection events in single-photon detectors was exploited to realize the detector-backflash attack~\cite{pinheiro2018eavesdropping}. A countermeasure against the THA is to add proper isolations and consider the leaking information in privacy amplification, which will be reviewed in Section~\ref{sec:4:THA}.

Besides the above attacks, Lamas-Linares and Kurtsiefer demonstrated that the timing information revealed during public communicating can be exploited to attack the entanglement-based QKD system~\cite{lamas2007breaking}. In two-way QKD system such as the ``plug-and-play" structure, Sun \emph{et al.} studied the imperfections of Faraday mirror and proposed the Faraday-mirror attack~\cite{Sun2011}; Jain \emph{et al.} experimentally demonstrated that the calibration routine of a commercial ``plug-and-play" system can be tricked into setting a large detector efficiency mismatch, and proposed an attack strategy on such a compromised system with a QBER less than 7\% \cite{Jain2011}. Moreover, Li et al.~\cite{Li2011} studied the imperfection of a practical beam splitter and demonstrated a wavelength-dependent beam-splitter attack on top of a polarization-coding QKD system. The detector dead-time issue was widely studied in~\cite{rogers2007detector} and demonstrated in~\cite{Wei2011}. Andun \emph{et al.} and Makarov \emph{et al.} demonstrated the laser damage attack by using a high-power laser to damage the SPDs~\cite{Audun2014,Vadim2016}. Recently, Huang \emph{et al.} show that the decoy states can be distinguishable if they were generated by modulating the pump current of a semiconductor laser diode~\cite{huang2018quantum}, and Wei \emph{et al.} exploited the efficiency mismatch in the polarization degree of freedom to hack SNSPD~\cite{wei2019implementation}.

Most of the imperfections reviewed so far are in fiber-based QKD systems. There are also quantum attacks reported for free-space QKD systems~\cite{nauerth2009information,sajeed2015security,chaiwongkhot2019eavesdropper}. For instance, imperfect encoding methods result in side channels from which encoded states are partially distinguishable~\cite{nauerth2009information}. The imperfection due to non-single-mode quantum signals is a crucial issue in free-space QKD. Eve can exploit this imperfection and launch the spatial-mode attack against a free-space QKD system. This problem has been carefully studied in~\cite{sajeed2015security,chaiwongkhot2019eavesdropper}, following an earlier discussion on the origins of detection efficiency mismatch in~\cite{fung2009security}. Besides DV-QKD, the practical security of CV-QKD also deserves future investigations, which will be reviewed in Section~\ref{cvqkd:hacking}.

More generally, as noted in~\cite{curty2017quantum}, in principle, there are simply too many side channels for Alice and Bob to close. This is because Eve might , in principle, attack Alice's and Bob's system via X-ray, neutrons, neutrinos or even gravitational waves. And, whatever detection systems Alice and Bob have will probably have limited ranges of responses. Moreover, classical post-processing units pose a serious threat to the security of QKD. Most QKD security framework assumes without proofs that classical post processing units are secure. However, in conventional security, it is well known that hardware Trojans and software Trojans are commonly used to compromise the security of conventional cryptographic system. It was proposed in~\cite{curty2017quantum} to use redundancies in QKD units and classical post-processing units to achieve security through e.g., verifiable secret sharing.

\section{Source Security}\label{sec:4}
In this section, we review various approaches to resolve the security issues of practical sources. On one hand, the imperfections in quantum state preparation, including multi-photon components of laser, nonrandomized phases, encoding flaws and so forth, need to be carefully quantified and taken into account in security analysis. In particular, we will discuss the decoy state QKD protocol in more details. On the other hand, practical countermeasures are required to prevent the Trojan horse attacks on the source. Note that we focus on the BB84 protocol, but most of techniques can be extended to other protocols.

\subsection{Decoy-state method} \label{sec4:decoy}
Decoy-state method is a common way to combat with source imperfection by introducing extra sources for better channal characterization. In the decoy-state method, the user randomly modify the source states during the quantum stage; after that, he reveal which state is used in each turn. Eve cannot modify her attack to different source states, but in the postprocessing the users can estimate their parameters conditioned on that knowledge. The decoy-state method is used mostly to bound the multi-photon components in a practical photon source.

In practical photon sources, multi-photon components are inevitable. As reviewed in Section~\ref{sec:3:PNS}, Eve can split a multi-photon pulse and save one photon from it for later hacking. Since Alice and Bob cannot tell whether a detection comes from a single-photon component or multi-photon component and Eve controls the channel, they have to pessimistically assume all the multi-photon states cause clicks with 100\% efficiency. All the losses come from the single-photon states. In order to reduce the effects of multi-photon components, Alice has to use very low intensity optical pulses. In the case of coherent state photon source, it has been shown that the optimal intensity used is close to the channel transmittance $\eta$ \cite{lutkenhaus2000security,Ma2006low},
\begin{equation} \label{eq:nodecoyoptmu}
\begin{aligned}
\mu_{opt} \approx \eta,
\end{aligned}
\end{equation}
where $\eta$ includes channel transmission and detection efficiency. Then, final key rate will quadratically depend on the transmittance $\eta$, $R=O(\eta^2)$.

Various protocols~\cite{DPS:2002,hwang2003quantum,Scarani2004} have been proposed to over the key rate limit caused by the PNS attacks, among which the most effective one is the decoy state method~\cite{hwang2003quantum,lo2005decoy,wang2005beating}. In the decoy-state QKD scheme, instead of using one intensity for encoding, Alice employs a few additional intensities of optical pulses (as decoy states) in order to monitor the transmittance of different photon number components. After Bob detects the signals, Alice announces the intensities she uses for each pulse. With detection rate for decoy states, Alice and Bob can bound tightly the number of detections from single-photon components. If Eve simply changes the transmittance for different photon number states as adopting in the PNS attacks, she will inevitably change the detection rates for signal and decoy states differently. Without Alice's intensity information ahead, Eve has to let a significant amount of single-photon states passing in order to maintain the ratio of detection rates among signal and decoy states. The decoy state idea was first proposed by Hwang~\cite{hwang2003quantum} who considered using a strong decoy signal with intensity around 2 photons as a decoy state.

\begin{figure*}[hbt]
\centering \resizebox{13cm}{!}{\includegraphics{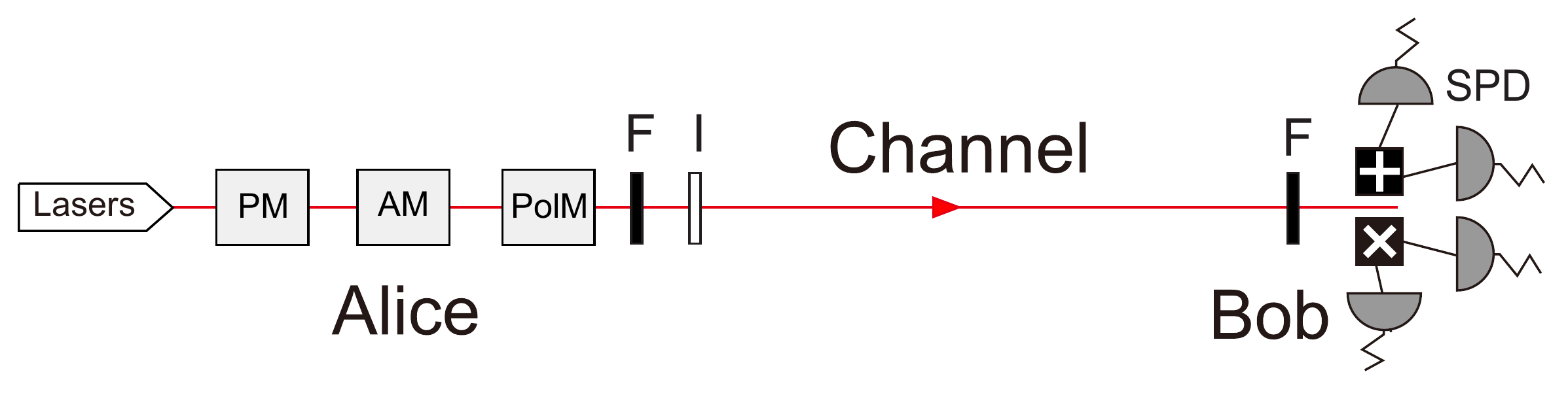}}
\caption{Schematic diagram of decoy-state QKD. In a decoy-state BB84 transmitter, the optical pulses are normally generated with phase-randomized laser pulses. Decoy states are prepared using an amplitude modulator (AM). In the figure: (PM) phase modulator for phase randomization, (PolM) polarization modulation for encoding, (F) optical filter and (I) optical isolator. } \label{fig:decoyimplement}
\end{figure*}

The security proof of the decoy-state method is given later in \cite{lo2005decoy}, where a photon number channel model \cite{Ma2008PhD} is employed. With an infinite number of decoy state, Alice and Bob can estimate the detections from all photon number components accurately. After adopting the GLLP security analysis, reviewed in Section~\ref{sub:GLLP}, one can show that the optimal intensity of optical pulses can be increased to $O(1)$, which results in a key rate having a linear dependence of transmittance, $O(\eta)$ \cite{lo2005decoy}. The decoy-state method significantly booming the performance of practical QKD. The schematic diagram of the decoy-state method is shown in Fig.~\ref{fig:decoyimplement}.

In the meantime, practical decoy-state methods with only a vacuum and \emph{weak} decoy states were proposed \cite{Lo2004Vacua,Ma2004PhD} and tight bounds were derived later \cite{wang2005beating,ma2005practical}. In the original security proof, continuous phase randomization is assume to decohere phases between different photon number components. As discussed later in Section \ref{Sc:NonRandom}, phase randomization is necessary but can be relaxed to discrete phase randomization \cite{Cao2015Discrete}. In fact, the uniformly discrete phase randomization with discrete phase number $m=10$ can already achieve a good approximation of continuous phase randomization.

\subsubsection{Theory}
For the source with different photon number components, one can assume a photon number channel model~\cite{Ma2008PhD}. The decoy state method is a tomography to the photon number channel model, providing tighter estimations on single photon component~\cite{lo2005decoy}. In the decoy state method, the source is operated at different photon number distributions, leading to different measurement outcome statistics. The communication partners can estimate the channel parameters of yield $Y_n$ and QBER $e_n$ for each photon number component. One crucial assumption in the decoy-state QKD is that the signal state and decoy states are \emph{identical} except for their average photon numbers. This means after Eve's photon-number measurement, she has no way of telling whether the resulted photon number state is originated from the signal state or decoy states. Hence, the yield $Y_n$ and QBER $e_n$ can depend on only the photon number, $n$, but not which distribution (decoy or signal) the state is from. That is,
\begin{equation}
\begin{aligned}
Y_n(signal) &= Y_n(decoy), \\
e_n(signal) &= e_n(decoy).
\end{aligned}
\end{equation}

The implementations of decoy state method can be divided into active ones and passive ones. In the active decoy state method, the user perpares the source signals with different intensities to change the probability distributions of each photon number component. A simple solution for decoy state preparation, as shown in Fig.~\ref{fig:decoyimplement}, is to use an amplitude modulator (AM) to modulate the intensities of each WCP to the desired intensity level. This is indeed the implementations reported in most of decoy state QKD experiments. Another solution for decoy state implementation is to use multiple laser diodes of different intensities to generate different states~\cite{Peng2007}. In the passive decoy state method, heralded single photon source are often applied \cite{mauerer2007quantum,Adachi2007Simple,Ma2008PDC}. The probability distribution is changed by observing different measurement outcomes of the heralded photons.

A popular source for decoy state method is the phase-randomized weak coherent state source, as shown in Eq.~\eqref{eq:coherentFock}. To apply the active decoy method, Alice randomly adjusts the intensity $\mu$ of the coherent state, which is related to different Poisson distribution $P_\mu(n)$. Alice estimates the single photon yield $Y_1$ and error $e_1$ by solving the equation provided by the observed gain $Q_\mu$ and quantum bit error rate (QBER) $E_\mu$ related to different intensity $\mu$
\begin{equation} \label{eq:QEmu}
\begin{aligned}
Q_\mu &= \sum_{n=0}^{\infty} P_\mu(n) Y_n, \\
E_\mu Q_\mu &= \sum_{n=0}^{\infty} P_\mu(n) e_n Y_n,
\end{aligned}
\end{equation}
where $P_\mu(n)=\mu^n e^{-\mu}/n!$ for the coherent state case.

Following the GLLP security analysis, Eq.~\eqref{eq:GLLP}, the key rate is given by
\begin{equation} \label{eq:decoyR}
\begin{aligned}
R &\ge -Q_\mu H(E_\mu) + Q_1[1-H(e_1)],
\end{aligned}
\end{equation}
where $Q_1=Y_1 \mu e^{-\mu}$. Here, the gain $Q_\mu$ and QBER $E_\mu$ can be directly obtained from experiment, and the signal intensity $\mu$ is set by Alice. With a tight estimation on $Y_1$ and $e_1$ by solving the linear equations in the form of Eq.~\eqref{eq:QEmu}, the key rate can be improved from $O(\eta^2)$ to $O(\eta)$.

In practice, only several different intensities are enough to make an accurate estimation. The most popular practical decoy state method is vacuum and weak decoy state method \cite{Lo2004Vacua,Ma2004PhD}. That is, Alice randomly generates coherent states with three different intensities $\{0,\nu,\mu\}$, where states with intensity $\mu$ is the signal states for key generation, and states with intensity $\nu<\mu$ and vacuum state with intensity $0$ is for parameter estimation. The two parameters need to estimate in Eq.~\eqref{eq:decoyR} can be bounded by \cite{ma2005practical},
\begin{equation}\label{Practical:Y1Le1U}
\begin{aligned}
Y_1 &\ge Y_1^L = \frac{\mu}{\mu\nu-\nu^2}\big(Q_\nu e^{\nu}-Q_\mu e^{\mu}\frac{\nu^2}{\mu^2} -\frac{\mu^2-\nu^2}{\mu^2}Y_0\big) \\
e_1 &\le e_1^U = \frac{E_\nu Q_\nu e^\nu-e_0Y_0}{Y_1^L\nu}.
\end{aligned}
\end{equation}
A similar result is also derived in \cite{wang2005beating}.

For finite-data size effect, Ma \emph{et al.} took the first step to analyze the statistical fluctuations using standard error analysis, which essentially assumes i.i.d.~channel behavior \cite{ma2005practical}. The idea is that instead of directly using $Q_\mu$ and $E_\mu$ obtained from the experiment directly, one assumes these parameters fluctuates according to a normal distribution. Then, in Eq.~\eqref{Practical:Y1Le1U}, one can substitute the upper and lower bounds of $Q_\mu$ and $E_\mu$. The failure probability for this estimation would link to the number of standard deviations used for bounds.

The finite data size effect has been discussed in a more rigorous manner in \cite{lim2014concise,Zhang2017Improved}. It turns out that the formulas used in the standard error analysis approach \cite{ma2005practical} can be directly applied with a different value of failure probabilities in parameter estimation, as presented in Table \ref{tab:deviations}.

\begin{table} [htb]
\centering
\caption{The failure probability as a function of the fluctuation deviations, measured by the number of standard deviations, $(\chi-\mathbb{E}^L[\chi])/\mathbb{E}^L[\chi] = (\mathbb{E}^U[\chi]-\chi)/\mathbb{E}^L[\chi]=n\sigma$, where $\chi$ is counts in experiment. Here, $\varepsilon_{G}$,  $\varepsilon_{\infty}$, $\varepsilon_{10000}$, and $\varepsilon_{70}$, respectively, denote failure probabilities for the bounds in the Gaussian approximate analysis, the rigourous method with large data size limit, and a data size of 10000 and 70. [A similar table is also presented in \cite{Zhang2017Improved}]}\label{tab:deviations}
\begin{tabular}{cccccc}
\hline
\hline
\textbf{Deviation} & \textbf{$\varepsilon_{G}$} & \textbf{$\varepsilon_{\infty}$} & \textbf{$\varepsilon_{10000}$} & \textbf{$\varepsilon_{70}$} \\ \hline
$2\sigma$ & $10^{-1.34}$ 	& $10^{-0.57}$ 	& $10^{-0.57}$ 	& $10^{-0.57}$  \\
$3\sigma$ & $10^{-2.57}$ 	& $10^{-1.65}$ 	& $10^{-1.65}$ 	& $10^{-1.54}$  \\
$4\sigma$ & $10^{-4.20}$ 	& $10^{-3.17}$ 	& $10^{-3.17}$ 	& $10^{-2.65}$  \\
$5\sigma$ & $10^{-6.24}$ 	& $10^{-5.13}$  & $10^{-5.09}$ 	& $10^{-3.92}$  \\
$6\sigma$ & $10^{-8.70}$ 	& $10^{-7.52}$  & $10^{-7.43}$ 	& $10^{-5.36}$  \\
$7\sigma$ & $10^{-11.59}$	& $10^{-10.34}$ & $10^{-10.13}$ & $10^{-6.95}$  \\
$8\sigma$ & $10^{-14.91}$ 	& $10^{-13.60}$ & $10^{-13.18}$ & $10^{-8.67}$  \\
$9\sigma$ & $10^{-18.65}$ 	& $10^{-17.29}$ & $10^{-16.60}$	& $10^{-10.50}$ \\
$10\sigma$ &$10^{-22.82}$ 	& $10^{-21.41}$ & $10^{-20.36}$	& $10^{-12.38}$ \\
\hline
\hline
\end{tabular}
\end{table}

Other than weak coherent state photon sources, one can also use PDC source or thermal source, as reviewed in Section \ref{Sub:Source}. As long as the photon number distribution of the source is different from Poisson distribution, one can employ the passive decoy state scheme \cite{Adachi2007Simple,Ma2008PDC}, where Alice splits the pulses with a beam splitter, detects one arm as triggers, and uses the other arm for QKD encoding. Conditioned on the detection of triggering signals, the photon number distribution of the encoding arm is different. Alice can announce her local detection after Bob's detection. Then, they can have linear gains for photon pulses with different (conditional) photon number distributions for the decoy-state method analysis. It turns out that even with phase randomized coherent states, one can employ the passive decoy state method \cite{Curty2010PassiveDecoy}.

\subsubsection{Experiment} \label{sec5:decoyexp}
\begin{figure*}[hbt]
\centering \resizebox{16cm}{!}{\includegraphics{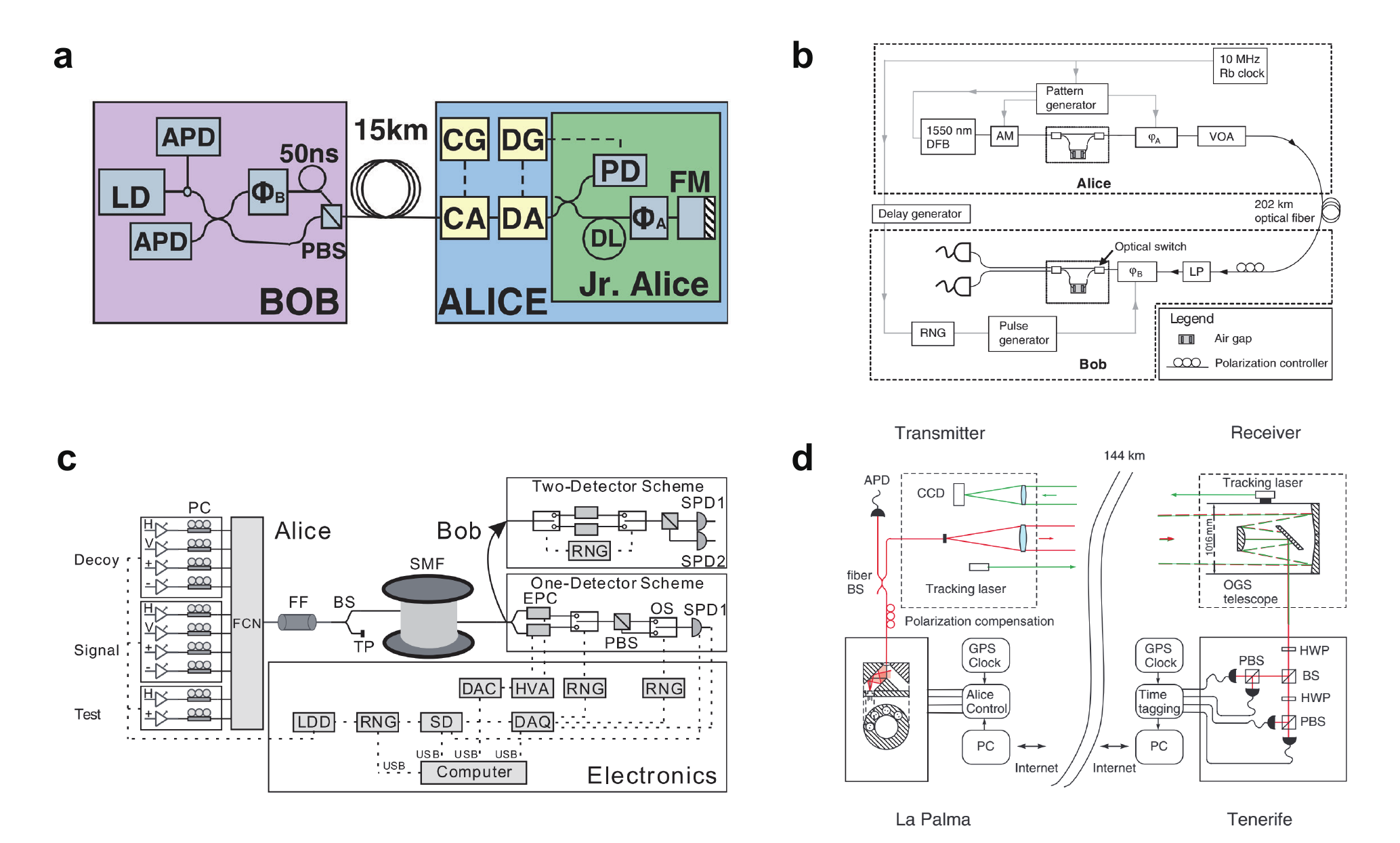}}
\caption{(Color online) Decoy state QKD experiments. \textbf{a}, Experiment on a commercial plug-and-play QKD system~\cite{Yi2006}. CA, compensating AOM; CG, compensating generator; DA, decoy AOM; DG, decoy generator; LD, laser diode; $\phi$, phase modulator; PD, classical photo detector; DL, delay line; FM, faraday mirror. \textbf{b}, Phase-encoding experiment~\cite{rosenberg2007long}. DFB, distributed feedback laser; VOA, variable optical attenuator; AM, amplitude modulator; LP, linear polarizer. \textbf{c}, Polarization-encoding experiment~\cite{Peng2007}. FCN, fiber coupling network; FF, fiber filter; EPC, electric polarization controller; DAC, digital-to-analog converter. \textbf{d}, Free-space experiment~\cite{Tobias2007}. BS, beam splitter; PBS, polarizing beam splitter; HWP, half-wave plate; APD, avalanche photo diode. [Figures reproduced from~\cite{Yi2006,rosenberg2007long,Peng2007,Tobias2007}].} \label{fig:decoy100}
\end{figure*}

The decoy state methods have been widely implemented in different QKD systems. The decoy-state experiments are summarized in Table~\ref{Tab2}. Fig.~\ref{fig:decoy100} shows the four initial decoy-state QKD experiments. Zhao~\emph{et al.} reported decoy state experiments \cite{Yi2006,zhao2006simulation} up to 60-km fiber on top of a commercial plug-and-play QKD system; Peng \emph{et al.}~\cite{Peng2007} implemented decoy-state QKD over 102-km fiber using a one-way polarization-encoding QKD system; Rosenberg \emph{et al.}~\cite{rosenberg2007long} implemented decoy-state QKD over 107-km fiber using a one-way phase-encoding QKD system; Schmitt-Manderbach \emph{et al.} achieved 144 km decoy state QKD in free space~\cite{Tobias2007}. These experiments demonstrated that decoy state BB84 was secure and feasible under real-world conditions.

\begin{figure}[hbt]
\centering \resizebox{7.8cm}{!}{\includegraphics{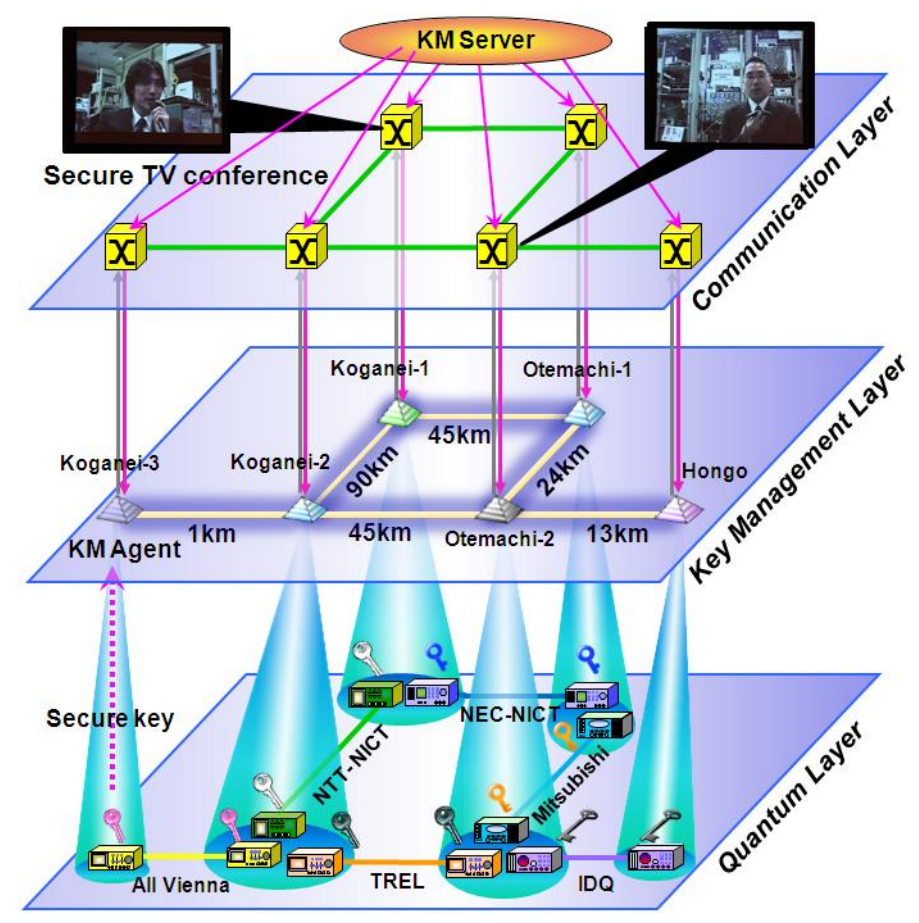}}
\caption{(Color online) Architecture of the Tokyo QKD Network~\cite{sasaki2011field}.  [Figure reproduced from~\cite{sasaki2011field}].} \label{fig:tokyo}
\end{figure}

Since then, more and more experimental efforts have been made to QKD deployments in labs and field tests. In 2007, Yuan \emph{et al.} realized a stabilized one-way phase-encoding decoy state QKD system~\cite{yuan2007unconditionally}. Later, Dixon \emph{et al.}~implemented decoy state QKD with a high clock rate of 1 GHz~\cite{Dixon2008} and Liu \emph{et al.} extended decoy state QKD to long distance of 200-km fiber~\cite{liu2010decoy}. Importantly, a number of field QKD networks with the decoy-state implementation have been built in Europe~\cite{peev2009secoqc}, Japan~\cite{sasaki2011field}, China~\cite{chen2009field,chen2010metropolitan,wang2010field} and so forth. An illustration of the Tokyo QKD network is shown in Fig.~\ref{fig:tokyo}.

In the mean time, Wang et al.~experimentally implemented decoy state with a PDC source~\cite{Qin2008}. Also, the passive decoy-state method has also been demonstrated \cite{Sun2014Passive}. Recently, the decoy state experiment has been extended to a record-breaking distance of 1,200 km in free space~\cite{Liaosate} and 421 km in ultra low-loss optical fiber~\cite{boaron2018secure}. Due to its convenient implementation and remarkable enhancement on performance, the decoy-state method becomes a standard technique in current QKD implementations.

\subsection{Source flaws}

\subsubsection{Basis-dependent source}
In practice, there is often some difference between $\rho_x$ and $\rho_z$, i.e., Eq.~\eqref{eq:basisindep} might not be fulfilled. Then, in the worst case scenario, we should assume Eve is capable of distinguishing the basis choice and hence she can attack two basis states separately. This kind of source is called basis-dependent source. Obviously, the more state dependence on the basis, the easier for Eve to distinguish the bases and hence a lower key rate.

Without loss of generality, we take $Z$-basis as example. The general Shor-Preskill's key rate formula is \cite{shor2000simple}
\begin{equation} \label{eq:ShorPreskillRbeta}
\begin{aligned}
r \ge 1-H(e_{Z})-H(e_{Z}^p), \\
\end{aligned}
\end{equation}
where $e_{Z}^p$ is the $Z$-basis phase error rate, defined in Eq.~\eqref{eq:biterror}.

For a basis-dependent source, $e_{Z}^p \neq e_{X}$ since $\rho_Z \neq \rho_X$. However, if $\rho_X$ is close to $\rho_Z$, we can still bound $e_{Z}^p$ from measured $e_X$. In the GLLP security analysis framework~\cite{gottesman2004security}, the basis dependence is quantified by a bias,
\begin{equation} \label{eq:SourceBias}
\begin{aligned}
\Delta=\frac{1-F(\rho_X,\rho_Z)}{2}, \\
\end{aligned}
\end{equation}
where $F(\rho_X,\rho_Z)=\sqrt{\sqrt{\rho_Z}\rho_X\sqrt{\rho_Z}}$ is the fidelity between the two states. Given this bias, the phase error rate used in the key rate formula can be bounded by \cite{koashi2009simple,Lo2007Nonrandom},
\begin{equation} \label{eq:ephasebound}
\begin{aligned}
e_z^p \le e_j^b+4\Delta(1-\Delta)(1-2e_x^b)
\\+4(1-2\Delta)\sqrt{\Delta(1-\Delta) e_x^b(1-e_x^b)}. \\
\end{aligned}
\end{equation}

For the practical photon sources presented in Section \ref{Sub:Source}, Alice and Bob have more information than the bias in Eq.~\eqref{eq:SourceBias}. For example, in principle, they can measure the photon number $n$, with which they can tag each quantum signal. Then, in phase error correction of entanglement distillation process, which would be reduced to privacy amplification for prepare-and-measure schemes, they could take advantage of these tagging. With tagging, the GLLP key rate formula can be written as \cite{gottesman2004security},
\begin{equation} \label{eq:GLLPR}
\begin{aligned}
r \ge -H(E)+(1-\Delta)[1-H(e_z^p)], \\
\end{aligned}
\end{equation}
where $E$ is the total QBER, $\Delta$ is the ratio of tagged signals, and $e_z^p$ is the phase error rate of the untagged signals. Here, we use the same notation of the bias $\Delta$ in Eq.~\eqref{eq:SourceBias}.

\subsubsection{Nonrandom phase}\label{Sc:NonRandom}
A general example of source flaw is to use the weak coherent states with nonrandom phases to encode the basis and key information \cite{Lo2007Nonrandom}. Their difference is treated as source flaw, i.e, a basis-dependence of the source. The encoded state $\ket{\psi_{\beta\kappa}}_B$ is
\begin{equation}
\ket{\psi_{\beta\kappa}}_B = \ket{\alpha}_R \ket{\alpha e^{i\pi(\kappa + \frac{1}{2} \beta)}}_S,
\end{equation}
where $\alpha$ is constant and $\mu=2|\alpha|^2$ is the intensity. In this case, the basis dependence $\Delta$ is
\begin{equation}
\Delta = \dfrac{1}{2}\left( 1 - e^{\mu/2}( \cos(\mu/2)+\sin(\mu/2) ) \right) = \mu/8 + O(\mu^3).
\end{equation}

Note that in the practical QKD experiment, we will post-select the clicked signals. In this case, to calculate the basis dependence, we have to take the channel transmittance $\eta$ into account. In a worst-case scenario, the channel loss is caused by Eve's selection on the transmitted signals. To clarify this, we can consider Eve performs a unambiguous state discrimintation (USD) attack \cite{Duifmmode2000unambiguous}, where Eve performs USD to discriminate $\rho_X$ and $\rho_Z$. If the discrimination is successful, Eve can learn the basis and key, then he generates the same state $\rho_{\beta\kappa}$, and sends it to Bob; if the discrimination fails, Eve partially blocks the signal as loss. In this case, the basis dependence $\Delta^\prime$ of left signals will be amplified by $\eta$
\begin{equation} \label{eq:Deltaprime}
\Delta^\prime = \Delta/(\eta\mu) \approx \mu/(8\eta).
\end{equation}

From Eqs.~\eqref{eq:ShorPreskillRbeta},~\eqref{eq:ephasebound},and~\eqref{eq:Deltaprime}, we can calculate the key rate. However, the achievable key generation rate scales only quadratically with the transmittance $\eta$ in the channel, i.e., $r = O(\eta^2)$. This question can be potentially solved using the scheme of discrete phase randomization~\cite{Cao2015Discrete}.

\subsubsection{Encoding flaws}
Another example of source flaw is the encoding flaws in the phase and polarization encoding due to the device imperfections in the encoding devices. This will also make the source basis-dependent. Although GLLP allows the security proof to consider th encoding flaws, the key rate drops dramatically~\cite{gottesman2004security}. This is because GLLP has a pessimistic consideration by assuming that the encoding flaws are in arbitrary dimensions. To address this issue, a loss-tolerant protocol was proposed in~\cite{tamaki2014loss}, which makes QKD tolerable to channel loss in the presence of source flaws~\cite{yin2014mismatched}.

On the basis of the assumption that the single-photon components of the states prepared by Alice remain inside a two-dimensional Hilbert space, it was shown that Eve cannot enhance state preparation flaws by exploiting the channel loss and Eve’s information can be bounded by the rejected data analysis. The intuition for the security of loss-tolerant QKD protocol~\cite{tamaki2014loss} can be understood in the following manner. By assuming that the state prepared by Alice is a qubit, it becomes impossible for Eve to perform an unambiguous state discrimination (USD) attack. Indeed, in order for Eve to perform a USD attack, the states prepared by Alice must be linearly independent; but by having three or more states in a two-dimensional space, in general the set of states prepared by Alice is linearly dependent, thus making USD impossible. The above loss-tolerant protocol has been further developed and demonstrated experimentally for decoy-state BB84~\cite{xu2014experimental,boaron2018secure} and MDI-QKD~\cite{tang2016experimental}.

\subsection{Leaky source} \label{sec:4:THA}
As discussed in Section~\ref{sec:3:otherattacks}, the source is vulnerable under the Trojan-horse attack (THA). In particular, Eve could inject bright light pulses into Alice's transmitter and then measure the back-reflected light to extract information about Alice's state preparation process. This problem has been analyzed in~\cite{lucamarini2015practical}. The authors evaluated the security of a QKD system in the presence of information leakage from Alice's phase modulator (PM), which is used to encode the bit and basis information of the generated signals. A key observation is that, the joint state of Alice's transmitted signals and Eve's back-reflected light from her THA is not basis-independent but it depends on Alice's basis choice. The security of the system can be analyzed by quantifying Eve's information and considering this information in privacy amplification, based on the techniques introduced in~\cite{Lo2007Nonrandom}. Recently, these seminal results have been generalized to prove the security of decoy-state QKD in the presence of arbitrary information leakage from both the PM and the intensity modulator (IM)~\cite{tamaki2016decoy,WangNJP2018}. Here the IM is normally used to select the intensity setting for each emitted signal. Consequently, it is possible to quantify the amount of device isolation, against THA, so as to achieve a certain performance with a realistic leaky QKD system.

\section{Detection Security}\label{sec:5}
In this section, we review the various approaches to address the detection security of practical QKD. We will review the MDI-QKD protocol and its extensions in more details.

\subsection{Countermeasures against detection attacks}
Many approaches have been proposed to defeat the attacks at detection. The first one is \emph{security patch}. That is, once one discovers a new type of attack, a corresponding countermeasure against this attack can be proposed and realized in an existing QKD system. This approach usually only requires modifying the software or the hardware of a current system. For instance, the time-shift attack introduced in the previous section can be avoided by simply shifting the gating window of the detectors at random~\cite{Qi2007}. The detector blinding attack could, in principle, be avoided by monitoring the detector's photocurrent for anomalously high values~\cite{yuan2011resilience,da2012real} or by randomly varying the detector efficiency~\cite{lim2015random}. Although security patch can defeat certain attacks, the patched countermeasures themselves might open other loopholes. This could as a result, introduce one more layer of security risk~\cite{Shihan2015,huang2016testing,qian2019robust}. Furthermore, the major issue associate with the security patch is that they only prevent the known attacks. For potential and \emph{unknown} attacks, the countermeasures may fail. Therefore, security patch is only ad-hoc, which abandons the information-theoretic security framework of QKD.

The second approach is to \emph{fully characterize} the devices using in a QKD system and precisely describe the devices in mathematical models. Then the models can be included in the security proof to estimate the real secure key rate based on an imperfect setup. A well-known example is the GLLP security proof~\cite{gottesman2004security}. While this approach seems straightforward, developing models to fully match the practical behavior of various QKD devices is rarely possible because the components are complex. Even so, there are several ongoing theoretical efforts to consider as many imperfections as possible into the security proof~\cite{fung2009security,maroy2010security,tamaki2014loss,lucamarini2015practical,tamaki2016decoy}. Nevertheless, this approach is limited by our understanding of the devices and a complete knowledge of the devices is rather challenging. Hence full characterization is still \emph{ad-hoc}.

The third approach is device-independent QKD (DI-QKD), which will be reviewed in Section~\ref{sec:diQKD}. Note that there are also proposals for semi-device-independent QKD, where one party's measurements are fully characterized while the other's
are unknown~\cite{PhysRevA.85.010301,pawlowski2011semi,smith2012conclusive}.

The final approach is MDI-QKD protocol, which closes \emph{all} detection attacks and is practical with current technology. In below, we will review MDI-QKD in detail.

\subsection{Measurement-device-independent scheme} \label{sec:5:mdi}
MDI-QKD generates secret keys based on the ``time-reversed" entanglement protocol and leaves all the single photon detections to a public \emph{untrusted} relay (Eve).

\subsubsection{Time-reversed EPR QKD}

\begin{figure}[hbt]
\begin{center}
 \includegraphics[scale=0.66,angle=0]{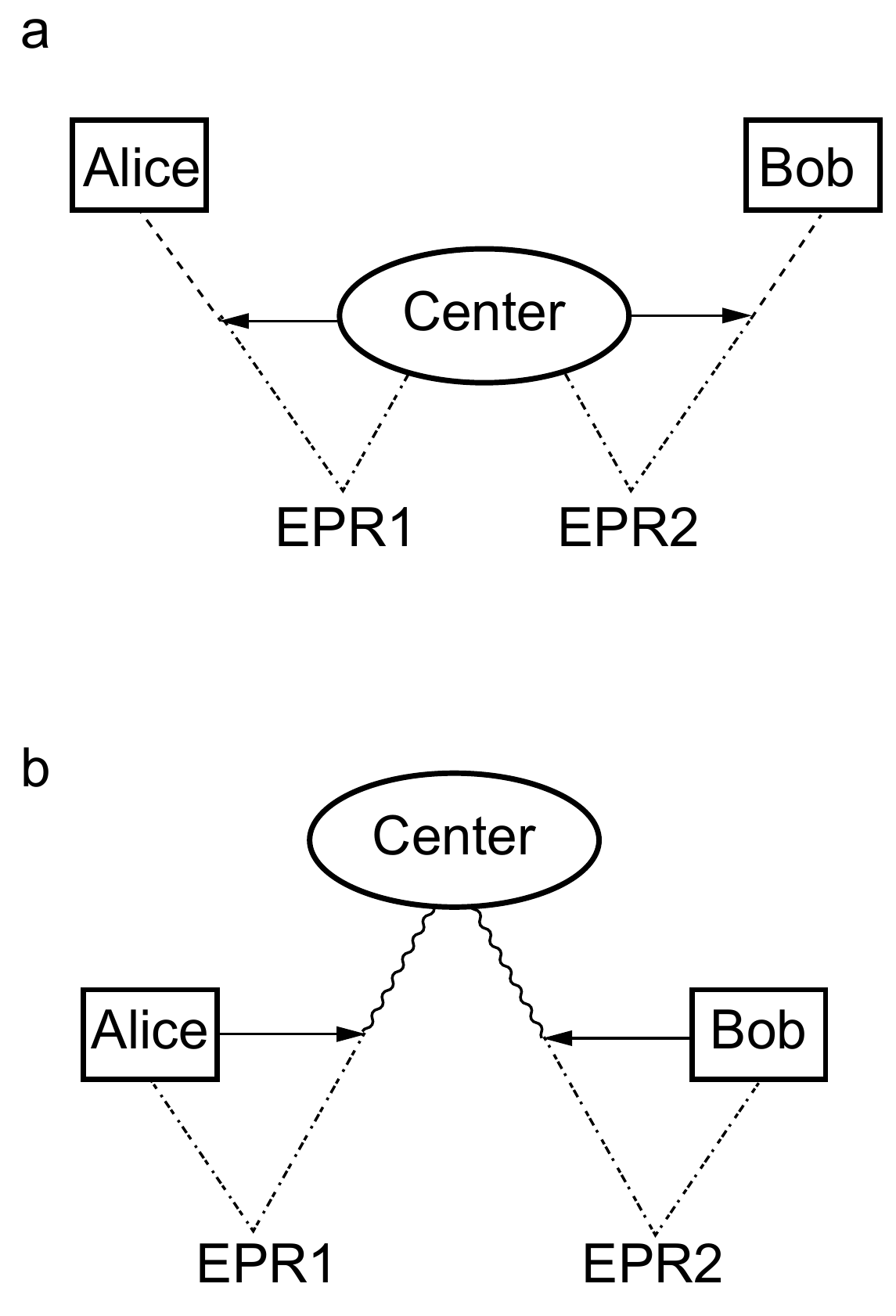}
 \end{center}
 \caption{Einstein-Podolsky-Rosen (EPR) based QKD protocol~\cite{biham1996quantum}. One particle of each EPR correlated pair, denoted by dashed lines, is sent to the center, who performs a Bell state measurement (BSM). The second particles are sent to Alice and Bob, respectively, who project them onto the BB84 states. \textbf{a}, Original EPR QKD. The first measurement is done by the center and the particles arriving at Alice and Bob are therefore in the Bell state, which can be used to do QKD as in the EPR based protocol~\cite{ekert1991quantum}. \textbf{b}, Time-reversed EPR QKD. The first measurement is performed by Alice and Bob and each particle sent to the center is therefore in one of the BB84 states, which forms the concept behind MDI-QKD~\cite{lo2012measurement}. [Figure reproduced from~\cite{biham1996quantum}].
\label{Fig:eprprotocol}}
\end{figure}

The idea of measurement-device-independent (MDI) is inspired by the Einstein-Podolsky-Rosen (EPR) based QKD protocol~\cite{ekert1991quantum,bennett1992quantum}. This is illustrated in Fig.~\ref{Fig:eprprotocol}~\cite{biham1996quantum}. In the initial EPR-based protocol (Fig.~\ref{Fig:eprprotocol}a), Alice and Bob individually prepare an EPR pair at each side and send one photon from each pair to an untrusted center party, Charles. Charles then performs a Bell state measurement (BSM) for entanglement swapping. The measurement result is announced. Once the BSM is finished, Alice and Bob measure the other photon of the EPR pairs locally by choosing between the $X$ and $Z$ basis randomly. Comparing a subset of their measurement results allows Alice and Bob to know whether Charles is honest. Then Alice and Bob can generate the secret by using the BBM92 protocol \cite{bennett1992quantum}.

Note that the EPR protocol can also work in a \emph{time-reversal} version, as shown in Fig.~\ref{Fig:eprprotocol}b. That is, Alice and Bob can measure their local photons first, instead of waiting for Charles' measurement results. This order of preparation and measurement is equivalent to that of the prepare-and-measurement QKD scheme, in which Alice and Bob prepare BB84 states and send them to Charles to perform the BSM. After that, the Charles' honesty can be checked by comparing a part of Alice's and Bob's results. Importantly, Charles' BSM is only used to check the \emph{parity} of Alice's and Bob's bits, and thus, it does not reveal any information about the individual bit values. This time-reversal EPR protocol forms the main concept behind MDI-QKD.

This time-reversed EPR QKD protocol has been first proposed in~\cite{biham1996quantum}. Later, Inamori provided a security proof~\cite{inamori2002security}. Nevertheless, these two important works offered very limited performance and, therefore, they have been largely forgotten by the QKD community. For instance, the scheme in~\cite{biham1996quantum} requires perfect single-photon sources and long-term quantum memories, which renders it unpractical with current technology. Inamori's scheme~\cite{inamori2002security} uses practical weak coherent pulses (WCPs) but it does not include decoy states, since it was proposed long before the advent of the decoy-state protocol. Moreover, the two early papers~\cite{biham1996quantum,inamori2002security} were not specifically considering the side channel problem in QKD at all. In 2012, Braunstein and Pirandola performed a general security analysis of the time-reversed EPR QKD approach~\cite{braunstein2012side} and proved that detector side-channel attacks can be eliminated by using teleportation where any incoming quantum signals are excluded from access to the detectors. Note that the idea of using teleportation for the specific purpose of removing side channels was first discussed in footnote 21 of~\cite{lo1999unconditional}.

\subsubsection{MDI-QKD protocol}

\begin{figure*}[hbt!]
\centering \resizebox{15cm}{!}{\includegraphics{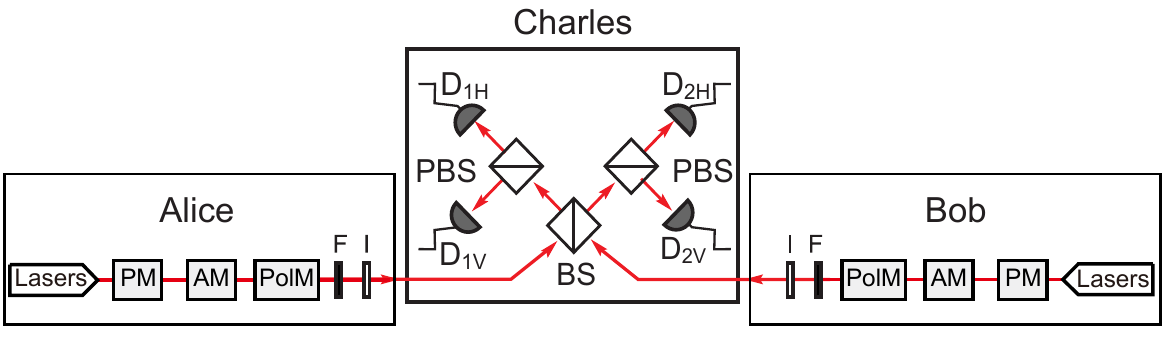}}
\caption{Schematic diagram of MDI-QKD proposed in~\cite{lo2012measurement}. Alice and Bob prepare BB84 polarization states using a decoy-state BB84 transmitter, same as the one illustrated in Fig.~\ref{fig:decoyimplement}. They send BB84 states to an \emph{untrusted} relay Charles/Eve. The relay is supposed to perform a Bell state measurement (BSM) that projects Alice's and Bob's signals into a Bell state. }\label{Fig:mdiqkd}
\end{figure*}

\begin{table}[!ht]\center
\begin{tabular}{l|l|l}
  \hline \hline
   \ & \textbf{Singlet state $\ket{\psi^{-}}$} & \textbf{Triplet state $\ket{\psi^{+}}$}  \\
     \hline
  Coincident clicks & $D_{1H}$ \& $D_{2V}$ or & $D_{1H}$ \& $D_{1V}$ or  \\
                  \ & $D_{2H}$ \& $D_{1V}$    &  $D_{2H}$ \& $D_{2V}$ \\
   \hline
  Rectilinear basis &  Bit flip & Bit flip \\
  \hline
  Diagonal basis & Bit flip & -- \\
  \hline
   \hline
\end{tabular}
\caption{Post-selection for MDI-QKD~\cite{lo2012measurement}. Alice and Bob post-select the events where the relay outputs a successful result and they use the same basis in their transmission. Moreover, either Alice or Bob flips her/his bits, except for the cases where both of them select the diagonal basis and the relay outputs a triplet}~\label{Tab:mdipost}
\end{table}

The MDI-QKD proposal~\cite{lo2012measurement} [see also~\cite{braunstein2012side}] builds on the time-reversed EPR QKD. In particular, the main merits of the proposal, introduced by Lo, Curty and Qi~\cite{lo2012measurement}, are twofold: first, it identified the importance of the results in~\cite{biham1996quantum,inamori2002security} to remove all detector side-channels from QKD implementations; second, it significantly improved the system performance with practical signals by including decoy states. The protocol can be summarized in four steps:
\begin{enumerate}
  \item Alice and Bob randomly and individually prepare one of four BB84 states using phase-randomized WCPs together with decoy signals. Then they send the states to an untrusted party, Charles.
  \item An honest Charles performs a BSM that makes Alice's and Bob's states interfere with each other, generating a Bell state. An example of a BSM implementation with linear optics in shown in Fig.~\ref{Fig:mdiqkd}: Charles interferes the incoming pulses at a 50:50 beam-splitter (BS), which has on each end a polarizing beam-splitter (PBS) that projects the photons into either horizontal ($H$) or vertical ($V$) polarization states. A ``click'' in the single-photon detectors $D_{\rm 1H}$ and $D_{\rm 2V}$, or in $D_{\rm 1V}$ and $D_{\rm 2H}$, indicates a projection into the singlet state $\ket{\psi^{-}}=(\ket{HV}-\ket{VH})/\sqrt{2}$, while a ``click'' in $D_{\rm 1H}$ and $D_{\rm 1V}$, or in $D_{\rm 2H}$ and $D_{\rm 2V}$, implies a projection into the triplet state $\ket{\psi^{+}}=(\ket{HV}+\ket{VH})/\sqrt{2}$. Other detection patterns are considered unsuccessful.
\item
Whether Charles is honest or not, he announces the outcome of his claimed BSM using a classical public channel when he claims to obtain a successful measurement.
\item
Alice and Bob keep the data that correspond to Charles' successful measurement events and discard the rest. Next, similar to the sifting in BB84 protocol, Alice and Bob announce their basis choices for sifting the events and keep the events using same bases. Based on Charles' measurement result, Alice flips part of her bits to guarantee the correct correlation with those of Bob. The post-selection strategy is illustrated in Table~\ref{Tab:mdipost}. Finally, they use the decoy-state method to estimate the gain and QBER of the single-photon contributions.
\end{enumerate}

In MDI-QKD, both Alice and Bob are senders, and they transmit signals to an untrusted third party, Eve, who is supposed to perform a Bell state measurement (BSM). Since the BSM is only used to post-select entanglement, it can be treated as an entirely \emph{black} box. Hence, MDI-QKD can remove all detection side-channels. The assumption in MDI-QKD is that the source should be trusted. The security assumptions of MDI-QKD, together with DI-QKD (see Section~\ref{sec:diQKD}), are summarized in Table~\ref{Tab:assumptions}. A comparison of practical security between MDI-QKD and DI-QKD, as commented by Charles H. Bennett in QCrypt 2018\footnote{See slide 6 of Charles H. Bennett's talk in Lightning Talks session of QCrypt 2018: http://2018.qcrypt.net/}, is summarized in Box~\ref{tab:bennett}.

\begin{tcolorbox}[title = {Box VI.B.2: A security remark about MDI-QKD and DI-QKD by Charles H. Bennett.}]
MDI-QKD at first sounds weaker than DI-QKD, but in fact it is stronger. In MDI-QKD, Eve's untrusted device remains outside Alice's and Bob's trusted enclosures. They need only trust themselves not to have inadvertently created a side channel to Eve through incompetent design of their do-it-yourself (DIY) light sources. By contrast, in DI-QKD they must trust Eve not to have deliberately created side channels from the untrusted devices to herself.
\end{tcolorbox}~\label{tab:bennett}

\begin{table}[!ht]\center
\begin{tabular}{l|l|l}
  \hline \hline
   \ & \textbf{DI-QKD} & \textbf{MDI-QKD}  \\
     \hline
  True random number generators & Yes & Yes \\
   \hline
  Trusted classical post-processing & Yes & Yes \\
  \hline
  Authenticated classical channel & Yes & Yes \\
 \hline
  No unwanted information leakage & \ & \  \\
  from the measurement unit & Yes & \textbf{No}  \\
  \hline
  Characterized source & \textbf{No} & Yes  \\
  \hline
   \hline
\end{tabular}
\caption{Security assumptions in DI-QKD and MDI-QKD. While DI-QKD has the advantage of being applicable to an uncharacterized source, it demands no unwanted information leakage from the measurement unit. MDI-QKD applies to {\it any} measurement units. This means that the measurement unit in MDI-QKD can be an entire black-box, purchased from untrusted vendors.}~\label{Tab:assumptions}
\end{table}

\subsubsection{Theoretical developments}

The decoy-state analysis is essential for MDI-QKD. The analysis is different from that of conventional decoy-state BB84 in that now both Alice and Bob send decoy signals to a common receiver (instead of only Alice sending decoy signals to Bob), which makes the mathematics slightly more complex. Fortunately, it has been shown that it is enough to obtain a tight estimation if Alice and Bob employ just a few decoy settings each. The authors of~\cite{ma2012statistical} and~\cite{wang2013three}, respectively, proposed a numerical method based on linear programming and an analytical approach based on Gaussian elimination. Both approaches assume that Alice and Bob can prepare a vacuum intensity. Following the similar analytical line, the authors of~\cite{Feihu:practical} studied the situation where none of the two decoy intensities are vacuum\footnote{A vacuum state is normally hard to realize in practice due to the finite extinction ratio of a practical intensity modulator}. A full parameter optimization method was proposed in~\cite{PhysRevA.89.052333}. Soon after, Yu et al proposed to use joint constraints for a better key rate~\cite{yu2015statistical}, and Zhou et al.~\cite{zhou2016making} proposed a four-intensity method, in which the key generation is conducted in $Z$ basis and the decoy analysis is performed only in $X$ basis. By doing so, the four-intensity method is efficient in the case of short data size. Recently, the four-intensity method was extend to seven-intensity method which can substantially enhance the key rate for MDI-QKD over asymmetric channels~\cite{wang2018enabling}. All these results provide experimentalists a clear path to implement MDI-QKD with a finite number of decoy states.

\begin{figure*}[!ht]
\centerline{\includegraphics[width=15cm]{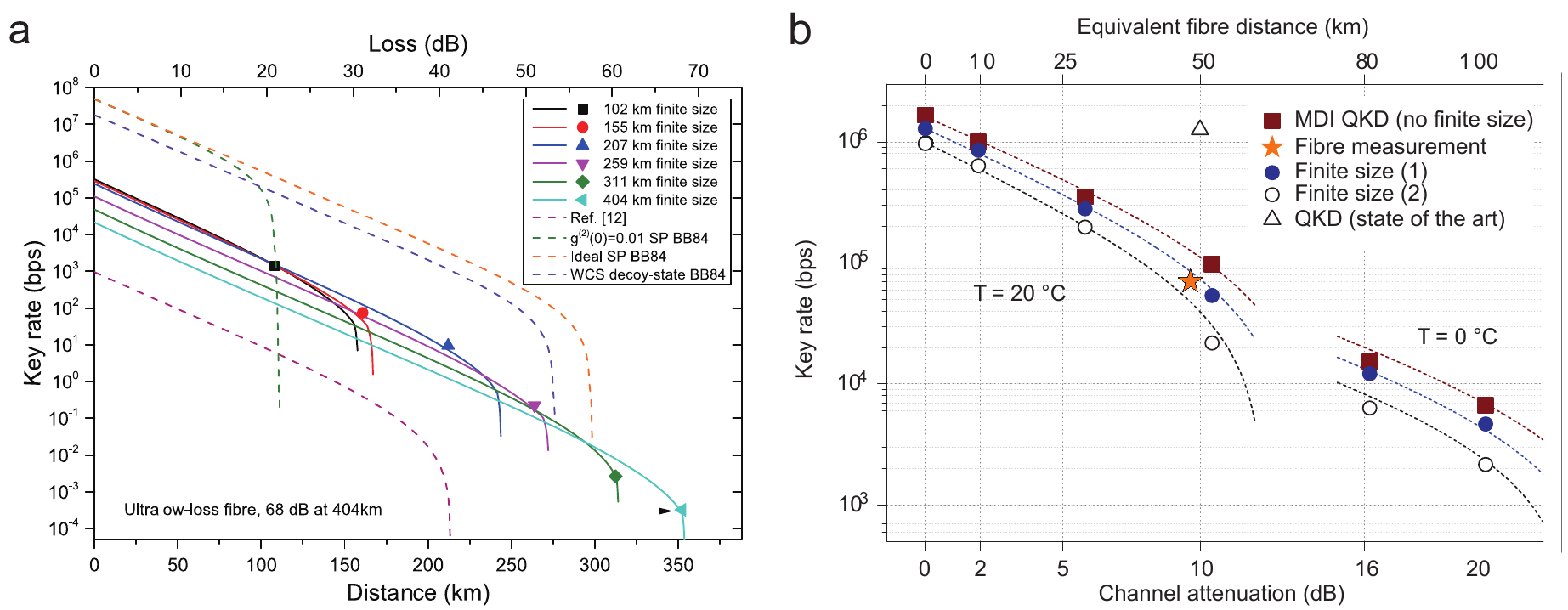}}
\caption{(Color online) Simulation and experimental secret rates of MDI-QKD demonstrated in~\cite{Yin2016,Comandar}. \textbf{a}, 404-km MDI-QKD~\cite{Yin2016}. The experimental results (symbols) agree well with the theoretical
simulations (solid lines). The dotted lines from upper to bottom show, respectively, the simulations for the balanced-basis passive BB84 protocol using ideal single-photon (SP) sources, the practical SP without the decoy-state method, the WCS with the decoy-state method, and the results of~\cite{yanlin2014}. \textbf{b}, 1-GHz MDI-QKD~\cite{Comandar}. Filled squares refer to key rates without the finite-size analysis. The star is the key rate obtained using two 25-km spools of fibre. The filled and open dots represent key rates with the finite-size analysis. The finite-size distillation methods are (1) standard error analysis~\cite{ma2012statistical}, and (2) composable security analysis~\cite{curty2014finite}. The dashed lines are simulations of the key rate for two different detector temperatures. [Figures reproduced from~\cite{Yin2016,Comandar}].}\label{Fig:mdirate}
\end{figure*}

For finite-key analysis, the authors of~\cite{ma2012statistical} provided an analysis that assumes a Gaussian distribution for the statistical fluctuations. Remarkably, Curty et al.~\cite{curty2014finite} presented a rigorous finite-key security proof against general attacks by using min-entropy analysis and Chernoff bound. In addition, this result satisfies the composable security definition. All these results confirm the feasibility of long-distance implementations of MDI-QKD within a reasonable time-frame of signal transmission. Simulations of the secret key rates with different kinds of decoy-state methods and finite-key analysis methods is shown in Fig.~\ref{Fig:mdirate}.

Other practical aspects were also extensively analyzed in theory. Besides polarization encoding in the original MDI-QKD protocol~\cite{lo2012measurement}, alternative schemes including phase encoding~\cite{tamaki2012phase} and time-bin encoding~\cite{Ma2012Alternative} were proposed and analyzed. To extend the transmission distance further, one could include quantum memories~\cite{abruzzo2014measurement,panayi2014memory}, entanglement sources~\cite{xu2013long} or adaptive operations \cite{azuma2015all2}. Moreover, a key security assumption in MDI-QKD is the source should be trusted. Recently, there have been efforts to prove the security of MDI-QKD when Alice's and Bob's encoding devices are flawed~\cite{tamaki2014loss,xu2014experimental}, or when their apparatuses are not fully characterized~\cite{yin2014mismatched}. Furthermore, a plug-play type of MDI-QKD was proposed in~\cite{xu2015measurement,choi2016plug} and experimentally demonstrated in~\cite{tang2016experimental}.

\subsubsection{Experimental developments}

Table~\ref{Tab3} summarizes the MDI-QKD experiments after its invention. The main experimental challenge of MDI-QKD is to perform a high-visibility two-photon interference between photons from two (Alice's and Bob's) independent laser sources~\cite{lo2012measurement}, which is not required in conventional QKD schemes. To do so, Alice's photons should be indistinguishable from those of Bob. Importantly, if one implements MDI-QKD over telecom fibres, it is necessary to include feedback controls to compensate the time-dependent polarization rotations and propagation delays caused by the two separated fibres. Note that in standard BB84 QKD systems, the requirement of compensating polarization rotations and propagation delays can be relaxed by using phase encoding, because the two optical pulses, which interfere with each other at the receiver's end, pass through the same optical fibre and thus experience the same polarisation rotation and phase change. Therefore, one can achieve high interference visibility without performing any polarization control. Nevertheless, this advantage of phase encoding (in comparison to other encoding schemes) cannot be directly translated to MDI-QKD, because the two pulses pass through two \emph{independent} quantum channels.

\begin{figure*}[ht!]
	\includegraphics[width=15.5cm]{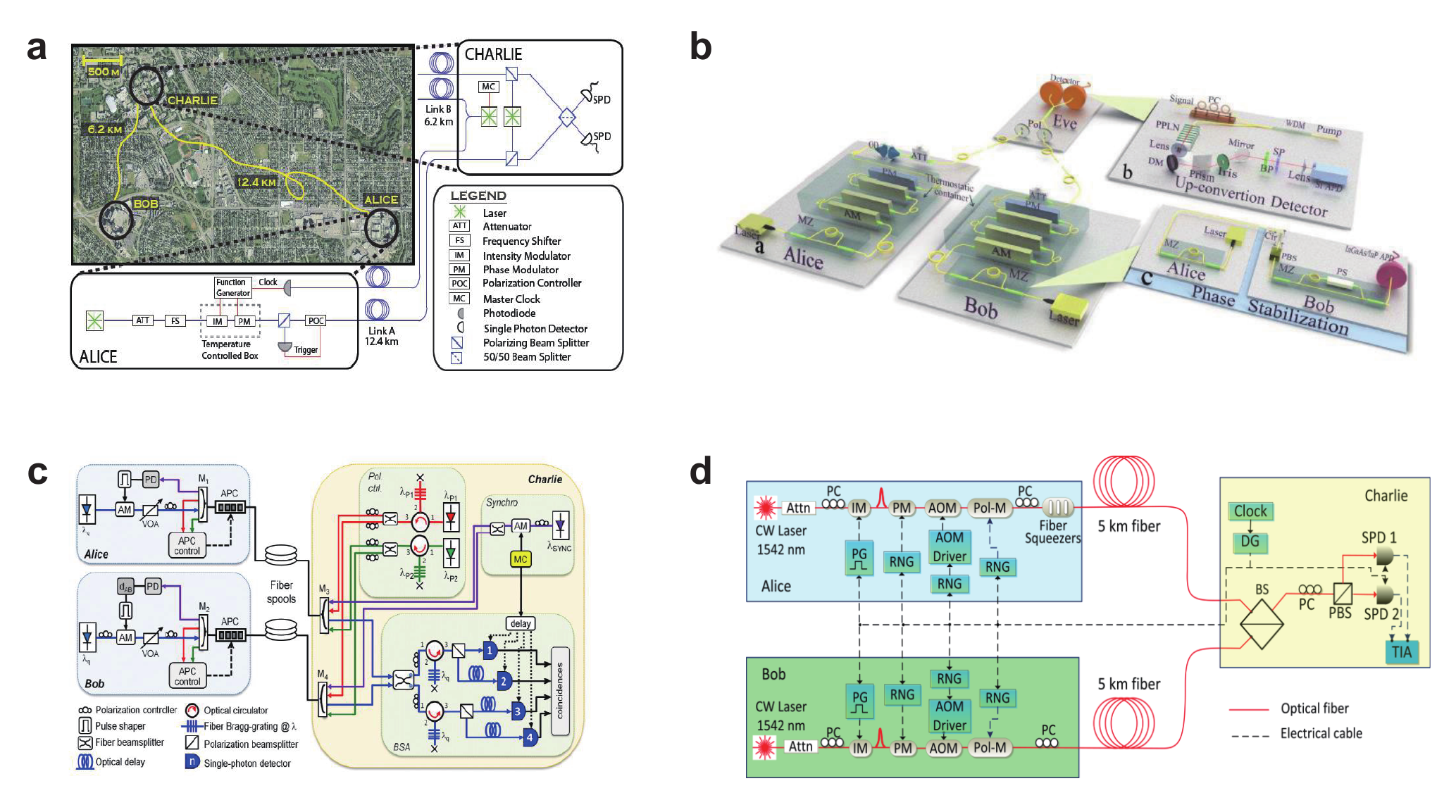}
	\caption{ (Color online) The four initial MDI-QKD experiments. \textbf{a}, Proof-of-principle MDI-QKD with time-bin encoding~\cite{Rub2013}. \textbf{b}, Full MDI-QKD implementation with random modulations of states and decoy intensities based on time-bin encoding~\cite{Liu2013}. \textbf{c}, Proof-of-principle MDI-QKD with polarization encoding~\cite{Silva2013}. \textbf{d}, Full MDI-QKD with random modulations of states and decoy intensities based on polarization encoding~\cite{Tang2014}. [Figures reproduced from~\cite{Rub2013,Liu2013,Silva2013,Tang2014}].}
	\label{Fig:exp4}
\end{figure*}

In 2013, several groups performed independent experimental study for MDI-QKD. Liu \emph{et al.}~\cite{Liu2013} reported the first demonstration of MDI-QKD with random modulation for encoding states and decoy states over 50 km fiber. Simultaneously, Rubenok \emph{et al.} were the first to demonstrate the feasibility of high-visibility two-photon interference between two independent lasers, passing through separate field-deployed fibres in a real world environment~\cite{Rub2013}. Later, Ferreira da Silva \emph{et al.} observed a similar interference using polarization encoding in the lab~\cite{Silva2013} and Tang \emph{et al.} reported a full demonstration of polarization encoding MDI-QKD with random modulation of encoding states and decoy states~\cite{Tang2014}. All these four initial experiments, when taken together, complete the cycle needed to demonstrate the feasibility of MDI-QKD using off-the-shelf optoelectronic devices. Their experiment diagrams are illustrated in Fig.~\ref{Fig:exp4}.

MDI-QKD is attractive not only because of its security against detection attacks, but also due to its practicality. It can resist high channel loss and reach long distance. Tang \emph{et al.}, implemented MDI-QKD over 200 km fiber~\cite{yanlin2014} and in field environment~\cite{Yanlin2015} by increasing the system clock rate from 1 MHz to 75 MHz, by developing an automatic feedback system, and utilizing superconducting single photon detectors (SNSPDs).

In 2016, two millstone MDI-QKD experiments that were subsequently reported. In the first one, Yin \emph{et al.}, extended the MDI-QKD distance to a record-breaking distance of 404 km by optimizing the implementation parameters and using a ultra-low loss fiber (0.16 dB/km)~\cite{Yin2016}. Importantly, the key rate achieved in the experiment at 100 km is around 3 kbps, which is sufficient for one-time-pad encoding of voice message. The results demonstrated in~\cite{Yin2016} are shown in Fig.~\ref{Fig:mdirate}a. In the second one, Comandar \emph{et al.}, increased the system clock rate of MDI-QKD to 1 GHz by exploiting the technique of optical seed lasers~\cite{Comandar}. The 1 GHz system demonstrated the feasibility for MDI-QKD to reach 1 Mbps key rate. The achieved secret rates in~\cite{Comandar} are shown in Fig.~\ref{Fig:mdirate}b.

Besides long distance and high rate, several research groups have analyzed the practical aspects in the implementation of MDI-QKD. For instance, Valivarthi \emph{et al.}, analyzed the trade-offs among complexity, cost, and system performance associated with the implementation of MDI-QKD~\cite{valivarthi2015measurement} and implemented a cost-effective system~\cite{valivarthi2017cost}. Wang \emph{et al.}, demonstrated a reference-frame-independent MDI-QKD that requires no phase reference between Alice and Bob~\cite{Wang2015,wang2017measurement} and this scheme was recently improved to a clock rate of 50 MHz by Liu et al.~\cite{liu2018polarization}. Tang \emph{et al.}, demonstrated MDI-QKD with source flaws \cite{tang2016experimental}. Roberts \emph{et al.}, reported a reconfigurable system to switch between QKD and MDI-QKD~\cite{roberts2017experimental}. Aside from the MDI-QKD demonstration with WCP sources, Kaneda \emph{et al.} demonstrated MDI-QKD using heralded single-photon source~\cite{kaneda2017quantum}. Furthermore, a continuous-variable version of MDI-QKD was also proposed and studied~\cite{pirandola2015high}, which will be reviewed in Section~\ref{sec:CVQKD}.

With all the above experimental efforts, MDI-QKD is ready for the applications in the future quantum network. Particularly, MDI-QKD is well suited to construct a centric star-type QKD network even with untrusted relays. Indeed, Tang \emph{et al.}, performed the first implementation of a field MDI-QKD network~\cite{Yanlin2016}, which has four nodes with one untrusted relay node and three-user nodes. Note that if the central relay is trusted, one can reconfigure the MDI-QKD network to allow many quantum communication protocols~\cite{roberts2017experimental}. Moreover, high-rate MDI-QKD over asymmetric fiber channels was demonstrated recently in~\cite{liu2018experimental}, based on the theoretical proposal in~\cite{wang2018enabling}. The asymmetric MDI-QKD is valuable to practical metropolitan network settings, where the channel losses are naturally asymmetric and the user nodes could be dynamically added or deleted. Furthermore, Wei et al., have implemented the first chip-based MDI-QKD at 1.25 GHz clock rate~\cite{wei2019high}. This is important to develop a low-cost and secure quantum network, where the expensive devices such as single-photon detectors can be placed in the central untrusted relay and each user requires only a simple Si chip.

\subsection{Twin-field QKD}
Fundamental bound~\cite{takeoka2014fundamental} and secret key capacity (SKC)~\cite{pirandola2017fundamental} have been obtained for the secure key rate vs distance of QKD. It was proven that, in the absence of relays, the key rate basically scales linearly with transmittance $O(\eta)$, where $\eta$ is the transmittance of the channel between Alice and Bob. This is called the linear bound (of the secret key rate of a lossy quantum channel). There are tremendous research interests towards developing a feasible scheme, known as quantum repeater~\cite{sangouard2011quantum}, to overcome the fundamental rate-distance limit. However, the deployment of quantum repeater is still beyond current technology.

\begin{figure}[!ht]
\includegraphics[width=7.5cm]{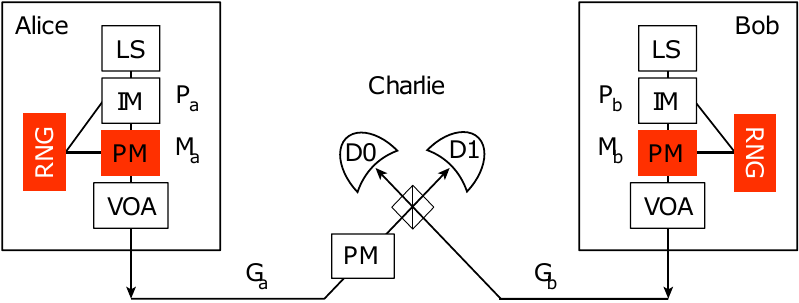}
\caption{Schematic diagram of TF-QKD~\cite{Lucamarini2018TF}. The light sources (LSs) at Alice’s and Bob’s stations generate pulses that are varied randomly by intensity modulators (IMs) to implement the decoy-state method. Phase modulators (PMs) are combined with random number generators (RNGs) to encode each light pulse with phases, which include bit and basis information as well as the random phases. The variable optical attenuators (VOAs) set the average output intensity of the pulses to bright (classical regime) or dim (quantum regime). The pulses travel along independent channels to then interfere on Charlie’s beam splitter and be detected by the single-photon detectors D0 and D1. Charlie uses the bright pulses in the classical regime and the phase modulator in his station to phase-align the dim pulses that are emitted in the quantum regime, which provide the bits of the key. [Figure adopted from~\cite{Lucamarini2018TF}.]}
\label{Fig:tfqkd}
\end{figure}

Remarkably, Lucamarini {\it et al.}~\cite{Lucamarini2018TF} have proposed a novel phase-encoding MDI-QKD protocol, called twin-field QKD (TF-QKD), which shows the possibility to overcome the SKC. In TF-QKD (see Fig.~\ref{Fig:tfqkd}), weak optical pulses are generated by two phase-locked laser sources, which are phase-randomized and then phase-encoded with secret bits and bases. The pulses are sent to Charlie for interference on a beam splitter; depending on which detector clicks, Charlie can infer whether the secret bits of the users (Alice and Bob) are equal or different, but cannot learn their absolute values. TF-QKD essentially uses single-photon interference~\cite{duan2001}, and the implementation requires only standard optical elements without the requirement of quantum memory~\cite{sangouard2011quantum}. The key goal of TF-QKD protocol is to achieve a quadratic improvement (i.e., scaling to  $O(\sqrt{\eta})$ to key rate as a function of channel transmittance. Unfortunately, in the original paper~\cite{Lucamarini2018TF}, such a quadratic improvement was only proven for a restricted class of attacks by Eve.

\begin{figure}[!ht]
\includegraphics[width=8cm]{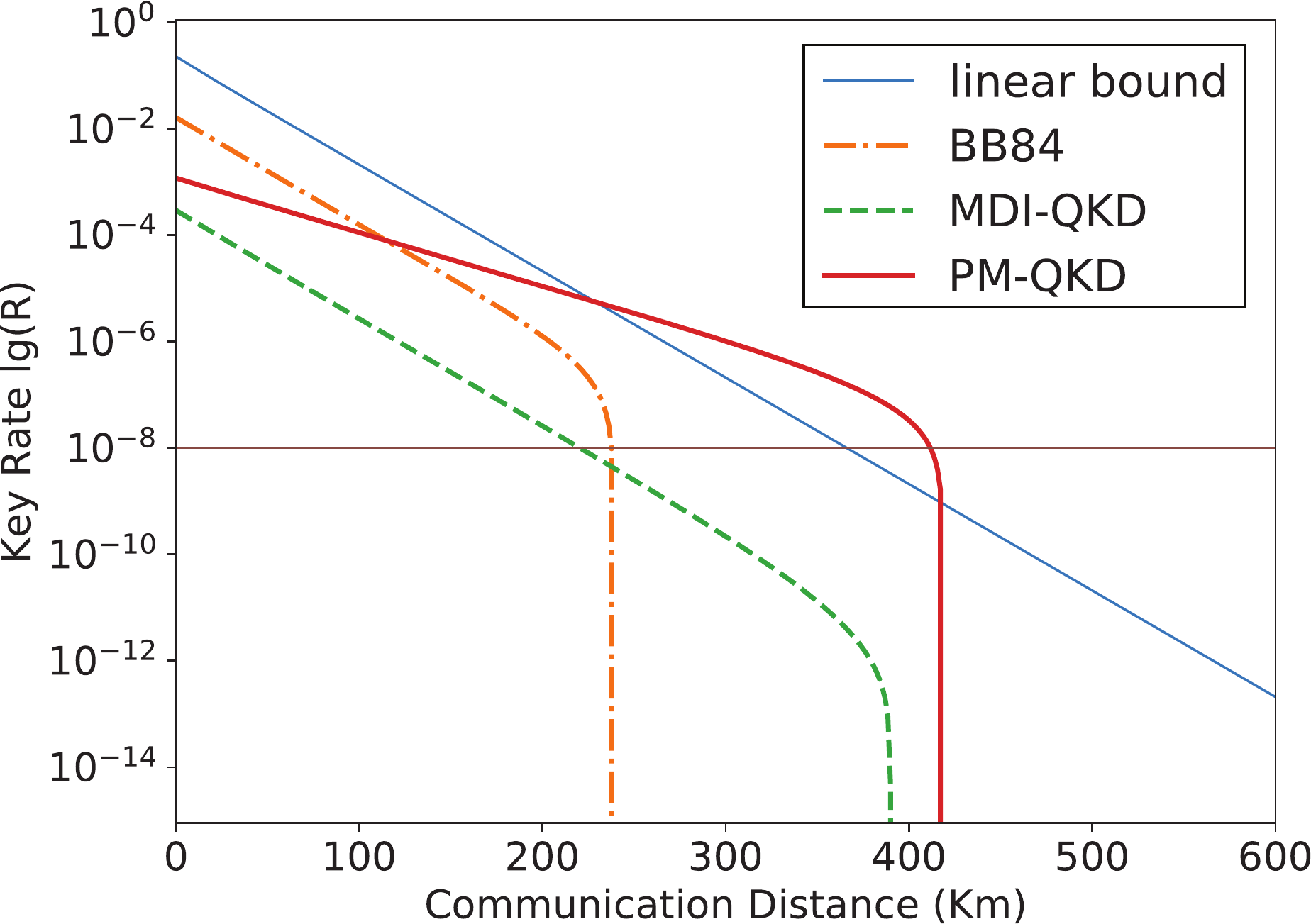}
\caption{Key rate of PM-QKD comparing to the theoretical SKC~\cite{pirandola2017fundamental} and other protocols~\cite{ma2018phase}. The key rate is shown to surpass the linear key rate bound when the communication distance $l>230$ km. The simulation uses realistic parameters: detector efficiency $14.5\%$, dark count rate $7.2\times 10^{-8}$, error correction efficiency $1.15$, channel misalignment error $1.5\%$ and number of phase slices $16$. [Figure reprinted from~\cite{ma2018phase}.]}
\label{Fig:pmqkd}
\end{figure}

Following the TF-QKD scheme~\cite{Lucamarini2018TF}, Ma \emph{et al.}~\cite{ma2018phase} proposed a protocol named phase-matching QKD (PM-QKD) and proved its unconditional security, inspired by the previous phase-encoding MDI-QKD protocol~\cite{tamaki2012phase}, and the MDI version of the Bennett1992 protocol~\cite{ferenczi2013security}. PM-QKD employs coherent states as information carriers directly and uses the decoy state method in an indirect way. In a sense, PM-QKD adopts a (discrete-modulation) continuous-variable encoding and discrete-variable single-photon detection. The performance of PM-QKD is shown in Figure \ref{Fig:pmqkd}. One can clearly see that its key rate can go beyond the linear SKC with certain realistic parameter settings.

On the other hand, with the BB84-type two-basis analysis, Tamaki \emph{et al.}~\cite{tamaki2018information} proposed a modified $X/Y$-basis protocol, and Wang \emph{et al.}~\cite{wang2018sending} proposed an $X/Z$-basis protocol, where the single-photon states are used regarded as the information carrier. Afterwards, the simplified coherent-state-based protocols without phase-randomization on the key generation mode have been proposed~\cite{curty2018simple,cui2018phase,lin2018simple} and analyzed in the infinite data size case. Later on, Maeda \emph{et al.}~\cite{maeda2019repeaterless} introduce an efficient parameter estimation method for the PM-QKD protocol and complete the finite-size analysis. All these recent theory works make the new TF-type MDI-QKD protocols important for the deployment of QKD over long distances.

Interestingly, the security of PM-QKD, different from the usual BB84-type two-basis protocol, is closely related to the symmetry of source state with respect to the encoding operation. To clearly establish the correlation between encoding symmetry and privacy, Zeng et al~\cite{Zeng2019PM} establish a symmetry-based security proof method for a general type of MDI-QKD protocols. For these MDI-QKD protocols, there exist symmetric source states, which promise perfect privacy, i.e., with no information leakage. Therefore, for a generic source state input, the privacy of the protocol only depends on the ratio of the symmetric component contained in it, irrelevant of the channel noise. As a result, this symmetry-based security proof allows higher error tolerance compared with the original complementarity-based proof. For example, PM-QKD is proved to be able to yield positive key even with high bit error rate up to $50\%$ and surpassing the linear key rate bound even with bit error rate of $13\%$~\cite{Zeng2019PM}.

From a technical point of view, the replacement of two-photon detection to single-photon detection is the key reason for the quadratic improvement, but single-photon interference with two remote independent lasers requires subwavelength-order phase stability for optical channels~\cite{duan2001}, which is more demanding in long-distance communication than achieving two-photon interference which does not require phase stability between the two photons. Nonetheless, TF-QKD protocols are expected to be feasible with the current techniques of active phase randomization, optical phase-locking and so forth. Indeed, in 2019, four research groups have reported the experimental demonstrations on the feasibility of TF-QKD respectively~\cite{LiuTF2019,wang2019beating,ZhongTF2019,minder2019experimental}. Very recently, the PM-QKD experiment has been realized with the random modulations and the consideration of finite-key effect~\cite{Fang2019surpassing}, whose key rate surpass the linear key rate bound via $302$ km and $402$ km commercial fibre; through a $502$ km ultra-low loss fibre with $87.1$ dB total loss, PM-QKD can yield key rate of $0.118$ bps with unconditional security. By using the ultra-stable cavity and optical phase lacking, the sending-or-not-sending version of TF-QKD protocol was demonstrated over $509$ km ultra-low loss fibre~\cite{chen2019sending}, where the achieved secure key rate is even higher than that a traditional QKD protocol running with a perfect repeaterless QKD device. A proof-of-principle experiment has demonstrated TF-QKD over optical channels with asymmetric losses~\cite{zhong2020proof}. Table~\ref{Tab:tfqkd} summarizes the recent TF-QKD experiments.


\section{Continuous-Variable QKD}\label{sec:CVQKD}

Broadly speaking, QKD can be divided into two classes, namely discrete variable (DV) or continuous variable (CV). Different from DV or qubit-based QKD, the secret keys in CV-QKD are encoded in quadratures of the quantized electromagnetic field and decoded by coherent detections~\cite{weedbrook2012gaussian}. Coherent detection is a promising candidate for practical quantum-cryptographic implementations due to its compatibility with existing telecom equipment and
high detection efficiencies without the requirement of cooling. CV-QKD protocols can be divided into several categories, according to the prepared state of coherent~\cite{gg02} or squeezed state~\cite{hillery2000quantum}, the modulation schemes of Gaussian~\cite{cerf2001} or discrete modulations~\cite{ralph1999continuous,hillery2000quantum,reid2000quantum}, the detection schemes of homodyne~\cite{gg02} or heterodyne detections~\cite{weedbrook2004quantum}, the error correction schemes of direct or reverse reconciliations~\cite{grosshans2003quantum}, and so forth.

In this section, we will primarily review the simplest and the most widely developed CV-QKD protocol -- Gaussian-modulated coherent state (GMCS) protocol~\cite{gg02,grosshans2003quantum} -- which is believed as the core of today's implementations. We will briefly discuss the security analysis and experimental developments of CV-QKD, together with a focus on the practical security aspects in its implementations, including the side channels and the advanced countermeasures. We will not cover too much about other CV-QKD protocols~\cite{silberhorn2002continuous,pirandola2008continuous,weedbrook2010quantum,usenko2015unidimensional}, which can be found in an earlier review~\cite{weedbrook2012gaussian}. We also refer the readers to two recent CV-QKD reviews for the security analysis and practical issues~\cite{diamanti2015distributing}, the issues of trusted noise~\cite{usenko2016trusted} and the models of implementation and noise~\cite{laudenbach2018continuous}.

\subsection{Protocol and security}

\subsubsection{Gaussian-modulated protocol}

The first Gaussian (continuous) modulated protocol ia a Gaussian-modulated squeezed state protocol~\cite{cerf2001}, where the key is encrypted in the displacement of a squeezed state. The random choice of the direction to squeeze is similar to the basis choice in the BB84 protocol. The squeeze state protocol was later extended to GMCS protocols~\cite{gg02,grosshans2003quantum} since coherent states are easier to prepare in practice. We summarize the prepare-and-measure version of a general GMCS protocol in Box~\ref{tab:gg02}. A difference from a DV-QKD protocol is that, in a coherent state protocol, the key information of both 'basis' is encrypted into the prepared state simultaneously per channel use. Therefore, Bob's measurement can be correlated with Alice's key in either basis or both bases.

\begin{tcolorbox}[title = {Box VII.A.1: GMCS QKD protocol.}]
(1) Alice produces two random numbers $x_A$ and $p_A$ from random numbers following a Gaussian distribution with a variance of $V_AN_0$, where $N_0$ is the vacuum noise unit. \\
(2) Alice prepares a coherent state $\ket{x_A+ip_A}$ and sends it to Bob through an untrusted quantum channel. \\
(3) Bob chooses homodyne (heterodyne) detection to measure $X$ and $P$ randomly (simultaneously) and obtains the outcomes $x_B$ and $p_B$. \\
(4) After repeating the above process $N$ times, Alice and Bob sift the measurement results by a classical channel, and obtain $N$ pairs of raw key, i.e, the correlated Gaussian variables, in the homodyne detection protocol ($2N$ pairs in heterodyne detection protocol).\\
(5) Alice and Bob perform the postprocessing on the raw key including parameter estimation, error correction and privacy amplification.
\end{tcolorbox} \label{tab:gg02}

In GMCS protocol~\cite{gg02,grosshans2003quantum}, the source is a mixture of coherent state $\ket{\alpha_j}=\ket{x_j+ip_j}$ with quadrature components $x_j$ and $p_j$ as the realizations for two i.i.d. random variables $X$ and $P$. These two random variables obey the same zero-centred Gaussian distribution $\mathcal{N}(0,V_m)$, where $V_m$ is the modulated variance. The total variance of the Gaussian modulated source is $V=V_s+V_m$, where $V_s$ is the intrinsic quadrature uncertainty of coherent state. Another type of GMCS scheme is coherent-state source mixed with trusted thermal noise~\cite{Weedbrook2010} whose total variance is $V=V_s+V_m+V_{th}$ with an additional thermal variance $V_{th}$. This type of protocol is also widely used due to its low cost in state preparation together with the feasibility for QKD in the wavelength longer than optical band. The decoding process is based on coherent detection measuring quadratures of optical fields. For CV-QKD schemes, coherent detection can be classified into homodyne detection and heterodyne detection, measuring quadratures of optical fields~\cite{weedbrook2012gaussian}.

Note that the coherent state protocol with homodyne detection can be modified to a no-switching protocol using heterodyne detection, which enables the communication partners to extract secure key from both quadrature measurements \cite{weedbrook2004quantum}. Post-selection \cite{silberhorn2002continuous} and two-way communication \cite{pirandola2008continuous} can also be applied to improve the performance. So far, the GMCS protocol is believed to be the best understood protocol in terms of security and implementation. Its implementation is also relatively simple, as it requires only standard technology in telecommunication.

\subsubsection{Discrete modulated protocol}
Besides Gaussian-modulated protocol, there exists a different type of protocol using discrete modulation. Here the key is encoded into the random phases of coherent states, and the source is a $N$-discrete randomized coherent state mixture. In fact, discrete modulated protocol was proposed earlier than Gaussian modulated protocol \cite{PhysRevA.61.010303,PhysRevA.73.012330,PhysRevA.62.062308}. However, due to its non-Gaussian nature, a complete security proof of discrete modulated protocol that gives a good key rate in practice is challenging. In a discrete modulated protocol, Alice prepares an alphabet of $N$ coherent states $\ket{\alpha_k}=\ket{|\alpha|e^{i2k\pi/N}}$, where $k$ is the secret key. Bob uses either homodyne or heterodyne detection to estimate $k$. Discrete modulated protocol is more practical, because (i) a real Gaussian modulation can never be perfectly implemented, and (ii) it can simplify the crucial step of error correction. Early proofs of discrete modulated protocol restrict attacks to be a linear quantum channel between Alice and Bob \cite{PhysRevLett.102.180504}. Though there are proofs for specific protocols where $N=2$ \cite{PhysRevA.79.012307} or $N=3$ \cite{PhysRevA.97.022310}, the key rate is quite pessimistic and cannot be generalized into multiple state cases.

Recently, the numerical method of security analysis has been proposed \cite{coles2016numerical}, where the security analysis is transformed into a convex optimization problem with the constraints that the statistics of certain observable should be compatible with experimental data. Following this line, there are two independent works analyzing the asymptotic security of the quadrature-phase-shift-keying (QPSK) protocol \cite{PhysRevX.9.021059,lin2019asymptotic}, i.e., $N=4$. With a photon number cut-off assumption on Bob's side, it is feasible to compute the target function and constraints as a function of Alice and Bob's two-mode state. Such a photon number cut-off assumption is valid since composable security proofs of CV-QKD usually require a projection onto a low-dimensional subspace of the Fock space, via some energy test \cite{PhysRevLett.102.110504}. These proofs can be generalized to multiple state cases, showing that the key rate converges to Gaussian modulated protocols when $N\rightarrow \infty$. Moreover, another security proof was reported lately by applying entropic continuity bounds and approximating a complex Gaussian probability distribution with a finite-size Gauss-Hermite constellation~\cite{kaur2019asymptotic}. Currently, to generalize those existing security proofs to the finite-size case remains an important open question.

\subsubsection{Security analysis}

Intuitively, the security of coherent state protocol comes from the fact that coherent states are non-orthogonal, which ensures the no-cloning theorem. To rigorously analyze the security, it is convenient to consider an entanglement-based protocol. Alice prepares an two-mode EPR state $\ket{EPR}_{AA^\prime}$. She keeps one mode $A$ in her lab and sends the other mode $A^\prime$ to Bob through a noisy channel $\mathcal{E}_{A^\prime\rightarrow B}$. Alice performs heterodyne detection on her mode and gets a coherent state output, which is identical to preparing a coherent state to Bob from Eve's point of view. We assume the worst case scenario where Eve holds a purification of $\rho_{AB}$, then the tripartite state shared by Alice, Bob and Eve is given by
\begin{equation}
\rho_{ABE}=(id_A\otimes\mathcal{U}_{A^\prime\rightarrow BE}(\ket{EPR}\bra{EPR}_{AA^\prime})),
\end{equation}
where $id_A$ denotes the identity map on Alice's mode $A$ and $\mathcal{U}_{A^\prime\rightarrow BE}$ is an isometry. Alice and Bob's secure information under collective attack in asymptotic limit, for reverse reconciliation, is given by the Devetak-Winter formula~\cite{Devetak2005Distillation},
\begin{equation}
K=I(A:B)-\sup\chi(B:E),
\end{equation}
where $\chi(B:E)$ is the Holevo bound~\cite{holevo1973bounds}. The supremum is computed over all possible quantum channels compatible with the statistics obtained in the parameter estimation step in implementation. The secure key can be distilled as long as Alice and Bob's mutual information is larger than the maximum of Bob's classical information accessible to Eve through the quantum channel between Bob and Eve.

Specifically, in the parameter estimation step, Alice and Bob exchange the statistics calculated from a subset of the sifted raw key, and estimate the covariance matrix of the two-mode state share by them,
\begin{equation}
\gamma_{AB}=
\left(
\begin{matrix}
V_A I_2 & Z \sigma_z \\
Z \sigma_z & V_B I_2
\end{matrix}
\right)
\end{equation}
where $V_A$ and $V_B$ are the variance of the quadratures, $I_2$ is the two-dimensional identity matrix, and $Z$ is the covariance calculated by the experimental data.

Thanks to the Gaussian optimality proved in \cite{PhysRevLett.97.190502,PhysRevLett.97.190503,PhysRevLett.96.080502}, the optimal collective attack Eve can implement is the one based on Gaussian operations, which result in a two-mode Gaussian state. Owing to the one-to-one correspondence between Gaussian states and covariance matrix, we can directly calculate the secure key rate under collective attack by the covariance matrix. Suppose the optimal attack is characterized by a Gaussian channel of transmittance $T$ and excess noise $\xi$, then there will be the following relations
\begin{equation}
\begin{aligned}
V_B&= T(V_A+\xi) \\
Z&=\sqrt{T(V_A^2-1)}.
\end{aligned}
\end{equation}
And the mutual information between Alice and Bob is given by
\begin{equation}
I(A:B)=\frac{\omega}{2}\log\frac{V+\xi}{\xi+1},
\end{equation}
where $\omega=1,2$ corresponds to Bob's homodyne detection and heterodyne detection, respectively. The Holevo bound is calculated by
\begin{equation}
\begin{aligned}
\chi(B:E)=S(E)-S(E|b)=S(AB)-S(E|b),
\end{aligned}
\end{equation}
where the second equation is because Eve holds a purification of $\rho_{AB}$ and $b$ is Bob's measurement result. Both $S(AB)$ and $S(E|b)$ can be calculated from the corresponding covariance matrix $\gamma_{AB}$ and $\gamma_{E|b}$ \cite{PhysRevLett.94.020505,PhysRevLett.94.020504}. The form of $V(E|b)$ depends on homodyne detection or heterodyne detection that Bob performs. Notice that to obtain a secret key, the two important parameters are the transmittance $T$ and excess noise $\xi$, which should be carefully estimated in the parameter estimation step.

The above security analysis is restricted to collective attacks in the asymptotic limit of infinitely long keys. On one hand, one needs to generalize the collective attacks to coherent (or general) attacks, which is a challenging problem in CV-QKD. Fortunately, it turns out that collective attacks are as efficient as coherent attacks assuming the permutation symmetry of the classical postprocessing \cite{PhysRevLett.102.110504}. The phase-space symmetries and the postselection technique \cite{PhysRevLett.102.020504} can also be exploited to perform a reduction from general to collective attacks~\cite{leverrier2013security}. Recently, a new type of Gaussian de Finetti reduction was proposed which confirms the belief that proving the security against Gaussian collective attacks in CV-QKD is sufficient obtain the security against coherent attacks \cite{leverrier2017security}.

On the other hand, the security analysis should be extend to \emph{finite-key case}. The finite-key rate will have a deviation from the asymptotic limit, which is due to the statistical fluctuations in parameter estimation. Moreover, other deviations arise when we assume Gaussian attacks and consider collective attacks instead of coherent attacks. These issues were well addressed in the literatures~\cite{furrer2012continuous,leverrier2013security,leverrier2015composable}. Based on postselection technique \cite{PhysRevLett.102.020504}, the security of GMCS CV-QKD was proven against general attacks in the finite-size regime~\cite{leverrier2013security}, but the security proof is not composable. For composable security proof, Furrer et al. provide the first proof for CV-QKD with squeezed states using the entropic uncertainty principle~\cite{furrer2012continuous}, whereas the analysis is only moderately tolerant to loss. For coherent-state protocols, Leverrier gives the first composable security proof against only collective attacks~\cite{leverrier2015composable}, and proposes a new type of Gaussian de Finetti reduction which shows the potential for finite-key security with small data sizes~\cite{leverrier2017security}. Nevertheless, the current proof techniques for composable security against coherent attacks still require rather large block sizes, e.g., $>10^{13}$~\cite{leverrier2017security}. The composable security of CV-QKD against coherent (or general) attacks in a realistic finite-size regime remains an outstanding open issue for the future study of improved proof techniques.

Beside coherent state, the squeezed state protocol has also been widely studied for CV-QKD. In a squeezed state protocol~\cite{hillery2000quantum,cerf2001}, Alice squeezes the $X$ quadrature of a vacuum state and displaces it by an amount $a$, which follows a Gaussian distribution of variance $V_A$. Then she adds a random phase of $0$ or $\pi/2$ on it, which is equivalent to randomly choosing a direction to squeeze. Finally she randomly displaces the output state along the other direction (not the squeezing direction) following another Gaussian variable of variance $V$. The two variance $V_A$ and $V$ should satisfy $V_A+V^{-1}=V$ such that Eve cannot distinguish which quadrature is squeezed. Bob randomly measures $X$ or $P$ quadrature. Alice and Bob perform the postprocessing after a certain rounds of measurements. The squeezed state protocol is more similar to the DV-QKD protocols than the coherent state protocol. Its security is based on an entropic uncertainty principle~\cite{cerf2001}. The composable security of finite size analyses have also been given in~\cite{furrer2012continuous,PhysRevA.90.042325}, together with experimental verifications~\cite{gehring2015implementation}.

\subsection{Experimental developments}

\begin{figure} [htbp]
\includegraphics[scale=0.62,angle=0]{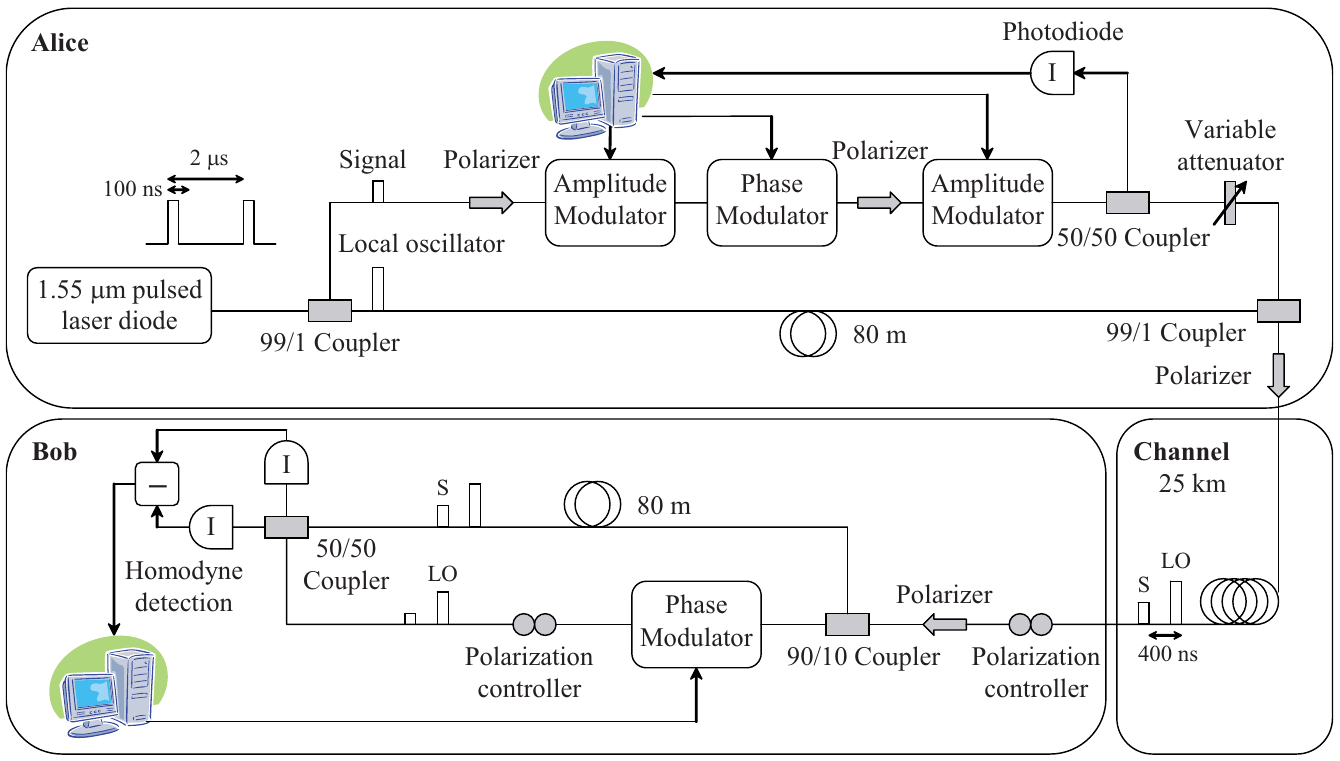}
\caption{An illustration of the implementation of GMCS CV-QKD. [Figure reproduced from~\cite{lodewyck2007quantum}].} \label{fig:cvexp}
\end{figure}

The widely implemented CV-QKD protocol is the GMCS protocol~\cite{gg02,grosshans2003quantum} (see Box~\ref{tab:gg02}) due to its simplicity in preparation, modulation and detection of coherent states. An illustration of the implementation is shown in Fig.~\ref{fig:cvexp}~\cite{lodewyck2007quantum}. Alice employs a laser diode to generate optical pulses, each of which is split into signal and local oscillator (LO) by a fiber-optic coupler. The signal pulses are modulated in amplitude and phase according to a Gaussian distribution, and attenuated to the desired modulation variance with a variable attenuator. The LO is time delayed and then combined with the signal at Alice's output. Bob passively demultiplexes the signal and LO using a coupler and then performs the measurement using a shot-noise-limited homodyne detector. Bob can select the quadrature to be measured by adjusting the measurement phase with a phase modulator placed in the LO path. An advanced feature of this implementation is that it consists entirely of standard fiber optics and telecommunication components.

Reverse reconciliation was introduced to the GMCS protocol in 2003~\cite{grosshans2003quantum}, which allows GMCS to beat the 3-dB loss limit. Moreover, a free-space experiment at visible light wavelength was also performed there. With telecom wavelength, GMCS was performed over practical distance of optical fibers (of 25km and 5km respectively) in~\cite{qi2007experimental,lodewyck2007quantum}. Meanwhile, the heterodyne detection~\cite{lance2005no} and Gaussian post-selection~\cite{symul2007experimental} were also demonstrated. Later, the feasibility of GMCS CV-QKD was also extensively tested in field environments~\cite{fossier2009field}. The secure distance was substantially extended to 80 km based on the improved efficiency of the post-processing techniques~\cite{jouguet2013experimental}. By controlling the excess noise, the distance was further extended to 100 km standard fiber~\cite{huang2016long}. Recently, state-of-the-art CV-QKD implementations were sequentially reported, such as high-rate demonstrations with a secret key rate up to 3.14 Mbits/s in the asymptotic limit over 25-km fiber~\cite{wang2018high}, a 4-node field network~\cite{huang2016field}, a field test over 50-km commercial fiber~\cite{zhang2019continuous}, a long-distance CV-QKD over about 200-km ultralow loss fiber~\cite{zhang2020long}, a Si photonic chip-based CV-QKD implementation~\cite{zhang2019integrated} and so forth. Although we focus on the GMCS implementations, we note several other important experiments, such as the squeezed-state protocols~\cite{gehring2015implementation} and a CV-QKD experiment with entangled states over 50-km fiber~\cite{wang2018long}. Some recent developments of CV-QKD are shown in Table~\ref{Tab:cvqkd}.

From a practical point of view, CV-QKD presents the key advantage that it only requires standard telecommunication technology which is compatible with classical optical communications, i.e., it uses the coherent detection techniques instead of single-photon detection technology as required in DV-QKD. Moreover, the LO in CV-QKD can serve as a built-in single mode filter, which makes it naturally resistant against background noises~\cite{qi2010feasibility}. This is particularly useful in the practical situations, including the coexistence of QKD with classical channels via DWDM~\cite{qi2010feasibility,kumar2015coexistence}, the daylight free-space CV quantum communication~\cite{heim2014atmospheric,peuntinger2014distribution,vasylyev2017free,zeng2019cv}. Nonetheless, CV-QKD systems are in general sensitive to losses, which restricts the secure distance, normally below 100 km fiber~\cite{jouguet2013experimental}. However, in theory, CV-QKD may provide higher key rates than DV-QKD at relatively short distances because of its high dimensionality~\cite{jouguet2014high}, while the exact rate in terms of bits/s depends on the technology of real implementation. High-rate CV-QKD requires high-speed and real-time implementations of several challenging techniques~\cite{wang2018high}, such as low-noise homodye detector, efficient error correction codes, precise parameter estimation etc. Also, the composable security proofs against general attacks still require a very large block sizes to allow a positive key in the finite-key regime~\cite{leverrier2015composable}, which cascades a challenge on the stability of the system. These issues are important subjects for future research.

\subsection{Quantum hacking and countermeasures} \label{cvqkd:hacking}
Similar to DV-QKD, the implementations of CV-QKD also suffer from side channels. On the source part, the Trojan horse attacks can probe Alice's modulators in CV-QKD systems~\cite{jain2014trojan}. Similar to DV-QKD, a countermeasure is to put an optical isolator and a monitoring detector at the output of Alice's setup. The imperfections in state preparation may also cause an increase of the excess noise and misestimate of the channel loss~\cite{liu2017imperfect}. On the detection part, the wavelength dependence of the beam splitter can be exploited by Eve to hack CV-QKD based on heterodyne detection~\cite{huang2013quantum,ma2013wavelength}. Qin et al. demonstrated the detector saturation attack~\cite{qin2016quantum} and blinding attack~\cite{qin2018homodyne} against homodyne detectors in CV-QKD by exploiting the nonlinear behavior of coherent detectors. A wavelength filter is effective against the first attack, and a proper monitor at detection may counter the second attack. A more general solution is to perform the real-time shot noise measurement as analyzed in~\cite{kunz2015robust,zhang2019one}.

To completely remove the detection attacks, a CV version of MDI-QKD was proposed~\cite{pirandola2015high}. See also~\cite{li2014continuous,ma2014gaussian} for a security analysis against restricted attacks. The concept is similar to the MDI-QKD protocol discussed in Section~\ref{sec:5:mdi}, but here Alice and Bob prepare coherent states with a Gaussian modulation and send them to Charlie. Charlie then mixes them on a balanced beam splitter, measures a different quadrature for both output modes and publicly announces his measurement results. The security of CV MDI-QKD can be analyzed by considering the entanglement-based version of the protocol~\cite{lupo2018continuous}. A proof-of-principle CV MDI-QKD experiment was demonstrated in free space with advanced detection techniques in 2015 \cite{pirandola2015high}. Nonetheless, afull implementation with practical lengths of optical fibers is still a great challenge which has not been reported in the literature yet, partly because of the requirement of high-efficiency detection~\cite{xu2015discrete}. Even so, CV MDI-QKD has the potential to provide slightly higher key rates, and it might be interesting for network communication over relatively short distances~\cite{pirandola2015high}.

\begin{figure} [htbp]
\includegraphics[scale=0.28,angle=0]{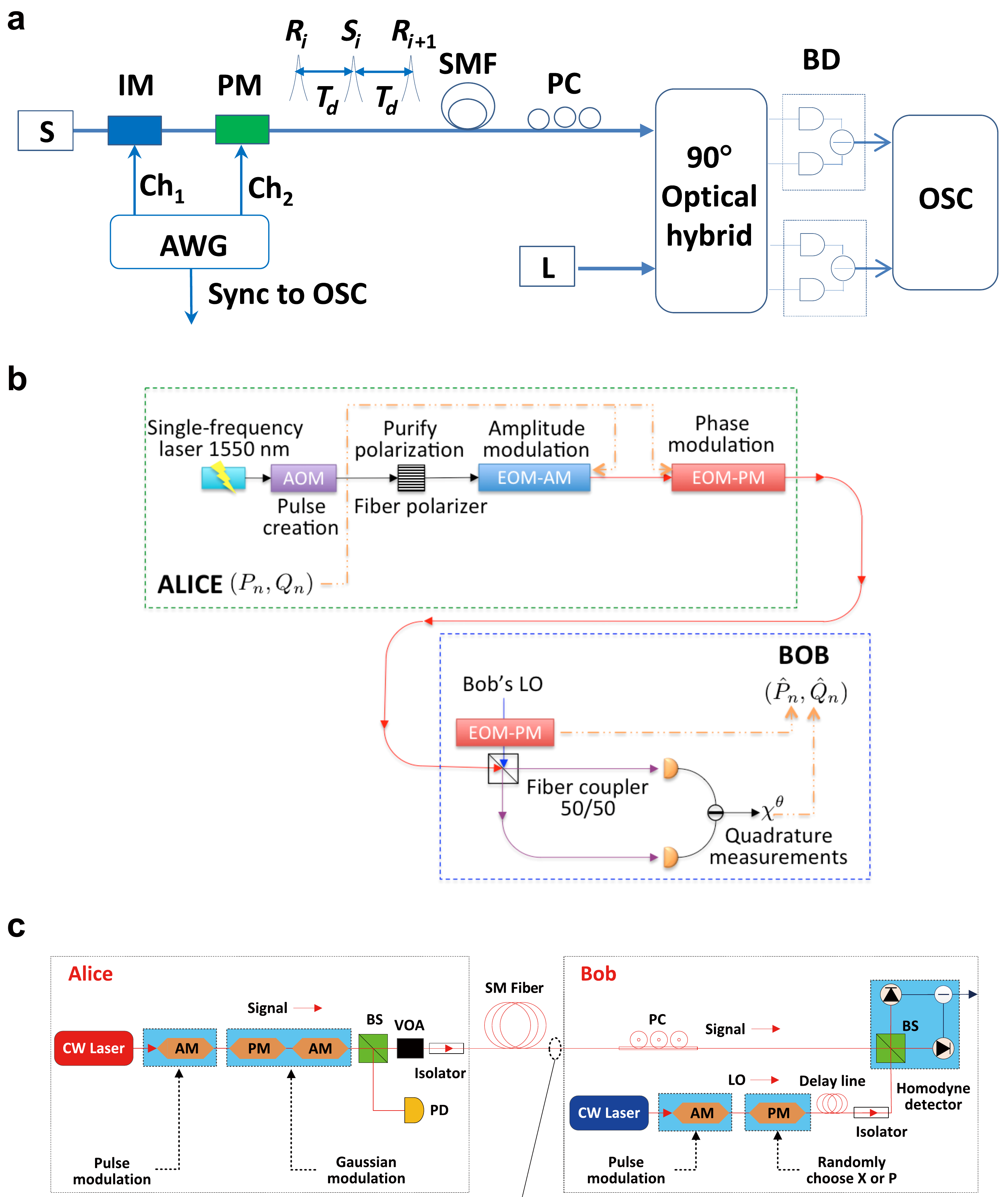}
\caption{(Color online) Initial local LO (LLO) CV-QKD experiments. \textbf{a}, LLO CV-QKD experiment with the pilot-aided feedforward data recovery scheme using commercial off-the-shelf devices~\cite{qi2015generating}. \textbf{b}, LLO CV-QKD experiment with self reference~\cite{soh2015self}. \textbf{c}, A high-speed LLO CV-QKD experiment~\cite{huang2015high}. [Figures reproduced from~\cite{qi2015generating,soh2015self,huang2015high}].} \label{fig:llocv}
\end{figure}

Besides, an additional threat for CV-QKD is the transmission of LO, which can be manipulated by Eve. The attacks by controlling the transmitted LO were proposed in~\cite{ma2013local}. Eve can also exploit a subtle link between the local oscillator calibration procedure and the clock generation procedure employed in practical setups~\cite{jouguet2013preventing}. A countermeasure for the LO attacks consists of implementing a rigorous and robust real-time measurement of the shot noise~\cite{kunz2015robust,zhang2019one}. A better solution is the locally LO (LLO) CV-QKD scheme~\cite{qi2015generating,soh2015self} (see Fig.~\ref{fig:llocv}), which can completely avoid the transmission of the LO through the insecure channel. In this scheme, Bob uses a second, independent laser to produce LO pulses locally for the coherent detection. A challenge here is how to effectively establish a reliable phase reference between Alice's and Bob's independent lasers, which require a careful synchronization of the frequencies and phases. This can be achieved by sending the reference or pilot-aided pulse along with the signal pulse from Alice. Bob can use his LO pulse to perform coherent detection for the reference pulse to estimate the relative phase between Alice's and Bob's lasers. A phase correction can thus be established on Alice and Bob's signal data in order to generate the secret key. Due to the enhanced security of LLO CV-QKD, there has been great attention in this scheme. In 2015, three groups have independently demonstrated LLO CV-QKD~\cite{qi2015generating,soh2015self,huang2015high}. These experiments are shown in Fig.~\ref{fig:llocv}. Afterwards, a LLO CV-QKD experiment with pilot and quantum signals multiplexed in the frequency domain was reported in~\cite{kleis2017continuous}, a comprehensive framework to model the performance of LLO CV-QKD was reported in~\cite{marie2017self}, a pilot-assisted coherent intradyne reception methodology for LLO CV-QKD was proposed and demonstrated in~\cite{laudenbach2017pilot}, and a high rate LLO CV-QKD was demonstrated in~\cite{wang2018high}.

\section{Other Quantum Cryptographic Protocols}\label{sec:6}

\subsection{Device-independent QKD} \label{sec:diQKD}
\begin{figure}[hbt]
\begin{center}
\includegraphics[scale=0.64,angle=0]{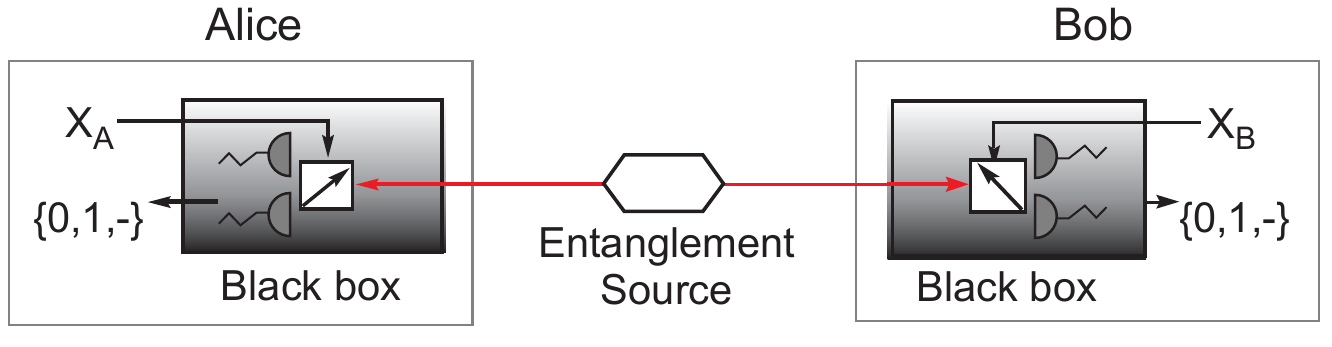}
\end{center}
 \caption{Schematic diagram of DI-QKD~\cite{mayers1998quantum,acin2007device}. Entangled photon pairs are distributed to Alice and Bob, who are supposed to perform some measurements. Alice and Bob see their quantum
devices as black boxes producing classical outputs, as a function of classical inputs $X_{A}$ and $X_{B}$. From the observed statistics, and without making any assumptions on the internal working of the devices, they should be able to conclude whether they establish a secret key. Alice and Bob assume giving the untrusted quantum devices tests that cannot be passed unless they carry out the QKD protocol securely, which can be check via the violation of a Bell's inequality~\cite{pironio2009device}.}\label{Fig:diqkd}
\end{figure}

A QKD protocol is device-independent if its security does not rely on trusting that the quantum devices used are truthful. A schematic illustration of device-independent QKD (DI-QKD) is shown in Fig.~\ref{Fig:diqkd}. DI-QKD~\cite{mayers1998quantum,barrett2005no,acin2007device} (hinted earlier by Ekert~\cite{ekert1991quantum}) relaxes all modelling assumptions about the quantum devices and allows the users to do QKD with uncharaterized devices. DI-QKD performs self-testing of the underlying devices, i.e., the devices cannot pass the test unless they carry out the QKD protocol securely. As a result, as long as certain necessary assumptions are satisfied, one can prove the security of DI-QKD based solely on a Bell nonlocal behaviour, typically the violation of a Bell's inequality, which certifies the presence of quantum correlations in a self-testing manner. Table~\ref{Tab:assumptions} lists a summary of the necessary assumptions of DI-QKD~\cite{pironio2009device}. The security proofs have required the assumption that the devices have no memory between trials, or that each party has many, strictly isolated devices~\cite{barrett2005no,acin2007device,pironio2009device,masanes2011secure}. If the devices have memory or the devices are reused, DI-QKD will suffer from the memory attacks~\cite{barrett2013memory} and covert channels~\cite{curty2017quantum}.

The security proof for DI-QKD is a challenging task, because in DI-QKD, both the quantum state (generated by the source) and the measurement operators (by the detection devices) are untrusted or under Eve's control. Fortunately, recent theory efforts have significantly advanced the developments of DI-QKD to be possible in a large quantum system~\cite{reichardt2013classical}, secure for a large class of protocols by independent measurements~\cite{masanes2011secure}, secure against general attacks~\cite{vazirani2014fully,miller2016robust} and robust against noise~\cite{Arnon2018Practical}. In the asymptotic case against collective attacks, the key rate formula can be expressed as a function of the Bell violation value. For a protocol where Alice and Bob carry out a CHSH-type Bell test for self-testing privacy, the key rate formula can be given by~\cite{pironio2009device},
\begin{equation}
    r \geq 1-h(E)-h\left(\dfrac{1+\sqrt{(S/2)^2-1}}{2}\right),
\end{equation}
where the quantum bit error rate $E$ determines the amount of randomness consumed for error correction, and the violation value $S$ of the CHSH inequality determines the amount of randomness for privacy amplification.

Though DI-QKD is remarkable in theory, unfortunately, it is hard to realize with current technology, because it needs almost perfect efficiency of single-photon detection~\cite{masanes2011secure}. In experiments, however, the emitted photons may not be detected due to the losses in the transmission or the limited detection efficiency of imperfect detectors. In addition, a faithful realization of DI-QKD requires that the Bell inequality is violated under the following two conditions: (i) The measurement settings are not correlated with the devices; (ii) The devices observe a no-signalling behaviour in generating the outcomes. To meet these conditions, a so-called loophole-free Bell test normally needs to be carried out. A key problem in a loophole-free Bell test is the limited detection efficiency which is referred to as efficiency loophole~\cite{pearle1970hidden}. It has been proved that for the simplest bipartite Bell inequalities with binary inputs and outputs, a detector efficiency of at least 2/3 is necessary for a faithful Bell inequality violation~\cite{eberhard1993background}. For the purpose of DI-QKD, a much higher efficiency is needed due to the requirement of information reconciliation. To guarantee the no-signalling behaviour between devices, i.e., the locality loophole, a space-like separated measurement set-up can be implemented~\cite{aspect1975proposed}. The requirement of uncorrelated inputs is referred to as the free-will loophole, which cannot be closed completely. Nonetheless, practical QRNGs can be used to overcome the problem to some extent.

Recently, an exciting news is that researchers have demonstrated the Bell's inequality which simultaneously closed the locality loophole and the detection loophole in the same experiment~\cite{hensen2015loophole,shalm2015strong,giustina2015significant,rosenfeld2017event,liu2018device}. This is a milestone result towards the realization of DI-QKD. In future, the advanced technologies might make DI-QKD more practical, and ideas such as qubit amplification~\cite{gisin2010proposal} might also be proved useful to increase the key rate and distance of DI-QKD, though the key rate might be relatively low~\cite{curty2011heralded,seshadreesan2016progress}. Recent theoretical progresses have proposed the two-way classical communication to enhance the noise tolerance~\cite{tan2020advantage}, and provided the detailed analysis towards the realization of DI-QKD~\cite{murta2019towards}. Overall, we do believe that DI-QKD is an important subject for future research.

\subsection{Some New QKD implementations}

Besides the efforts in the security with imperfect devices, quite a few new QKD protocols have been proposed and implemented during the past ten years, which are summarized in Table~\ref{Tab:otherqkd}.

\subsubsection{Round-robin DPS QKD}

In general, there exists a threshold of the error rate for each scheme, above which no secure key can be generated. This threshold puts a restriction on the environment noises. Especially, in the key rate formula, the bit error can be directly computed from the experimental data, whereas the phase error needs to be estimated or bounded. In the BB84 protocol with strong symmetries, both error rate are approximately the same in the long key length limit. In other protocols, normally there is a relation between the two error rates. In the end, when the bit error rate goes beyond some threshold level, no secure key can be generated. For example, BB84 cannot tolerate error rates beyond 25\% considering a simple intercept-and-resend attack \cite{bennett1984quantum}. This threshold puts a stringent requirement on the system environment, which makes some practical implementations challenging.

Round-robin differential phase shifted (RRDPS) QKD, proposed by Sasaki, Yamamoto and Koashi~\cite{sasaki2014practical} in 2014, essentially removes this restriction and can in principle tolerate more environment disturbance. In this protocol, Eve's information can be bounded \emph{only} by user's certain experiment parameters other than the error rates. In particular, the phase error rate $e_{p}$ is determined by the user's own settings rather than the channel performance, which makes the protocol fundamentally interesting and tolerate more errors.

\begin{figure}[!ht]
\centering
\resizebox{8cm}{!}{\includegraphics{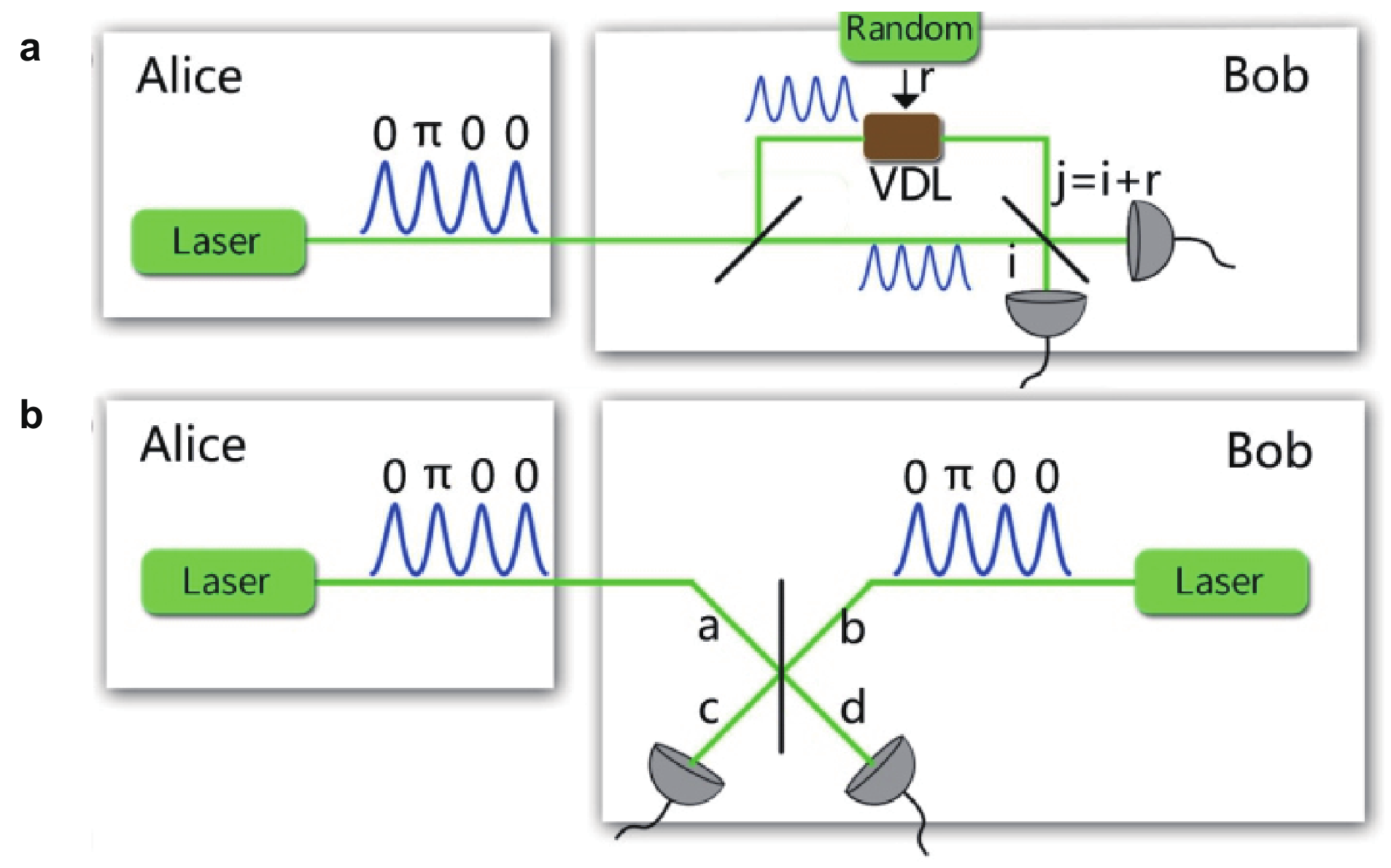}}\\
\caption{\textbf{a}, The original RRDPS scheme \cite{sasaki2014practical}. VDL means variable delay line. Bob splits the received signals into two paths and applies a variable delay $r$ to one of the paths. A click at $i$th place will indicate an interference between the pulses $i$ and $j=i+r$. \textbf{b}, The passive RRDPS scheme. Bob uses a local laser to generate an $L$-pulse reference, which interferes with Alice's $L$-pulse signal. Bob then records the coincidence clicks. [Figure reproduced from~\cite{guan2015experimental}].}
 \label{rrdps scheme}
\end{figure}

In the RRDPS QKD protocol, the sender Alice puts a random phase, chosen from $\{0, \pi\}$, on each of $L$ pulses, with an average photon number of $\mu$ in such an $L$-pulse signal. Upon receiving the block, the receiver Bob implements a single-photon interference with an Mach-Zehnder interferometer (MZI), as shown in Fig.~\ref{rrdps scheme}a. Bob can randomly adjust the length difference of the two arms of the MZI. After obtaining a detection click, Bob first identifies which two pulses interfere and then announces the corresponding indices $i$, $j$ to Alice. Alice can derive the relative phase between the two pulses as the raw key, and Bob can record the raw key from the measurement results. The phase error rate depends only on the number of photons in the $L$-pulse signal and $L$, not the bit error rate. By setting a larger $L$, the phase error tends to 0, and the scheme can tolerate a higher bit error rate.

Triggered by the original protocol, an alternative passive type of RRDPS QKD is proposed by Guan et al. \cite{guan2015experimental}. As is shown in Fig.~\ref{rrdps scheme}b, when Bob receives a block from Alice, he prepares a local $L$-pulse reference in plain phases, i.e., all phases are encoded at phase 0. This $L$-pulse reference interferes with the $L$-pulse signal sent by Alice on a beam splitter. For each block, Bob records the status of his two detectors with time stamps, $i$ and $j$. If Bob's reference is in phase with Alice's signal, i.e. Bob has a phase reference, the whole setup is essentially a huge Mach-Zehnder interferometer. Any detection signal at time slot $i$ will tell the phase difference between $i$ and the phase reference. Then the encoding bit value can be figured out by Bob. If the phase Bob's reference is random relative to Alice's signal, the interference is a Hong-Ou-Mandel type interference \cite{HOM:1987} instead of a MZI. Bob post-selects the block where there are exactly two detections and announces their positions i and j (if i = j, the detection result is discarded). The raw key is the relative phase between these two pulses in the L-pulse signal. Alice can derive this phase difference from her record. While Bob can infer the bit value depending on that the coincidence happens between two different detectors or one detector at two different time slots. The security proofs of the two protocols are beyond the scope of the paper and we refer the readers to the original papers~\cite{sasaki2014practical,guan2015experimental}.

\begin{figure}
\centering
\resizebox{8.2cm}{!}{\includegraphics{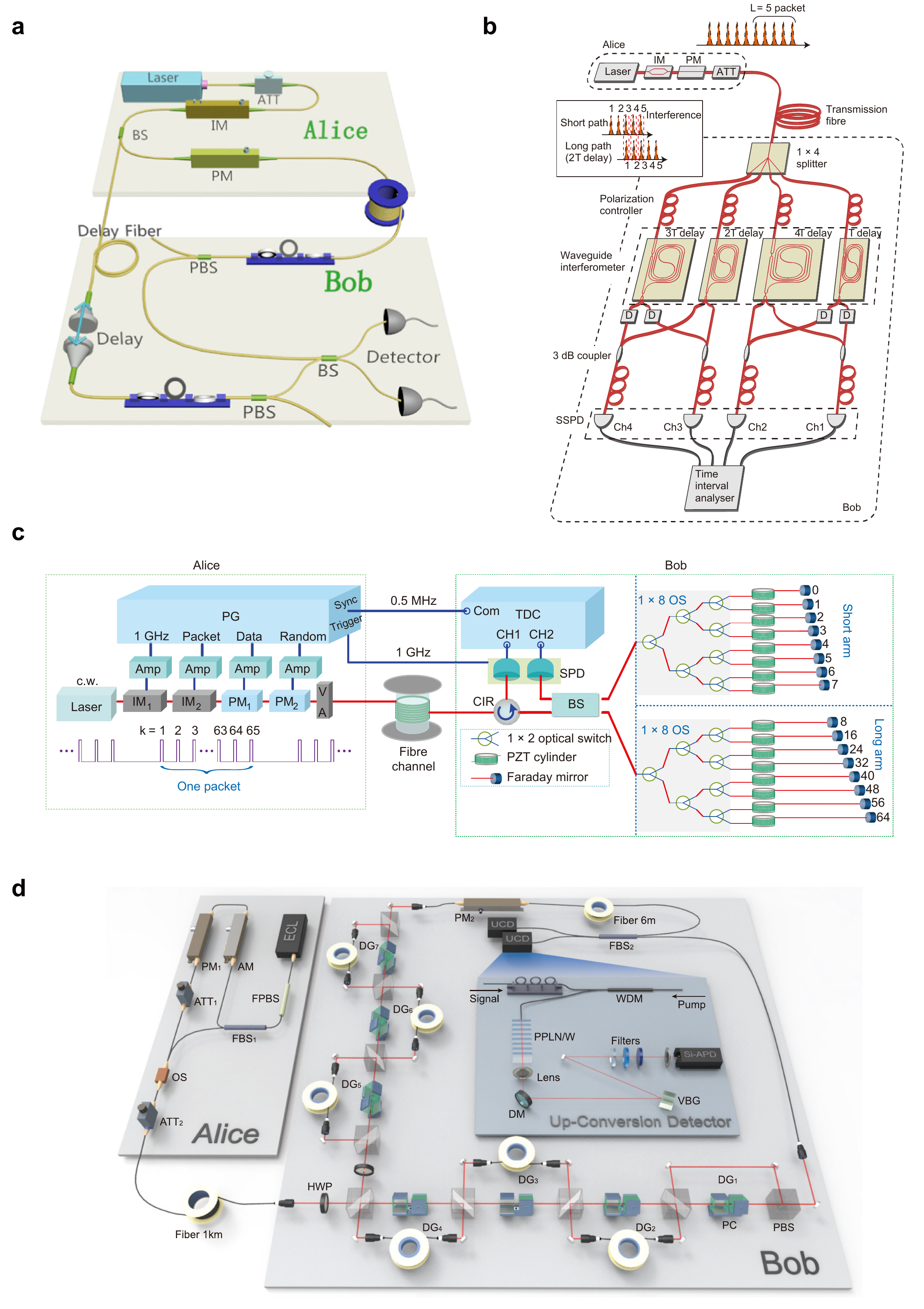}}\\
\caption{(Color online) The experimental setups for RRDPS QKD, reported in the references of \textbf{a},~\cite{guan2015experimental}; \textbf{b}, \cite{takesue2015experimental}; \textbf{c}, \cite{wang2015experimental}; \textbf{d}, \cite{li2016experimental}. [Figures reproduced from~\cite{guan2015experimental,wang2015experimental,li2016experimental}].}
 \label{rrdps exp}
\end{figure}

The first published experimental result is based on the passive protocol \cite{guan2015experimental}, as shown in Fig.~\ref{rrdps exp}a. Comparing to the original protocol, the passive one avoids randomly adjusting the length difference of the MZI. Based on the current technology, the main adjust-delay method is to utilize optical switches, which cannot provide both high speed and low insertion loss simultaneously. But meanwhile, it requires remote optical phase locking, which is challenging in the real deployment.

The key point for an active RRDPS is to realize the random time delay. Takesue et al. \cite{takesue2015experimental} exploits a one-input, four-output optical splitter followed by four silica waveguides based MZI with 0.5,1.0,1.5,2.0 ns temporal delays respectively, as is shown in Fig.~\ref{rrdps exp}b. Any two delays constitute a new MZI and the whole system realizes a $L=1-5$ variable delay. With this delay, the authors achieved secure key rate through 30 km fiber with an error rate of $18\%$. Later, Wang et al. \cite{wang2015experimental}, shown in Fig.~\ref{rrdps exp}c, combines a three-port circulator, a beam splitter, two $1\times8$ optical switches followed with two groups of fiber delays. The two optical switches shall actively choose different delays and achieve a $L$=1 to 64 bit variable-delay Faraday-Michelson interferometer. Based on the delay, Wang et al. distributed a secret key over a distance of 90 km fiber. In addition, Li et al. \cite{li2016experimental} exploited a different configuration, which can be seen in Fig.~\ref{rrdps exp}d. They put 7 MZI in series to achieve a 127-value variable delays. And each MZI is constructed of a Pockels cell, a fiber, or free-space link with specific length and two polarizing beam splitters (PBSs). The Pockels cell, controlled by a random number, may change the polarization of the photon and thus provide a delay. Very recently, the secure distance is extended to 140 km by increasing the bound on information leakage \cite{yin2018improved}.

\subsubsection{High-dimensional QKD}

\begin{figure}
\centering
\resizebox{8.2cm}{!}{\includegraphics{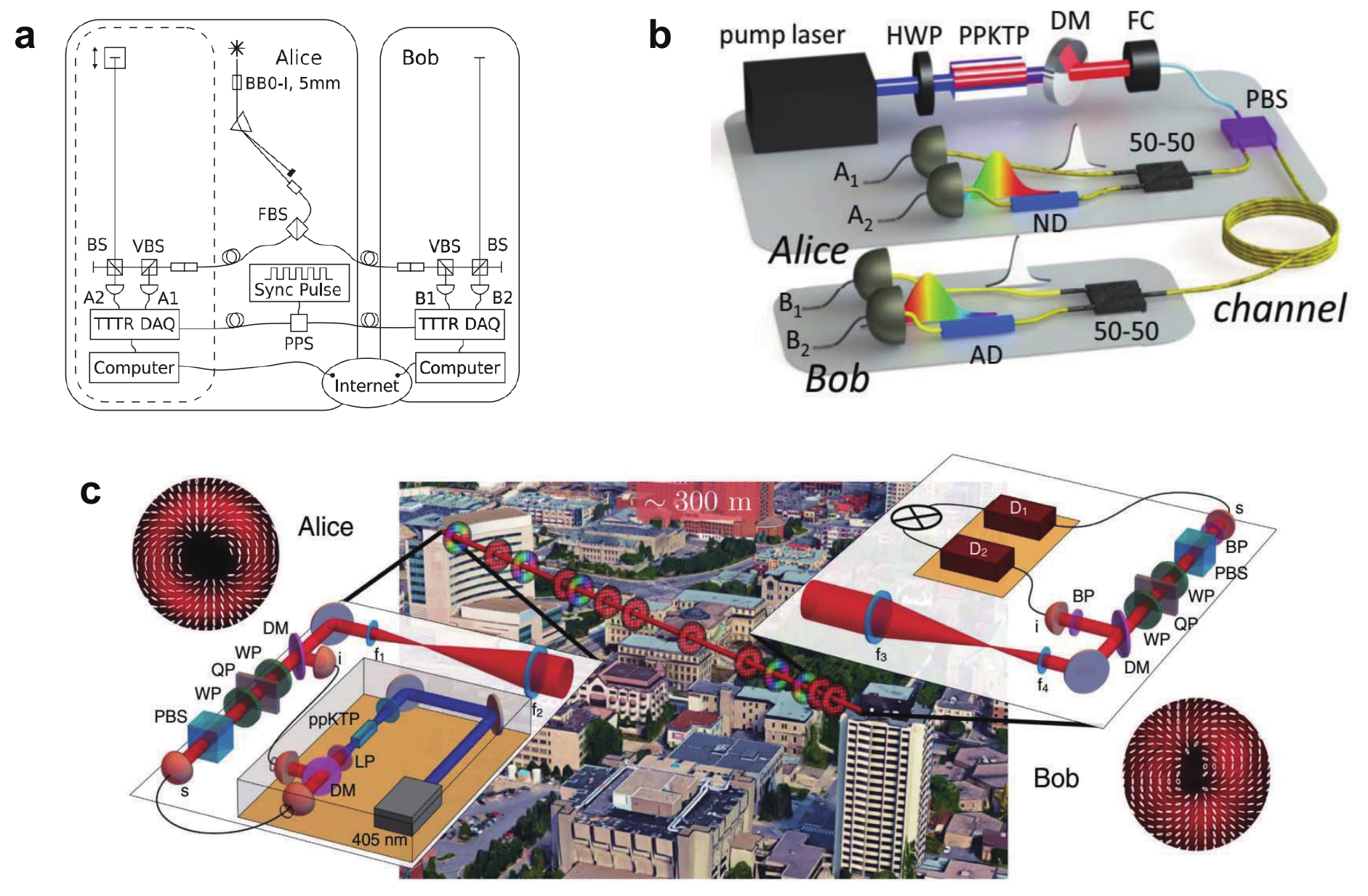}}\\
\caption{(Color online) The experimental setup for HD QKD. \textbf{a},HD QKD with time-energy entangled photon pairs \cite{Ali-Khan2007}. \textbf{b}, Dispersive-optics time-energy HD QKD \cite{lee2014entanglement}. \textbf{c}, A field test of OAM HD QKD in the city of Ottawa \cite{sit2017high}. [Figures reproduced from~\cite{Ali-Khan2007,lee2014entanglement,sit2017high}].}
 \label{HD QKD}
\end{figure}

Besides the qubit-based QKD, the secret keys can also be encoded with multi-level system, i.e. high-dimensional QKD (HD-QKD). HD-QKD can provide higher key rate per particle comparing to the qubit system \cite{Bourennane2001multilevel}, and it has a higher tolerance to noise \cite{Cerf2002}. A recent review on the subject can be seen in~\cite{xavier2020quantum}. The first experimental attempts of HD-QKD are using higher-order dimensional alphabets with spatial degrees of freedom of photons~\cite{walborn2006quantum} or energy-time entangled photon pairs \cite{Ali-Khan2007}. The later is shown in Fig.~\ref{HD QKD}a. With this setup, Ali-Khan et al. can generate a large-alphabet key with over 10 bits of information per photon pair, albeit with large noise. QKD with 5\% bit error rate is demonstrated with 4 bits of information per photon pair, where the security of the quantum channel is determined by the visibility of Franson interference fringes.

Zhang et al. \cite{zhang2014unconditional} reported a complete security proof of time-energy entanglement QKD using dual-basis interferometry. Mower et al. \cite{mower2013high} suggested to utilize dispersive optics to replace the Franson interferometer and demonstrated its security against collective attack. In this scheme, as shown in~Fig.~\ref{HD QKD}b~\cite{lee2014entanglement}, Alice or Bob utilize normal or abnormal group-velocity dispersive element to measure the frequency basis. The absolute group delays of their dispersive elements are matched such that the group-velocity dispersion is nonlocally canceled. Alice and Bob use time basis measurements for generating keys and frequency basis measurements for bounding Eve's maximum accessible information about the time basis measurements. This is based on the fact that the dispersion cancelation only happens with entanglement and any reduced entanglement visibility due to eavesdropping will bring a broaden time correlation measurement.

With the time energy entangled photon pairs, Zhong et al. observed a secure key rate of $2.7 Mbps$ after 20 km fiber transmission with a key capacity of $6.9$ bits per photon coincidence \cite{zhong2015photon}. Recently, high-rate QKD using time-bin qudits was reported in~\cite{islam2017provably}. Time energy type HD-QKD has advantage with a constant clock rate because it can utilize more time slots with high time resolution single photon detector. However, the advantage will be offset when the clock rate can be increased to the bandwidth of the single photon detector~\cite{Zhang:08}. One solution is to utilize a degree of freedom other than time, for example, the optical angular momentum (OAM). The first HD-QKD for OAM was published in 2006 \cite{Simon2006}. Qutrit entangled photon pairs were utilized to generate quantum key. In an E91-type protocol\cite{ekert1991quantum}, the violation of a three-dimensional Bell inequality verifies the security of the generated keys. A key is obtained with a qutrit error rate of approximately 10\%. Later, Etcheverry et al. report an automated prepare-and-measure HD-QKD with 16-dimensional photonic states~\cite{etcheverry2013quantum}; Mafu et al. exploit high-dimension OAM up to five dimensions for HD-QKD \cite{Mafu2013}; Mirhosseini et al. use the OAM of weak coherent state and the corresponding mutually unbiased basis of angular position~\cite{mirhosseini2015high}; Sit et al. \cite{sit2017high} implement a field test of OAM HD-QKD in the city of Ottawa, where 4-dimensional OAM HD-QKD was implemented and a QBER of 11\% was attained with a corresponding secret key rate of 0.65 bits per sifted photon (see Fig.~\ref{HD QKD}c). Recently, Cozzolino et al., demonstrated OAM HD-QKD over a 1.2-km-long multi-mode fiber~\cite{cozzolino2019orbital}. Different groups utilized spatial-division multiplexing optical fibres, such as multi-core fibres, to perform HD-QKD~\cite{canas2017high,ding2017high}.

Naively, one might think that since a HD-QKD system offers a higher key rate per signal than a qubit-based QKD system. It seems always better to use a HD-QKD system. One has to be very careful in making such a comparison, because key rate per signal may not be the best measure when the signal size itself is big. Key rate per second (certain period of time) can be a better merit for applications. In fact, a HD-QKD protocol uses many e.g. time-bins/modes for each signal. Now, if one were to use the many time-bins/modes separately and in parallel (with many sets of high-speed single photon detectors), one would actually get a higher key rate in such a multiplexed QKD system. At the end of the day, the private capacity per \emph{mode} of a simple prepare-and-measure QKD system is limited by fundamental bounds~\cite{takeoka2014fundamental,pirandola2017fundamental}. The key rate of HD-QKD is still limited by those fundamental bounds. Nonetheless, HD-QKD may be useful in a practical situation, where the single photon detector has long dead time or resetting time and it can not operate at high speed~\cite{zhong2015photon}. Overall, the practical advantages of HD-QKD in real-life applications remain to be seen in future.

\subsubsection{QKD with wavelength-division multiplexing}

Except for the new protocols, reducing the cost of QKD system is another important topic in the field. Wavelength-division multiplexing (WDM) technology, which enables the coexistence of QKD and telecom communication in a single fiber, is exploited to reduce the cost of the channel.

In order to protect ultra-weak QKD signals, most of previous QKD experiments are implemented in dark fibers. This implies dedicated fiber installations for QKD networks, which bears cost penalties in fiber leasing and maintenance, as well as limitations on the network scale. In classical optical communications, WDM technology has been widely exploited to increase the data bandwidth and reduce the requirement of fiber resource. Then, it is natural for QKD to coexist with classical optical communication based on WDM technology. The scheme of simultaneously transmitting QKD with conventional data was first introduced by Townsend in 1997 \cite{Townsend1997}. A series of QKD experiments integrating with various classical channels have been demonstrated \cite{Chapuran2009,Eraerds2010,Patel2012,patel2014quantum,Wang2015exp,dynes2016ultra}. Currently, by using spectral and temporal controls, state-of-the-art developments have been made to realize co-propagation of QKD with one 100 Gbps dense wavelength-division multiplexing (DWDM) data channel in 150 km ultra-low loss fiber at $-5 dBm$ launch power \cite{frohlich2017long}. By setting QKD wavelength to 1310 nm and inserting $100 GHz$ DWDM filters, Wang et al. implement QKD together with classical traffic with $11 dBm$ input power over 80 km fiber spools \cite{WANG2017}. A field trial of simultaneous QKD transmission and four 10 Gbps encrypted data channels was implemented over 26 km installed fiber at $-10 dBm$ launch power \cite{Choi:14}.

Recently, the coexistence of QKD and commercial backbone network of $3.6 Tbps$ classical data over 66 km fiber at 21 dBm launch power has been demonstrated \cite{Mao:18}. The system provides 3 kbps secure key rate with a 2.5\% quantum bit error rate. Note that in current backbone networks, the data traffic is around Tbps and the launch power is around 20 dBm. In that sense, the recent work \cite{Mao:18} demonstrate the possibility of coexistence of QKD with backbone network.

\subsubsection{Chip-based QKD} \label{sec:chipQKD}

\begin{figure*}[!ht]
\centering
\resizebox{13cm}{!}{\includegraphics{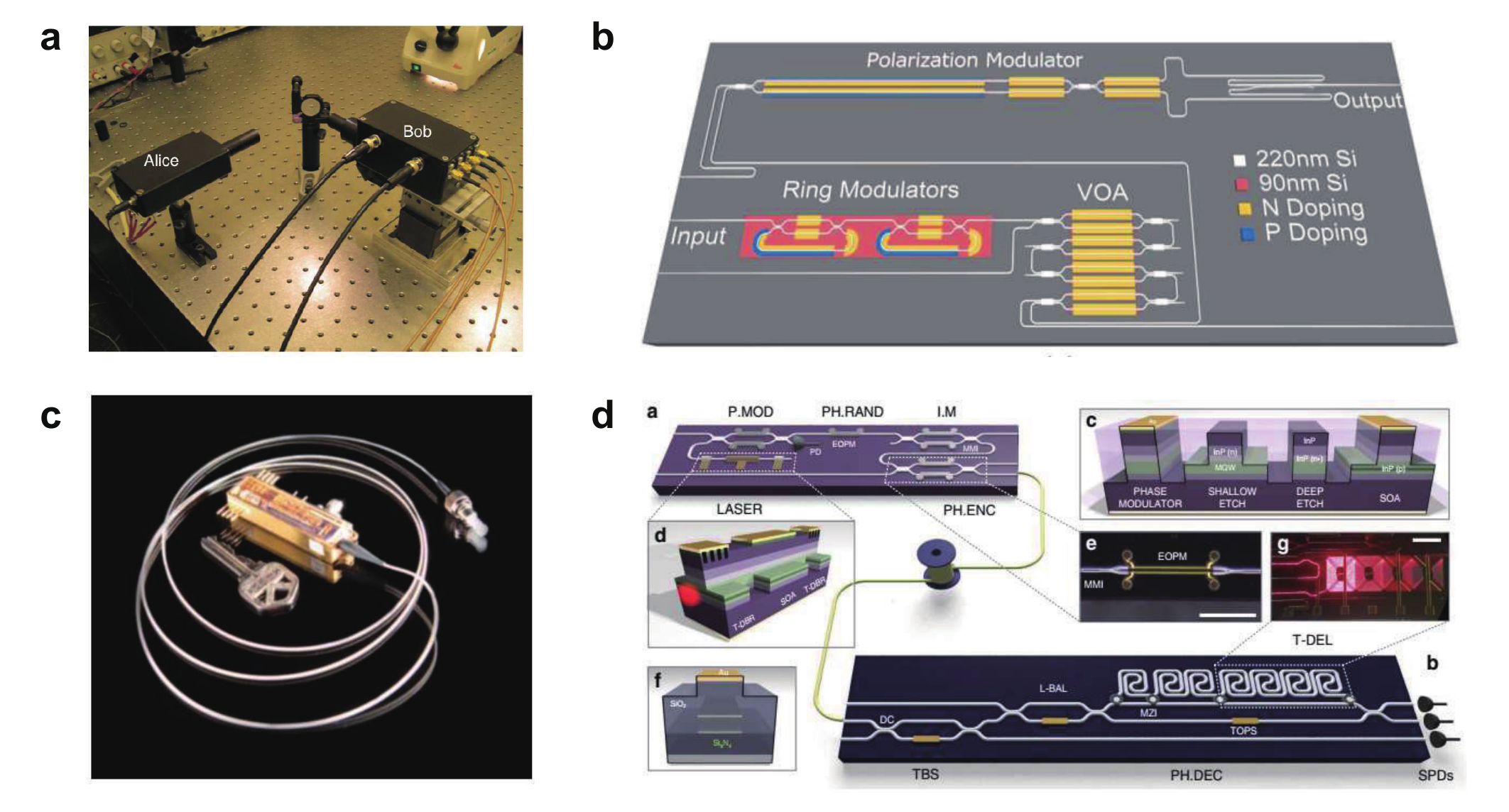}}\\
\caption{(Color online) The experimental layout for \textbf{a}, Low cost and compact QKD setup~\cite{duligall2006low}, \textbf{b}, Silicon photonic QKD emitter~\cite{ma2016silicon}, \textbf{c}, A compact QKD transmitter, QCard~\cite{hughes2013network}, \textbf{d}, InP based QKD sender and SiOxNy receiver chip~\cite{sibson2017chip}. [Figures reproduced from~\cite{duligall2006low,ma2016silicon,hughes2013network,sibson2017chip}].}
\label{Inte QKD}
\end{figure*}

Integrating QKD system has attracted more and more attention due to its advantage at compact size, low energy consumption and potential for low cost~\cite{orieux2016recent}. QKD, including optics and electronics, is a complicated system. Thus an integrated QKD system research should include integrating both optics and electronics. Fortunately, the integrated circuits (IC) is already commercialized and integrated optics is also well developed in industry. Table~\ref{Tab:chipqkd} summarizes a list of chip-based QKD experiments.

In 2005, a commercial unbalanced Mach-Zahnder interferometer made of planar lightwave circuits (PLC) based on silica-on-silicon technology was exploited for the first time in a QKD system \cite{takesue2005differential} to replace the fiber based interferometer. Comparing to its fiber counterpart, PLC interferometer is more stable and can maintain its phase for several hours without any feedback~\cite{takesue2007quantum,nambu2008quantum}. Meanwhile, IC is exploited in a research towards compact and low cost QKD system \cite{duligall2006low}. As is shown in Fig.~\ref{Inte QKD}a. Alice module uses off-the-shelf IC components in a driver circuit to control four AlInGaP LEDs to emit four polarized BB84 states. The channel is a several-meter free space link, which is supposed to find application in a future quantum based Automated Teller Machine and even in a smart phone \cite{pizzi2012affordable} according to the authors. Along this direction of research, Gwenaelle Vest \emph{et al.} demonstrated an integrated QKD sender \cite{vest2015design}, where an array of four vertical cavity surface emitting lasers (VCSELs) emit synchronized picosecond optical pulses, which are coupled to micro-polarizers generating the polarization qubits. The final size of the QKD device can be as small as $25 mm \times 2 mm \times 1 mm$, which makes the system a strong candidate for short distance free-space QKD applications.

On the metropolitan fiber network side, many individual users in the network will trust a central relay station. This is so called network-centric structure \cite{hughes2013network} or access network \cite{Bernd2013}. In such a structure, many users are all the senders and share only one central relay receiver. In that sense, the receiving station can have more space and expensive and bulky detection system can be used. Therefore, the community concentrates more integration efforts in the sending side. Hughes et al. provided a QCard~\cite{hughes2013network} in their pioneering paper, as is shown in Fig.~\ref{Inte QKD}b. The QCard has a similar size as an electro-optic modulator or a normal key. It incorporates a distributed feedback laser and modulator. The laser is attenuated into single photon level and  modulated into BB84 polarization-state with decoy state. The repetition frequency is 10 MHz at the wavelength of 1550 nm, the telecom band.

Recently, the size of the QKD sender has reduced dramatically. In 2014, Zhang \emph{et al.} put forward an on-chip LiNbO$_{3}$ polarization rotator and demonstrated the reference-frame-independent QKD protocol to overcome unstable fibre birefringence \cite{zhang2014reference}. In 2015,, the same group from University of Bristol implemented integration of QKD based on an indium phosphide transmitter chip and a silicon oxynitride receiver chip \cite{sibson2017chip}. This chip is shown in Fig.~\ref{Inte QKD}d. The authors exploited the chips in three different QKD protocols, namely BB84, coherent-one-way and differential-phase-shift QKD.

Later, researchers from University of Toronto \cite{ma2016silicon} and Bristol \cite{sibson2017integrated} exploited Silicon photonics to build QKD sender system, respectively. As shown in Fig.~\ref{Inte QKD}c, Ma et al.~\cite{ma2016silicon} fabricated the QKD sender chip with a standard Si photonic foundry process and integrated two ring modulators, a variable optical attenuator and a polarization modulator in a $1.3 mm \times 3 mm$ die area. Meanwhile, Sibson \emph{et al.} \cite{sibson2017integrated} demonstrated coherent one-way QKD, polarization encoded BB84, and time-bin encoded BB84 based on Si photonic devices. The authors achieve estimated asymptotic secret key rates of up to 916 kbps and QBER as low as 1.01\% over 20 km of fiber. The clock rate of later experiment is much higher than the former one. However, Ma et al. integrated more components on the chip, i.e., the whole QKD emitter.

Very recently, other research groups have demonstrated high-speed Si photonic chips for high dimensional QKD over multimode fiber~\cite{ding2017high}, transceiver circuit~\cite{cai2017silicon} and metropolitan QKD~\cite{bunandar2018metropolitan}. Moreover, CV-QKD is naturally suitable for photonic chip integration as its implementation is compatible with current telecom technologies (see Section~\ref{sec:CVQKD}). In particular, CV-QKD essentially uses the same devices as classical coherent communication, and only homodyne detector is required rather than the dedicated single-photon detector. Indeed, a recent experiment demonstrates Si photonic chips for CV-QKD, which integrates all the optical components (except the laser source)~\cite{zhang2019integrated}. Furthermore, based on the directly phase modulated light source~\cite{yuan2016directly,roberts2018direct}, a modulator-free QKD transmitter chip was demonstrated in~\cite{paraiso2019modulator}. This approach has the advantages that do not require conventional phase modulators, and it is versatile to accommodate several QKD protocols, such as BB84, COW and DPS, using the same optics.

\subsection{Other quantum cryptographic protocols} \label{sec:8}

So far, QKD is the most developed and mature subfield of quantum cryptography. Meanwhile, quantum cryptography has many other protocols~\cite{broadbent2016quantum}, which also have achieved quite remarkable progresses. A list of recent developments of other quantum cryptographic protocols is shown in Table~\ref{Tab5}. We will review a few examples.

\subsubsection{Quantum bit commitment} \label{sec:qbc}

Bit commitment is another important and fundamental cryptographic task that guarantees a secure commitment between two mutually mistrustful parties. Alice first commits her to a particular bit value $b$. After a period of time, Alice reveal the bit value to Bob. A success bit commitment requires that Bob can not learn $b$ before Alice reveals it, which is called concealing criterion. Meanwhile, Alice should not change $b$ once she made the commitment. This is called binding criterion. Bit commitment is a building block for many
cryptographic primitives, including coin tossing~\cite{brassard1991quantum}, zero-knowledge proofs \cite{goldwasser1989knowledge,goldreich1986proofs}, oblivious transfer \cite{bennett1992practical,unruh2010universally} and secure two-party computation \cite{kilian1988founding}.

In conventional cryptography, bit commitment is based on computational complexity assumptions similar to public key exchange protocols and might be vulnerable to quantum attacks. Unfortunately, it has been proven that information-theoretically secure bit commitment is impossible even if Alice and Bob are allowed to use quantum resources in the standard quantum circuit model by Mayers~\cite{mayers1996proceedings,mayers1997unconditionally} and by Lo and Chau~\cite{lo1997is}. Subsequently, such a no-go theorem has been further extended to case with superselection rules~\cite{kitaev2004superselection}. For an re-examination of this result, see e.g.~\cite{D'Ariano2007reexamination}. Furthermore, information-theoretic security of oblivious transfer and two-party secure computations are also proven to be impossible in~\cite{lo1997insecurity}.

Interestingly, if we take into account the signalling constraints implied by the Minkowski causality in a relativistic context, it has been shown that there are bit commitment protocols offering unconditional security \cite{kent2012unconditionally,kaniewski2013secure}. On the experimental side, two groups implemented the secure relativistic quantum bit commitment simultaneously in 2013. One followed the original protocol and utilized decoy state method in free-space channel \cite{liu2014experimental} and the other exploited a revised protocol with a plug and play system, in fiber link \cite{lunghi2013experimental}. Both experiments were secure against any quantum or classical attack. The commitment time is defined as the maximal time during which the commitment can be held. The commitment time in these two experiments, however is limited to 21 ms if all attendees are located on Earth considering the relativistic constrains. Later, new protocols with weaker security but longer commitment time was proposed \cite{lunghi2015practical,chakraborty2015arbitrarily}. A 24-hour committed experiment \cite{verbanis2016hour} was presented, which is secure against only classical attacks. Alternatively, secure quantum bit commitment can be achieved with some additional physical assumptions as, for example, that the attacker’s quantum memory is noisy \cite{ng2012experimental}.

\subsubsection{Quantum digital signature} \label{sec:QDS}
Comparing to the previous two-party protocols, digital signature has one sender and multiple recipients,  requiring that the messages cannot be forged or tampered with. Classical digital signature mainly
exploits the Rivest-Shamir-Adleman protocol \cite{rivest1978method}, the security of which is based on the mathematical complexity of the integer factorization problem. Based on the quantum physics, quantum digital signature (QDS) protocol was provided~\cite{gottesman2001quantum}, which could provided information-theoretical security \cite{martin2001introduction}. Although novel, this protocol needs nondestructive state comparison, long-time quantum memory,
and a secure quantum channel for real application. Thereafter, QDS has attracted a great deal of interest in both theory \cite{andersson2006experimentally,dunjko2014quantum,wallden2015quantum} and experiment \cite{clarke2012experimental,collins2014realization,donaldson2016experimental}. All the three requirements have been fixed sequentially \cite{clarke2012experimental,collins2014realization,donaldson2016experimental}. Later, more than 100 km QDS experiment has been demonstrated based on decoyed BB84 system \cite{yin2017experimental} and DPS QKD \cite{collins2016experimental}, which are also secure against PNS attack. Very recently, measurement-device-independent (MDI) QDS have been implemented in both lab \cite{roberts2017experimental} and field \cite{Yin2017}.

\subsubsection{Other protocols}

QKD has been assuming that the eavesdropper has unlimited power as long as it is not violated quantum physics. A protocol is said to be information-theoretically secure if it allows an adversary (e.g. an eavesdropper) to have unlimited quantum computing power as long as it does not violate quantum mechanics. As noted in Section~\ref{sec:qbc} above, information-theoretic security is not possible for quantum bit commitment, quantum oblivious transfer and two-party secure quantum computation. Naturally, restriction on adversary's power can expand the territory of quantum cryptography. Wehner et al. \cite{wehner2008cryptography} proposed one realistic assumption that quantum storage of qubits is noisy and demonstrated that an \emph{oblivious transfer} protocol is unconditionally secure for any amount of quantum-storage noise \cite{damgaard2008cryptography,konig2012unconditional}. Similar as bit commitment, oblivious transfer protocol is another primitive cryptograph protocol between two entrusted parties. The demonstration of the protocol was performed based on a modified entangled QKD system \cite{erven2014experimental}. The experiment exchanged a 1,366 bit random oblivious transfer string in 3 minutes and include a full security analysis under the noisy storage model, accounting for all experimental error rates and finite size effects.

Similar to bit commitment and oblivious transfer, a quantum protocol for \emph{coin flipping} \cite{blum1981advances} can be unconditionally secure when considering relativistic constrains. This also means that without relativistic designs, no bias coin flipping could not be unconditionally secure~\cite{lO1998why}. Nevertheless, a quantum protocol can limit the cheating probability strictly lower than $1/\sqrt{2}$ \cite{aharonov2000quantum,kitaev1999quantum}. The first experimental demonstration was provided with OAM qutrit entangled photon pairs, which shows the quantum advantage in coin flipping for the first time \cite{molina2005experimental}. As a proof of principle demonstration, this experiment does not consider the channel loss. Theoretical and experimental efforts have been attempted towards this direction \cite{nguyen2008experimental,berlin2011experimental,pappa2014experimental}. For instance, an implementation of the loss-tolerant protocol using an entangled-photon source was provided \cite{berlin2011experimental}. The secure distance was extended to 15 km with a modified plug and play system \cite{pappa2014experimental}.

Quantum \emph{data locking} \cite{divincenzo2004locking} allows one to lock information in quantum states with an exponentially shorter key, presenting an efficient solution to many resource-limited secure applications. However, the original quantum data-locking scheme may suffer from significant qubit loss. In 2013, Fawzi, Hayden, and Sen (FHS) developed a loss-tolerant quantum data-locking scheme \cite{Fawzi2013low}, in which the possible information leakage can be made arbitrarily small in a lossy environment while the unlocked information is significantly larger than the key size. This feature makes the protocol attractive also in secure communication \cite{seth2013quantum,lupo2014robust}. Two groups have implemented the loss tolerant protocols respectively \cite{liu2016experimental,lum2016quantum}.

Quantum \emph{secret sharing} was proposed to share a secret quantum state among multiple parties~\cite{cleve1999share} or to use quantum states to share classical secrets~\cite{hillery1999quantum,cleve1999share}. Moreover, secure multi-party computing has been extended to quantum computation with quantum inputs and circuits~\cite{crepeau2002secure}.

In \emph{distributed quantum computing}, quantum crypto protocols are still inevitable. Quantum computing is currently attracting tremendous interest from both academic and industry~\cite{mohseni2017commercialize}. However, due to its implementation complexity and cost, the future path of quantum computation is strongly believed to delegate computational tasks to powerful quantum servers on cloud~\cite{fitzsimons2017private}. Universal blind quantum computing (UBQC) \cite{broadbent2009universal} is an effective method for a common user, who has limited or no quantum computational power, to delegate computation to an \emph{untrusted} quantum server, without leaking any information about the user's input and computational task. The security or blindness of the UBQC protocol is unconditional, i.e., the server cannot learn anything about user's computation except its size. A proof of concept demonstration was reported in 2012~\cite{barz2012demonstration}. Recently, UBQC protocol with completely classical clients was proposed \cite{reichardt2013classical} and demonstrated in experiment~\cite{huang2017experimental}. UBQC with weak coherent states was proposed in~\cite{dunjko2012blind}, and adding the ingredient of decoy states, an efficient experimental demonstration with weak coherent states was made in~\cite{jiang2019remote}. Because of the developments in the field of quantum computing, we expect that BQC will play an important role in the future infrastructure of delegated quantum computation~\cite{fitzsimons2017private}.

\section{Concluding Remarks}\label{sec:9}

In this review, we have discussed the security aspects of practical QKD. These range from the security proofs of practical QKD (Section~\ref{Sc:Security}), the implementation (Section~\ref{Sc:Implement}), the practical vulnerabilities (Section~\ref{sec:3}), to the solutions of advanced QKD protocols (Sections~\ref{sec:4},~\ref{sec:5} and~\ref{sec:CVQKD}) and the advances of other quantum cryptographic protocols (Section~\ref{sec:8}).

Historically, QKD has been a concrete playground for concepts in quantum mechanics. The study of QKD often leads to unexpected insights in other areas of quantum information. For instance, apparently the concept of quantum teleportation was invented during a search for a security proof of QKD~\cite{bennett1993teleporting}. We expect that in future the study of QKD will continue to lead to many new insights in other subfields of quantum information.

Meanwhile, as a new technology stemming from the counterintuitive theory of quantum physics, QKD might not be easy to be understood and recognized by a general audience. For broad interest, in Appendix~\ref{app:general}, we summarize a few frequently asked questions/concerns on practical QKD, together with our views on how they can be overcome. Finally, we discuss the perspectives on the past, the present and the future about the developments of QKD.

Overall, during the past three decades, the theory and practice of QKD have developed extensively. These developments can be divided into several stages, which can be summarized as follows (with a focus on DV-QKD).

\begin{enumerate}
  \item
  \emph{Stage 1.} After the invention by Bennett-Brassard~\cite{bennett1984quantum} and Ekert~\cite{ekert1991quantum}, QKD was first demonstrated in the early 1990s~\cite{bennett1992experimental}, which started a series of theories and experiments~\cite{Townsend1993,doubleMZI:1994,Franson1995,muller1996quantum} to prove the possibility of QKD.

  \item
  \emph{Stage 2.} The implementation of QKD was extended from laboratory to outdoor environments, and various technical difficulties were studied~\cite{Townsend1997,buttler1998practical,Hughes2000quantum,ribordy2000fast,gobby2004quantum}. See ref.~\cite{gisin2002quantum} for a review on the developments in early experiments. Meanwhile, on the theory side, the security proof of QKD was a major challenge until a few papers appeared and solved the problem~\cite{mayers2001unconditional,lo1999unconditional,shor2000simple,biham2000proof}. These results put the security of QKD on a solid foundation.

   \item
  \emph{Stage 3.} With the security proofs for QKD under imperfect devices~\cite{gottesman2004security,hwang2003quantum,lo2005decoy,wang2005beating}, the feasibility of QKD was demonstrated from short range to long range, up to the scale of 100-km standard fiber~\cite{Yi2006,rosenberg2007long,Peng2007} and free space~\cite{Tobias2007}.

  \item
  \emph{Stage 4.} QKD was extensively deployed from point-to-point to small-scale metropolitan networks in field~\cite{elliott2005current,chen2009field,peev2009secoqc,sasaki2011field}. Meanwhile, the practical security loopholes, particularly those for detection devices, were identified~\cite{Lars2010} and then removed by the advanced MDI-QKD protocol~\cite{lo2012measurement} [see also~\cite{braunstein2012side}].

  \item
  \emph{Stage 5.} The feasibility of QKD was extended to long distances and high rates, such as in a scale of 400 km~\cite{Yin2016,boaron2018secure} over ultra low-loss fiber and 1200-km over free space~\cite{Liaosate}, and a secret key rate of over 10 Mbits/s with GHz QKD system~\cite{Yuan:18}.

  \item
  \emph{Stage 6.} QKD was implemented from small scale to large scale that covers a wide area~\cite{Yuao2019}. See Fig.~\ref{Fig:backbone} for an example of the QKD network which has more than 700 QKD links, and covers more than 2,000~km area. New TF-QKD protocols~\cite{Lucamarini2018TF} were proposed to enable secure QKD over even longer distances~\cite{Fang2019surpassing,chen2019sending}.
\end{enumerate}

In future, towards the ultimate goal of a global QKD network, we expect that more and more QKD networks will be built in different countries. Besides physics, communities of computer science, engineering, optics, mathematics and so forth may work together to realize this goal. We do believe that a revolutionized global QKD network for secure communication stemming from quantum physics will be deployed and find widespread applications in the near future. This review is concluded with a discussion on a few directions for future research.

\begin{enumerate}
  \item
  \emph{Quantum repeaters.}
  Quantum repeater can achieve an effective restoration of the quantum information without resorting to a direct measurement of the quantum state~\cite{BDCZ1998,duan2001}, enabling the realization of global quantum network in existing optical networks. Quantum repeater has received intense research efforts in recent years~\cite{kimble2008quantum,sangouard2011quantum,pan2012multiphoton,wehner2018quantum}. Nonetheless, the limited performance of quantum memory is still a major obstacle in realizing practical quantum repeaters without a future experimental breakthrough \cite{sangouard2011quantum,Yang2016nphoton}. New recent approaches manage to reduce the need for a quantum memory by using all-photonic quantum repeaters~\cite{azuma2015all,hasegawa2019experimental,li2019experimental}, but they require the resources of large-scale cluster states. Overall, we believe that quantum repeater is an important subject for future research. The first goal is to realize a practical quantum repeater that can beat the fundamental limits of direct quantum communication~\cite{takeoka2014fundamental,pirandola2017fundamental}.

  \item
  \emph{Standardization.}
  Towards the widespread applications, the commercial standards for QKD should be established. Important progress has been made in this direction, such as the efforts of ETSI, ISO/IEC, CCSA and ITU by several countries. One important direction is to include the practical security into the standardization process, by defining the best practices to operate QKD systems and standardizing those countermeasures to guarantee the security of a QKD setup. We encourage future research to establish the commercial standards for QKD.

  \item
  \emph{Battle-testing security.}
  We have provided a review on the practical vulnerabilities in Section~\ref{sec:3}, together with the solutions of advanced countermeasures and QKD protocols. However, the practical security issue has not been perfectly solved. For instance, as discussed in Section~\ref{sec:5:mdi}, a security assumption in MDI-QKD is that the source should be trusted without loopholes. It is important to verify this assumption in practice. Hence, the research in analyzing the practical security of QKD setup should continue. This includes the developments of practically-secure QKD systems building on the experience gained from the research on practical vulnerabilities and advanced countermeasures. It is highly important to battle-test existing QKD implementations, quantify and validate the security claims of real-world QKD systems, and design real-life QKD systems with testable security assumptions.

  \item
  \emph{Small-size, low-cost, long-distance system.}
  Recent developments of integrated QKD system have been reviewed in Section~\ref{sec:chipQKD}. These developments should continue to further reduce the costs and sizes of QKD, and to realize robust fully-integrated chip-based QKD systems. One important direction is to develop the star-type quantum access network~\cite{hughes2013network,Bernd2013}, in which the expensive devices such as single-photon detectors can be placed in the central relay and many users share this relay. Each user requires only a low-cost transmitter such as a compact QCard \cite{hughes2013network} or a simple Si chip~\cite{ma2016silicon,sibson2017chip}. Together with MDI-QKD, the central relay can be \emph{untrusted}. Wei et al., already implemented the first chip-based MDI-QKD at high secret key rates~\cite{wei2019high}. This is particularly valuable for star-type metropolitan QKD networks. Moreover, by using the new type of twin-field QKD~\cite{Lucamarini2018TF}, the distance can be further extended for intercity QKD. Therefore, we expect that MDI-type QKD networks will play an important role in the future global quantum network.

  \item
  \emph{QKD network with untrusted relays.}
  The previously deployed networks were based on trusted relays~\cite{elliott2005current,chen2009field,peev2009secoqc,sasaki2011field,Yuao2019}, which may raise the concern about the security properties of the relays. To remove this concern, it is important to develop QKD network with untrusted relays. In fact, MDI-QKD is naturally suitable for a star-type metropolitan network with an untrusted relay. Tang et al., already put forward the first implementation of a MDI-QKD network~\cite{Yanlin2016}. We expect that metropolitan MDI-QKD networks will be built soon. Besides, the TF-QKD can also be adopted to extend the transmission distance with an untrusted relay. Moreover, in entanglement-based QKD, the relay can be fully untrusted. A possible direction is to develop a entanglement-based QKD network, e.g., based on satellite~\cite{Yin2019}. For ultralong-distance QKD in fiber, it needs the quantum repeaters~\cite{sangouard2011quantum} to realize QKD networks with untrusted relays. We expect that with the technical improvements, quantum-repeater assisted QKD network may be achieved in the near future.

  \item
  \emph{Satellite-based QKD.}
  The reported satellite-based QKD was based on a low-earth-orbit (LEO) satellite of Micius~\cite{Liaosate,liao2018satellite}. To increase the coverage time and area for a more efficient satellite-based QKD network, one can launch higher-orbit quantum satellites and implement QKD in daytime. Important progress has been made in this direction~\cite{hughes2002practical,liao2017long}. An ultimate goal is to realize a satellite-constellation-based global quantum network.
\end{enumerate}

\section*{Acknowledgement}
We thank enlightening discussions with and helpful comments from the reviewers and the numerous colleagues including, K. Azuma, D. Bacco, C. H. Bennett, Y.-A. Chen, A. Huang, D.~Huang, M. Lucamarini, Y.~Liu, S. Pirandola, B. Qi, K. Tamaki, K.~Wei, G. B. Xavier, Z. Yuan P.~Zeng and H.~Zhou. F.~Xu, X.~Ma, Q.~Zhang and J.-W.~Pan were supported by the National Key Research and Development (R\&D) Plan of China, the National Natural Science Foundation of China, the Anhui Initiative in Quantum Information Technologies and the Chinese Academy of Sciences. H.-K. Lo was supported by NSERC, US Office of Naval Research, CFI, ORF, MITACS Accelerate, Royal Bank of Canada (RBC) and Huawei Canada.

\appendix

\section{General questions to QKD} \label{app:general}

We summarize a few frequently asked concerns on QKD and \emph{our} views on how they can be overcome.

\begin{enumerate}
  \item
  \emph{Concern 1.} Since RSA is secure under current computational power, we do not need QKD now.

  \emph{Our view.} Some important data such as government secrets and health data need to kept secret from decades, i.e., \emph{long-term security}. RSA cannot guarantee long-term security, because one can record the encrypted information and later on decrypt it when the quantum computer comes up or new advanced algorithm is discovered. In contrast, QKD can provide everlasting security, which is independent of all future hardware advances. Hence, QKD is required today for the transmission of top-secret data.

  \item
  \emph{Concern 2.} QKD vs post-quantum cryptography.

  \emph{Our view.} QKD and post-quantum cryptography are two parallel research directions. They go hand in hand with each other. It is not an ``either-or" situation. Post-quantum cryptography has the advantages of being compatible with existing crypto infrastructure, but it has the drawback that its security cannot be proven or it is only secure against known quantum attacks. In contrast, QKD has the advantage of proven security based on the laws of quantum physics, but it is a symmetric-key algorithm, which can not replicate all the functionalities of public-key cryptography. In future, we believe that QKD is likely to be combined with the post-quantum cryptography to jointly form the infrastructure of quantum-safe encryption scheme.

  \item
  \emph{Concern 3.} QKD does not address large parts of the security problem.

  \emph{Our view.} The secure keys generated from QKD have widespread applications, such as encryption and authentication. Note that in QKD, authentication is only required in a short period, and once it is done, QKD can be employed for encryption in a rather long period\footnote{As an example, one can even use public key based authentication scheme in the initial authentication of a QKD session. Provided that the public key based authentication scheme is secure during the short time for initial authentication, the generated QKD key will be secure forever. Therefore, post-quantum cryptography and QKD may go hand in hand.}. Moreover, with the developments on high key-generation rate, QKD is also suitable for some of the future challenges such as securing the Internet of Things, big data, or cloud applications. Furthermore, as mentioned in Section~\ref{sec:QDS}, there exists quantum digital signature schemes with information-theoretical security.

  \item
  \emph{Concern 4.} Distance limitation.

  \emph{Our view.} In fiber, even without quantum repeater, the feasibility of QKD has been proved in experiments over long ranges of 400-500 km~\cite{Yin2016,boaron2018secure,Fang2019surpassing,chen2019sending}. Using trusted relays, the distance has been extend to 2000 km fiber~\cite{Yuao2019}. Using quantum satellite, QKD has been demonstrated 7600-km~\cite{liao2018satellite}. Moreover, with the help of quantum repeaters~\cite{BDCZ1998,duan2001}, QKD is feasible over arbitrarily long distance even with untrusted relay nodes. Important progress has been made in the developments of quantum repeaters \cite{pan2012multiphoton,munro2015inside}.

  \item
  \emph{Concern 5.} Cost limitation.

  \emph{Our view.}  The recent developments of integrated QKD, such as compact transmitter \cite{hughes2013network} and Si photopic chip-based QKD systems\cite{ma2016silicon,sibson2017chip,wei2019high}, demonstrated already the possibility of low-cost hardware for QKD. Hence QKD is very likely to be cost-effective. See Section~\ref{sec:chipQKD} for detail.

  \item
  \emph{Concern 6.} Point-to-point limitation.

  \emph{Our view.} Small-scale metropolitan QKD networks were intensively deployed in field by several countries~\cite{elliott2005current,chen2009field,peev2009secoqc,sasaki2011field}. A large-scale network which covers a wide area was established lately~\cite{Yuao2019}. These networks already enable secure QKD for multiple users instead of point-to-point. Furthermore, the recent discovery of MDI-QKD protocols~\cite{lo2012measurement} and TF-QKD protocols~\cite{Lucamarini2018TF} work well in a star-type network setting~\cite{xu2015discrete} by sharing a single detection system between multiple users. A prototype of MDI-QKD network has already been implemented in 2016~\cite{Yanlin2016}. Therefore, these QKD networks and advanced QKD protocols enable QKD for network settings beyond point-to-point.

  \item
  \emph{Concern 7.} Trusted-relay limitation.

  \emph{Our view.} The discovery of MDI-QKD protocols~\cite{lo2012measurement} and TF-QKD protocols~\cite{Lucamarini2018TF} enable QKD with \emph{untrusted} relays. Moreover, entanglement-based QKD works well with untrusted relays, and it has been demonstrated between two ground stations separated by a distance over 1120 kilometers~\cite{Yin2019}. Furthermore, quantum repeaters~\cite{BDCZ1998,duan2001} enable secure QKD over arbitrarily long distance even with untrusted relay nodes. Hence trusted node is not a true limitation in QKD.

  \item
  \emph{Concern 8.} Hardware patches are expensive.

  \emph{Our view.} MDI-QKD already enables secure QKD with untrusted measurement devices, in which the expensive measurement devices do not need to be recalled/replaced once they are installed. Moreover, chip-based QKD makes the patches for hardware at a low-cost and simple manner. We believe that a star-type of MDI-QKD network, together with chip-based transmitter, is promising to realize a low-cost and practical QKD for applications.

  \item
  \emph{Concern 9.} Security loopholes in practical QKD.

  \emph{Our view.}  Researchers in the filed of QKD have extensively understood and managed the security loopholes. All quantum attacks reported in the literature have been reviewed in Section~\ref{sec:3}. Those crucial loopholes have been eliminated by designing advanced countermeasures (Sections~\ref{sec:4},~\ref{sec:5} and~\ref{sec:CVQKD}). In particular, MDI-QKD has removed the weakest security link, i.e., the detection, in a standard QKD system~\cite{lo2012measurement}. Secret sharing ideas have been proposed to foil covert channels and malicious classical post-processing units~\cite{curty2017quantum}. Advanced technology in the future might make DI-QKD feasible~\cite{hensen2015loophole}. Therefore, the gap between theory and practice of QKD has been reduced remarkably, and a number of loopholes have been completely removed. These achievements have made QKD a robust solution for secure communication.

  \item
  \emph{Concern 10.} Denial of service (DoS) attack.

  \emph{Our view.}  One solution for DoS attack is to use alternative channel links by designing suitable network architectures. For instance, a circle type of QKD network has been implemented in the Beijing metropolitan network (see Fig.~\ref{Fig:backbone}). Moreover, Tokyo~\cite{sasaki2011field} and SECOQC~\cite{peev2009secoqc} QKD networks have demonstrated robustness against DoS attacks already. The other solution is that the secure communication can be done offline. One can load the secret keys, generated from QKD, to USB or mobile phones. The secure communication via mobile phone will be immune to DoS attack. This method already finds commercial use, see e.g., the QUKey\footnote{www.quantum-info.com/English/product/2017/1007/394.html}.
\end{enumerate}


%

\end{document}